\documentclass[12pt]{iopartn}

\usepackage{url, hyperref}
\usepackage{color,soul}

\usepackage{graphicx}
\usepackage{subfigure}
\usepackage{mathtools}
\usepackage{natbib}
\usepackage{cleveref}

\usepackage{iopams}
\usepackage{dcolumn}
\usepackage{bm}
\usepackage[latin1]{inputenc}
\usepackage[T1]{fontenc}
\usepackage{rotating}
\usepackage{tikz} 
\usetikzlibrary{shapes.geometric, arrows}
\usepackage[T1]{fontenc}
\usetikzlibrary{shapes,arrows,calc,fit,backgrounds}
\usetikzlibrary{positioning,decorations.pathreplacing}
\usepackage{tikzpagenodes}
\newcommand{\newblock}{}
\newcommand{\apj}{ApJ}
\newcommand{\prd}{Phys. Rev. D}
\newcommand{\prl}{Phys. Rev. L}
\newcommand{\nat}{Nature}
\usepackage{citesort}

\begin{document}

\title[Solutions for bound trajectories around a Kerr black hole]{Astrophysically relevant bound trajectories around a Kerr black hole}

\author{Prerna Rana$^{1,2}$ and A. Mangalam$^{1,3}$}

\address{$^{1}$Indian Institute of Astrophysics, Sarjapur Road, 2nd block Koramangala, Bangalore, 560034, India}
\ead{prernarana@iiap.res.in$^{2}$, mangalam@iiap.res.in$^{3}$}
\vspace{10pt}
\date{\today}

\begin{abstract}

We derive alternate and new closed-form analytic solutions for the non-equatorial eccentric bound trajectories, $\{ \phi \left( r, \theta \right)$, $\ t \left( r, \theta \right),\ r \left( \theta \right) \}$, around a Kerr black hole by using the transformation $1/r=\mu \left(1+ e \cos \chi \right)$. The application of the solutions is straightforward and numerically fast. We obtain and implement translation relations between energy and angular momentum of the particle, ($E$, $L$), and eccentricity and inverse-latus rectum, ($e$, $\mu$), for a given spin, $a$, and Carter's constant, $Q$, to write the trajectory  completely in the ($e$, $\mu$, $a$, $Q$) parameter space. The bound orbit conditions are obtained and implemented to select the allowed combination of parameters ($e$, $\mu$, $a$, $Q$). We also derive specialized formulae for equatorial, spherical and separatrix orbits. A study of the non-equatorial analog of the previously studied equatorial separatrix orbits is carried out where a homoclinic orbit asymptotes to an energetically bound spherical orbit. Such orbits simultaneously represent an eccentric orbit and an unstable spherical orbit, both of which share the same $E$ and $L$ values. We present exact expressions for $e$ and $\mu$ as functions of the radius of the corresponding unstable spherical orbit, $r_s$, $a$, and $Q$, and their trajectories, for ($Q\neq0$) separatrix orbits; they are shown to reduce to the equatorial case. These formulae have applications to study the gravitational waveforms from extreme-mass ratio inspirals (EMRIs) using adiabatic progression of a sequence of Kerr geodesics, besides relativistic precession and phase space explorations.  We obtain closed-form expressions of the fundamental frequencies of non-equatorial eccentric trajectories that are equivalent to the previously obtained quadrature forms and also numerically match with the equivalent formulae previously derived. We sketch non-equatorial eccentric, separatrix, zoom-whirl, and spherical orbits, and discuss their astrophysical applications.

\end{abstract}


\submitto{\CQG}

\section{Introduction}
It has now been established with observational evidence that black holes with masses ranging from $4$-$20 M_{\odot}$ in X-ray binaries, to $10^{6}$-$10^{9} M_{\odot}$ in galactic nuclei, are ubiquitous. One among such important evidences is the recent detection of gravitational waves from the black hole binary merger \citep{Abbott2016} and more such events are awaited to be detected by the planned LISA mission \citep{Glampedakis2005}. One of the major objectives of the LISA mission is the detection of gravitational waves from EMRIs, most probably to be sourced from the compact objects spiralling and finally plunging onto the super-massive black hole (SMBH) in galactic nuclei. The dynamics of EMRIs is widely accepted as representative of test-particle motion, evolving adiabatically, in the spacetime of a rotating black hole. Understanding of such strong gravity regimes involves them using the Kerr metric \cite{Kerr1963}, which is a vacuum solution of Einstein's equation for a rotating black hole.  The study of trajectories around the black holes is critical to our understanding of the physical processes and their observational consequences \cite{Narayan2005}.

The trajectories in Kerr and Schwarzschild \cite{Schwarzschild} geometries have been studied extensively. Some of these results are covered in a pioneering work by \cite{C1983book} in an elegant manner. The key idea that the general trajectory in Kerr background can be expressed in terms of quadratures, was first given in \cite{Carter1968}. In \cite{Bardeen1972}, the energy, $E$, and angular momentum, $L$, were expressed in terms of the circular orbit radius and the spin parameter $a$; the specific solution for the radius of the innermost stable circular orbit (ISCO) was also derived. The necessary conditions for bound geodesics for spherical orbits  and the dragging of nodes along the direction of spin of a black hole was discussed \cite{Wilkins1972}. The formulae have proved to be extremely useful in predicting observables in astrophysical applications like accretion around the black holes. For example, a general solution for a star in orbit around a rotating black hole was expressed in terms of quadratures \cite{Vokrouhlicky1993} using the formulation given by \cite{Carter1968}; the resulting integrals have been calculated numerically. The general expression in terms of quadratures for fundamental orbital frequencies $\nu_{\theta}$, $\nu_{\phi}$ and $\nu_{r}$, for a general eccentric orbit, were first derived by \cite{Schmidt2002}, where different cases for circular and equatorial orbits are also discussed but complete analytic trajectories were not calculated. An exact solution for non-spherical polar trajectories in Kerr geometry and an exact analytic expression for $t(r)$ for eccentric orbits in the equatorial plane were derived \cite{Kraniotis2007}. These were used to obtain the expressions for the periapsis advance and Lense-Thirring frequencies.  The time-like geodesics were expressed in terms of quadratures involving hyper-elliptic, elliptic and Abelian integrals for Kerr and Kerr-(anti) de Sitter spacetimes including cosmological constant \cite{Kraniotis2004} and applied in a semi-analytic treatment of Lense Thirring effect.

A time-like orbital parameter $\lambda$ called Mino time \cite{Mino2003} was introduced to decouple the $r$ and $\theta$ equations, which was then used to express a wider class of trajectory functions in terms of the orbital frequencies $\nu_{\theta}$, $\nu_{\phi}$ and $\nu_{r}$ \cite{Drasco2004}. These methods are applied to calculate closed-form solutions of the trajectories and their orbital frequencies \cite{Fujita2009}, using the roots of the effective potential. However, the solutions are expressed in terms of Mino time. The commensurability of radial and longitudinal frequencies, their resonance conditions for orbits in Kerr geometry, and their location in terms of spin and orbital parameter values were studied using numerical implementations of Carlson's elliptic integrals \cite{Brink2015}. Considering the problem of the precession of a test gyroscope in the equatorial plane of Kerr geometry, the analytic expressions to transform energy, angular momentum of the orbiting test particle, and spin of the black hole ($E$, $L$, $a$) to eccentricity, inverse-latus rectum of the bound orbit ($e$, $\mu$, $a$) parameters were obtained \cite{Bini2016a}. The expressions for radial and orbital frequencies are obtained to the order $e^{2}$ for bound orbits and analytically for the marginally bound homoclinic orbits \cite{Bini2016b}. The dynamical studies of an important family of Kerr orbits called separatrix or homoclinic orbits are important for computing the transition of spiralling to plunge in EMRIs emitting gravitational waves \citep{Levin2009,Glampedakis2002}. The test particles (compact objects in this case) transit through an eccentric separatrix orbit in EMRIs, while progressing adiabatically, before they merge with the massive black hole.

This paper is an expanded version of the published article \citep{RMCQG2019}. In this paper, we study the generic bound trajectories, which are eccentric and inclined, around a Kerr black hole, and then we investigate the non- equatorial separatrix orbits as a special case. We have solved the equations of motion and produce alternate and new closed-form solutions for the trajectories in terms of elliptic integrals without using Mino time, $\{\phi \left( r, \theta\right) , \ t \left( r, \theta\right), r\left(\theta \right) \ $ $\mathrm{or} \ \theta\left(r \right) \}$, which makes them numerically faster. We also implement the essential bound orbit conditions to choose the parameters ($e$, $\mu$, $a$, $Q$) of an allowed bound orbit, derived from the essential conditions on the parameters for the elliptic integrals involved in the trajectory solutions. We find that the $e-\mu$ space is more convenient for integrating the equations of motion as the turning points of the integrands are naturally specified in terms of the bound orbit conditions. The  exact solutions for the trajectories are found in terms of not overly long expressions involving elliptic functions. We implement the translation formulae between ($E$, $L$) and ($e$, $\mu$) parameters that help us to express the trajectory solutions completely in the ($e$, $\mu$, $a$, $Q$) parameter space which we call the conic parameter space. We then study the case of non-equatorial separatrix trajectories in the conic parameter space. First, we write the essential equations for the important radii like innermost stable spherical orbit ($ISSO$),  marginally bound spherical orbit ($MBSO$), and spherical light radius. Similar to the equatorial separatrix orbits, the non-equatorial separatrix or homoclinic trajectories asymptote to an energetically bound unstable spherical orbit, where the spherical orbit radius can vary between $MBSO$ and $ISSO$. We show that the formulae for ($e$, $\mu$) for the non-equatorial separatrix orbits can be expressed as functions of the radius of the corresponding spherical orbit, $r_s$, $a$, and $Q$, which also reduce to their equatorial counterpart \cite{Levin2009} by implementing the limit $Q\rightarrow 0$. These formulae are obtained by using the expressions of $E$ and $L$ for the spherical orbits. Next, we derive the exact solutions for the non-equatorial separatrix trajectories by reducing our general eccentric trajectory solutions to this case. These solutions are important for investigating the behaviour of gravitational waveforms emitted by inspiralling and inclined test objects near non-equatorial separatrix trajectories in the case of EMRIs.

 The ab-initio specification of the allowed
geometry of bound orbits in the parameter space is crucial for the calculation of the orbital trajectories and its frequencies. These criteria are used in building, studying and sketching different types of trajectories around a Kerr black hole: for instance, spherical, non-equatorial eccentric, non-equatorial separatrix and zoom-whirl orbits, using our closed-form expressions for trajectories are constructed. We also derive closed-form analytic expressions for the fundamental frequencies of the general non-equatorial trajectories as functions of elliptic integrals around the Kerr black holes. We use a time-averaging method on the first-order equations of motion to derive these frequencies and show that our closed-form analytic expressions of frequencies match with the formulae given by \cite{Schmidt2002} which were left in quadrature forms. We also reduce the general forms to the equatorial case, which is also a new form that is easier to implement and faster by a factor of $\sim 20$.

This paper is organized as follows (see Fig. \ref{concepts}): in \S \ref{inofmttn}, we review the basic equations describing $\{ r, \ \theta, \ \phi, \ t \}$ motion around the Kerr black hole using Hamiltonian dynamics. In \S \ref{trans}, we write the translation formulae from $(e, \ \mu, \, a, \ Q)$ to $(E, \ L, \ a, \ Q)$ parameter space. In \S \ref{analyticsoln}, we derive the exact closed-form solutions for the trajectories by solving all involving integrals and writing them in terms of elliptic integrals, $\{\phi \left( r, \theta\right) , \ t \left( r, \theta\right), r\left(\theta \right) \ $ $\mathrm{or} \ \theta\left(r \right) \}$. In \S \ref{bndcnd}, we give the essential bound orbit conditions on $(e, \ \mu, \, a, \ Q)$ parameters applicable to the astrophysical situations. In \S \ref{equatorialtrajec}, we reduce the analytic solutions to the case of equatorial plane. In \S \ref{ELforsphorbits}, we derive the formulae for $E$ and $L$ for spherical orbits as a function of radius $r_s$, $a$, and $Q$. In \S \ref{emuseparatrix}, we write the equations for the radii $ISSO$, $MBSO$, and spherical light radius. We then derive the exact expressions for $e$ and $\mu$ for the non-equatorial separatrix trajectories. In \S \ref{separatrixtrajec}, we derive the trajectory solutions for the non-equatorial separatrix orbits. In \S \ref{trajec}, we sketch and discuss various bound trajectories around the Kerr black hole. In \S \ref{frequencies}, we derive the closed-form expressions of the fundamental frequencies in terms of elliptic integrals by the long time averaging method without using Mino time. In \S \ref{reduction}, we conduct consistency checks by reducing the separatrix trajectories to the equatorial case, and also match the azimuthal to polar frequency ratio, $\nu_{\phi}/ \nu_{\theta}$, with the spherical orbits case derived by \cite{Wilkins1972}. We discuss possible applications of our trajectory solutions and frequency formulae in \S \ref{app}. We summarize and conclude our results in \S \ref{summary} and \S \ref{disc} respectively. In Table \ref{gloss} a glossary of symbols is provided.

\begin{table}
\large
\begin{center}

\scalebox{0.62}{
\begin{tabular}{|c l | c l|}
\hline
{\bf Boyer Lindquist coordinates } & & & \\
\hline
$t$& Time coordinate & $r$ & Radial distance from the black hole\\
$\theta$ & Polar angle & $\phi$& Azimuthal angle\\
 $\rho^{2}$ & $r^{2}+a^{2}cos^{2}\theta$ & $a$ & Spin of the black hole \\
 \hline
  {\bf Common physical parameters } &&& \\
\hline
$u$& $\frac{1}{r}$  & $\tau$ & Proper time \\
$r_{+}$& Horizon radius  & $Q$& Carter's constant \\
  $E$ & Energy per unit rest mass of the test particle & $L$ & z component of Angular momentum per unit  \\
  & & & rest mass of the test particle \\
  $p_t$ & Generalized momentum for $t$ coordinate & $p_\phi$ & Generalized momentum for $\phi$ coordinate \\
  $p_r$ & Generalized momentum for $r$ coordinate & $m_0$ & =0 for photon orbits and $=1$ for particle orbits \\
$V_{eff}$ & Radial effective potential for an eccentric test & $\mathcal{H}$ &Relativistic Hamiltonian for the geodesic motion\\
  & particle trajectory &  & \\
$r_1$ & apastron distance ($=r_a$) & $r_2$ & periastron distance ($=r_p$)\\
$r_3$ & Third turning point of the test particle & $r_4$ & Innermost turning point of the test particle  \\
 $e$ & eccentricity parameter & $\mu$ & inverse latus-rectum parameter \\
\hline
{\bf Integrals of motion } & &&\\
\hline
$\chi$ & defined by $u=\mu\left(1+ e \cos \chi \right) $ & $\psi$ & $\frac{\chi}{2}-\frac{\pi}{2}$ \\
$y$ & $1+ e \cos \chi $ & $I$ & Terminology used for radial integrals \\
$H$ & Terminology used for $\theta$ integrals & & \\
\hline
{\bf Spherical and separatrix orbits} & &&\\
\hline
 $r_s$ & radius of spherical orbit & $r_c$ & radius of circular orbit \\
$e_s$ & eccentricity of the separatrix orbits & $\mu_s$ & inverse latus-rectum of the separatrix orbits\\
$Z$ & ISCO radius & $X$& Light radius \\
\hline
{\bf Fundamental frequencies} & &&\\
\hline
$\nu_{\phi}$ & Azimuthal frequency & $\nu_{r}$ & Radial frequency  \\
$\nu_{\theta}$ & Vertical oscillation frequency & & \\
\hline
\end{tabular}
}
\end{center}
\caption{Glossary of symbols used.}
\label{gloss}
\end{table}
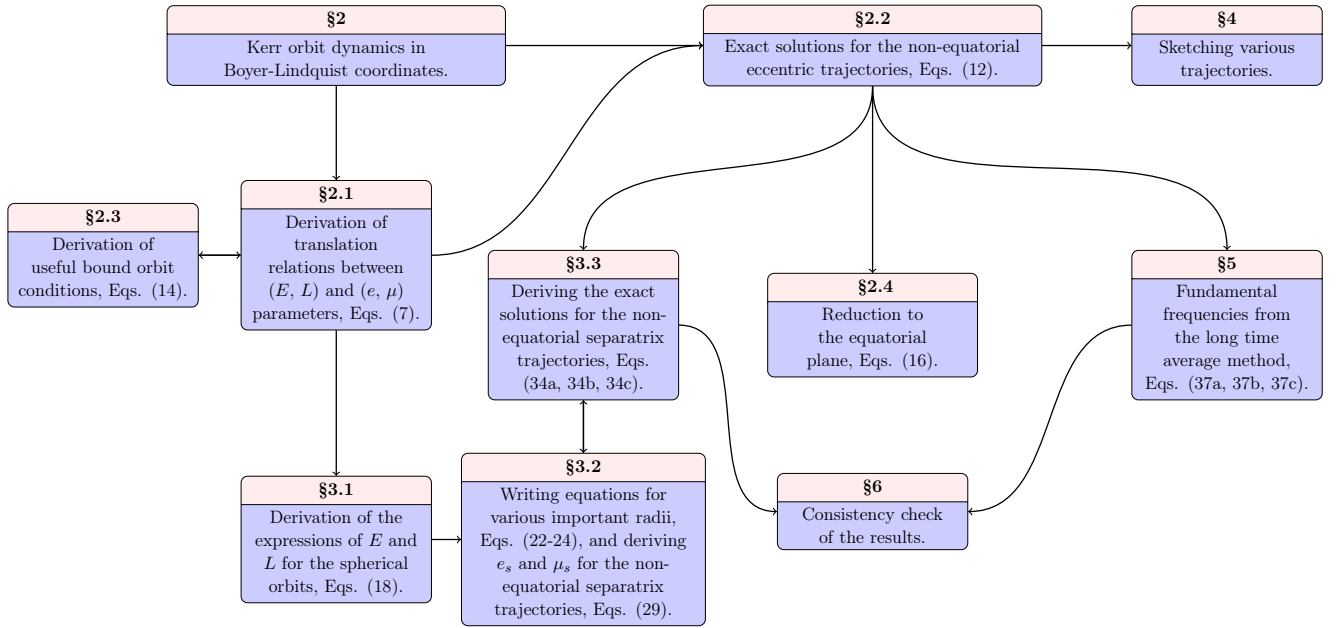
\begin{figure}
\scalebox{0.62}{
\tikzstyle{decision} = [diamond, draw,  
    text width=6em, text badly centered, node distance=3cm, inner sep=0pt]
\tikzstyle{block} = [rectangle, draw, 
 minimum width=4.5cm, minimum height=1cm,  text width=10em, text centered, rounded corners]
\tikzstyle{line} = [draw, -latex']
\tikzstyle{process} = [rectangle, draw, minimum width=3cm, minimum height=1cm,   text width=10em, text centered, rounded corners ]
\tikzstyle{cloud} = [draw, ellipse,node distance=3cm,
    minimum height=2em]
\begin{tikzpicture}[node distance = 2cm, auto]
\small
\node [block, text width=7cm, node distance=1cm, rectangle split,rectangle split parts=2,rectangle split part fill={pink!30,blue!20}] (intro) {\textbf{\S \ref{inofmttn}}\nodepart{second} Kerr orbit dynamics in\\
Boyer-Lindquist coordinates.}; 

\node [block, right of=intro, text width=7cm, node distance=11.5cm, rectangle split,rectangle split parts=2,rectangle split part fill={pink!30,blue!20}] (Analyticsol) {\textbf{\S \ref{analyticsoln}}\nodepart{second}Exact solutions for the non-equatorial eccentric trajectories, Eqs. \eqref{final}.}; 

\node [block, below of=Analyticsol, text width=3cm, node distance=6.0cm, rectangle split,rectangle split parts=2,rectangle split part fill={pink!30,blue!20}] (equatorial) {\textbf{ \S \ref{equatorialtrajec}}\nodepart{second} Reduction to the equatorial plane, Eqs. \eqref{finaleq}. };

\node [process, right of=Analyticsol, node distance=7.6cm, rectangle split, rectangle split parts=2,rectangle split part fill={pink!30,blue!20},text width=10em] (traject) {\textbf{\S \ref{trajec}}
            \nodepart{second}Sketching various trajectories.};

\node [process, below of=traject, node distance=6cm, rectangle split, rectangle split parts=2,rectangle split part fill={pink!30,blue!20},text width=10em] (freq) {\textbf{\S \ref{frequencies}}
            \nodepart{second}Fundamental frequencies from the long time average method, Eqs. (\ref{nur}, \ref{nutheta}, \ref{nuphi}).};
            
            \node [process, below of=equatorial, node distance=4cm, rectangle split, rectangle split parts=2,rectangle split part fill={pink!30,blue!20},text width=10em] (cons) {\textbf{\S \ref{reduction}}
            \nodepart{second}Consistency check of the results.};
            
       \node [process, left of=equatorial, node distance=6.2cm, rectangle split, rectangle split parts=2,rectangle split part fill={pink!30,blue!20},text width=10em] (sepQ) {\textbf{\S \ref{separatrixtrajec}}
            \nodepart{second}Deriving the exact solutions for the non-equatorial separatrix trajectories, Eqs. (\ref{phisep}, \ref{tsep}, \ref{rthetasep}).};
            
              \node [process, below of=sepQ, node distance=4.6cm, rectangle split, rectangle split parts=2,rectangle split part fill={pink!30,blue!20},text width=13em] (sepemu) {\textbf{\S \ref{emuseparatrix}}
            \nodepart{second}Writing equations for various important radii, Eqs. (\ref{ISSOequation}-\ref{ISOrad}), and deriving $e_s$ and $\mu_s$ for the non-equatorial separatrix trajectories, Eqs. \eqref{emusepfinal}.};

 \node [process, below of=intro, node distance=4.5cm, rectangle split, rectangle split parts=2,rectangle split part fill={pink!30,blue!20},text width=10em] (translation) {\textbf{\S \ref{trans}}
            \nodepart{second}Derivation of translation relations between ($E$, $L$) and ($e$, $\mu$) parameters, Eqs. \eqref{transfinal}.};
            
            \node [process, left of=translation, node distance=5.0cm, rectangle split, rectangle split parts=2,rectangle split part fill={pink!30,blue!20},text width=10em] (bound) {\textbf{\S \ref{bndcnd}}
            \nodepart{second}Derivation of useful bound orbit conditions, Eqs. \eqref{deltaregion}.};
            
             \node [process, below of=translation, node distance=6.1cm, rectangle split, rectangle split parts=2,rectangle split part fill={pink!30,blue!20},text width=10em] (ELspherical) {\textbf{\S \ref{ELforsphorbits}}
            \nodepart{second}Derivation of the expressions of $E$ and $L$ for the spherical orbits, Eqs. \eqref{ELsphfinal}.};
            
\draw[->,thick] (intro) -- (Analyticsol);
\draw[->,thick] (intro) -- (translation);
\draw[->,thick] (Analyticsol) -- (traject);
\draw[->,thick] (Analyticsol.south) to [out=270,in=90]  (freq.north);
\draw[->,thick] (Analyticsol) -- (equatorial);
\draw[->,thick] (Analyticsol.south) to [out=270,in=90]  (sepQ.north);
\draw[->,thick] (sepemu) -- (sepQ);
\draw[->,thick] (sepQ) -- (sepemu);
\draw[->,thick] (translation.east) to [out=0,in=180]  (Analyticsol.west);   
\draw[->,thick] (translation) -- (bound);
\draw[->,thick] (bound) -- (translation);
\draw[->,thick] (translation) -- (ELspherical);
\draw[->,thick] (ELspherical) -- (sepemu);
\draw[->,thick] (sepQ.east) to [out=0,in=180]  (cons.west);
\draw[->,thick] (freq.west) to [out=180,in=0]  (cons.east);
\end{tikzpicture}
}
\caption{The flow chart of the paper is shown with the sections labeled at top of the box where the concept is presented.}
\label{concepts}
\hspace{-3cm}

\end{figure}

\section{Integrals of motion and bound orbits around Kerr black hole}
\label{inofmttn}

In this section, we first set up the basic equations defining the integrals of motion of the general eccentric orbit with $Q\neq0$ around a Kerr black hole. We then write the formulae defining the transformation from conic parameters ($e$, $\mu$) to dynamical parameters ($E$, $L$) for bound orbits and also provide the conditions for the selection of the parameters ($e$, $\mu$, $a$, $Q$) for the bound orbits. These results are essential for expressing the integrals of motion in ($e$, $\mu$, $a$, $Q$) space for bound orbits. Finally, we derive and present an alternate and simple form of the analytic solutions for the integrals of motion in terms of standard elliptic integrals using the transformation $1/r=\mu \left( 1+ e \cos \chi \right)$. Such transformations lead to a compact and useful trajectory solution for the non-equatorial and eccentric orbits around a rotating black hole. Later, we reduce these results to a simpler form for equatorial eccentric trajectories.

 Considering the Kerr metric \cite{Kerr1963} for a rotating black hole of mass $M$ in the Boyer-Lindquist coordinates, $x^{\alpha}=(t, r, \theta, \phi)$, in geometrical units $G=c=1$
\begin{eqnarray}
{\rm d}s^{2}= -\left( 1-\frac{2  r}{\rho^{2}}\right){\rm d}t^{2} - \frac{4ar  \sin^{2}\theta}{\rho^{2}} \ {\rm d}\phi \ {\rm d}t \ + \frac{\rho^{2}}{r^{2}-2 r+a^{2}}\ {\rm d}r^{2} \ + \rho^{2}{\rm d}\theta^{2} \nonumber \\ 
 +\left( r^{2}+a^{2}+\frac{2r a^{2}\sin^{2}\theta}{\rho^{2}}\right) \sin^{2}\theta \ {\rm d}\phi^{2}, 
\label{kerrmetric}
\end{eqnarray}
where $a=J/M^2$ is the specific angular momentum of the black hole and $\rho^{2}=r^{2}+a^{2}\cos^{2}\theta$.  We have written the variables $\rho, r$, and $t$ are in units of $M$.
The relativistic and conservative Hamiltonian for the geodesic motion of a test particle in Kerr spacetime \cite{Carter1968,MTW,FrolovNovikov}:
\begin{eqnarray}
\mathcal{H}=&&\frac{1}{2}g^{\mu\nu}p_{\mu}p_{\nu} \equiv -\frac{1}{2}m_0^2,\nonumber \\
=&&-\frac{\left(r^2 + a^2\right) ^{2}-\left(r^{2}-2 r+a^{2} \right)a^2 \sin^2 \theta}{2 \left( r^{2}-2 r+a^{2}\right) \rho^{2}} p_{t} ^{2}-\frac{2 a  r}{\rho^{2}\left(r^{2}-2 r+a^{2} \right) }p_{t} p_{\phi} \nonumber \\
&& +\frac{\left( r^{2}-2 r+a^{2}\right) - a^2 \sin^2 \theta}{2 \left( r^{2}-2 r+a^{2}\right) \rho^{2} \sin^2 \theta}p_{\phi} ^{2}+ \frac{\left( r^{2}-2 r+a^{2}\right) }{2 \rho^{2}}p_{r} ^{2}+\frac{1}{2 \rho^{2}}p_{\theta} ^{2},
\label{kerrtonian}
\end{eqnarray}
 where $m_0$ is the particle's rest mass,  $p_{\beta}$ are the conjugate momenta associated with particle's coordinates. To derive the complete set of constants of motion, a canonical transformation, ($q^{\alpha}$, $p_{\beta}$)$\rightarrow$($Q^{\alpha}$, $P_{\beta}$), can be found such that the Hamiltonian is cyclic and the new set of momenta, $P_{\beta}$, are conserved along the world-line of the particle. A characteristic function is obtained to generate such transformation and Hamilton's equations are used to obtain the first-order equations of motion \cite{Carter1968,MTW,FrolovNovikov}:
\begin{subequations}
\begin{eqnarray}
m_0\rho^{2}\frac{{\rm d}t}{{\rm d} \tau}=&&\frac{r^2+a^2}{\left( r^2+ a^2- 2 r\right) }\left[ E\left(r^2+a^2 \right) -a L\right] -a \left( a E \sin^2 \theta - L\right), \label{Eeq}\\
 m_0\rho^{2}\frac{{\rm d}r}{{\rm d} \tau}=&&\pm\sqrt{R}, \label{Req}\\
 m_0\rho^{2}\frac{{\rm d}\theta}{{\rm d} \tau}=&& \pm\sqrt{\Theta}, \label{Thetaeq}\\
 m_0\rho^{2}\frac{{\rm d}\phi}{{\rm d} \tau}=&&\frac{a}{\left( r^2+ a^2- 2 r\right)}\left[ E\left(r^2+a^2 \right) -a L\right]-a E + \frac{L}{\sin^2 \theta},
\label{phieqn}
\end{eqnarray} 
 where 
\begin{eqnarray}
R=&&\left[\left( r^2 +a^2 \right)E -a L \right] ^{2}- \left(r^2+a^2 -2 r \right) \left[m_0^2 r^2+\left(L-a E \right)^{2}+Q  \right], \\
\Theta =&& Q - \left[\left(m_0^2 - E^2 \right)a^2+\frac{L^2}{\sin^2 \theta}  \right] \cos^{2}\theta.
\end{eqnarray}
\end{subequations}
We have written the variables $\rho, r$, and $t$ in the units of $M$.
The integrals of motion have also been derived to be \cite{Carter1968,Schmidt2002}
\begin{subequations}
\begin{eqnarray}
\tau - \tau_{0}=&&\int_{r_{0}}^{r}\frac{{r^{'}}^2 dr^{'}}{\sqrt{R}}+ \int_{\theta_{0}}^{\theta}\frac{a^2 \cos^2 \theta^{'}{\rm d}\theta^{'}}{\sqrt{\Theta}}, \\
\phi - \phi_{0}=&& -\frac{1}{2}\int_{r_{0}}^{r} \frac{1}{\Delta \sqrt{R}}\frac{\partial R}{\partial L} {\rm d} r^{'} -\frac{1}{2}  \int_{\theta_{0}}^{\theta} \frac{1}{\sqrt{\Theta}}\frac{\partial \Theta}{\partial L}{\rm d}\theta^{'}= -\frac{1}{2} I_{1} -\frac{1}{2}  H_{1}, \label{phiintmtn}\\
t - t_{0}=&& \frac{1}{2}\int_{r_{0}}^{r} \frac{1}{\Delta \sqrt{R}}\frac{\partial R}{\partial E} {\rm d} r^{'}  + \frac{1}{2}\int_{\theta_{0}}^{\theta} \frac{1}{\sqrt{\Theta}}\frac{\partial \Theta}{\partial E}{\rm d}\theta^{'}= \frac{1}{2} I_2 + \frac{1}{2} H_2, \label{tintmtn} \\
\int_{r_{0}}^{r} \frac{{\rm d} r^{'}}{\sqrt{R}}=&& \int_{\theta_{0}}^{\theta} \frac{{\rm d}\theta^{'}}{\sqrt{\Theta}}\Rightarrow I_8=H_3, \label{Rthetaint}
\end{eqnarray}
where $\Delta={r^{'}}^2-2r^{'}+a^2$ and $I_1$, $I_2$, $H_1$, $H_2$ are integrals defined above and solved in \S \ref{analyticsoln}.
\end{subequations}

The equation for the radial motion around the Kerr black hole, Eq. \eqref{Req}, can be expressed in the form 
\begin{equation}
\frac{\left( E^2 - m_0^2 \right)}{2}=\frac{m_0^2 \rho^{4}}{2 r^4}\left( \frac{{\rm d}r}{{\rm d} \tau}\right)^2 + V_{eff}\left( r, a, E, L, Q \right), \label{radialeqn1}
\end{equation}
where the term on the LHS represents the total energy, the first term on the RHS represents the radial kinetic energy and the second term on the RHS represents the radial effective potential given by
\begin{equation}
V_{eff}\left( r, a, E, L, Q \right)=-\frac{m_0^2}{r}+ \frac{L^2-a^2\left( E^2 -m_0^2 \right)+Q }{2 r^2}-\frac{\left( L- aE\right)^2 +Q }{r^3}+\frac{a^2 Q}{2r^4}.\label{Qeffpot}
\end{equation}

\subsection{\underline{Translation relations between ($E$, $L$) and ($e$, $\mu$)} }
\label{trans}
We present the transformation of energy, angular momentum, and Carter's constant ($E$, $L$, $Q$) space of the test particle to the eccentricity, inverse-latus rectum ($e$, $\mu$, $Q$) space of its corresponding bound orbit. These relations can be derived if $R(r)$ is factorized and the periastron $r_p$ and apastron $r_a$ of the orbit are substituted as $1/\mu\left(1+e \right)$ and $1/\mu\left(1-e \right)$ respectively. Hence, the formulae connecting ($E$, $L$) and ($e$, $\mu$) parameters for bound orbits (a derivation
of these formulae is given in \ref{trnsltnbndder}) are 
\begin{subequations}
\begin{equation}
 E\left( e, \mu, a, Q\right)= \left[ 1-\mu^3 \left( 1-e^2\right)^2 \left( \mu a^2 Q -Q -x^2\right) - \mu  \left( 1-e^2\right) \right] ^{1/2}, \label{Esqr}
 \end{equation}
where $x=L-aE$ and it can be written in terms of conic parameters as
\begin{equation}
 x^2\left( e, \mu, a, Q\right) =\frac{-S-\sqrt{S^2 -4 P R}}{2P}, \label{xsqr}
 \end{equation}
 where 
 \begin{eqnarray}
 P\left( e, \mu, a, Q\right)=&&\frac{1}{4a^2}\left[ \left(3+e^2 \right) \mu -1 \right]^2 -\mu^3 \left(1-e^2 \right)^2, \label{Pvar}
 \end{eqnarray}
 \begin{eqnarray}
 S\left( e, \mu, a, Q\right)=&&\mu \left( 1-e^2\right)+\mu^3 \left(1-e^2 \right)^2 \left( \mu a^2 Q -Q\right)-1+ \frac{1}{2a^2}\left[ \left(3+e^2 \right) \mu -1 \right]\cdot \nonumber \\
 && \left[\frac{1}{\mu}-a^2 -Q +a^2 Q \mu^2 \left( 1-e^2 \right) - \mu \left( 3+e^2 \right)   \left( \mu a^2 Q - Q\right)   \right], \label{Svar} \\
 R\left( e, \mu, a, Q\right)=&& \frac{1}{4a^2}\left[\frac{1}{\mu}-a^2 -Q +a^2 Q \mu^2 \left( 1-e^2 \right) - \mu \left( 3+e^2 \right)   \left( \mu a^2 Q - Q\right)   \right]^2. \label{Rvar}
 \end{eqnarray} 
 \label{transfinal} 
 \end{subequations}
 These expressions are used to derive analytic results for the integrals of motion, given in \S \ref{analyticsoln}, completely in the ($e$, $\mu$, $a$, $Q$) parameter space.

\subsection{\underline{Analytic solutions of integrals of motion}}
\label{analyticsoln}
Next, we solve for the integrals of motion, i.e. Eqs. (\ref{phiintmtn}-\ref{Rthetaint})
and reduce them to a simple form involving elliptic integrals. We first derive the expressions for the radial integrals $I_1$ and $I_2$. We assume the starting point of the radial motion to be apastron point of the bound orbit, $r_{0}=r_{a}$. 
The steps taken to obtain the reduced form of the radial integrals are as follows:
\begin{enumerate}
\item We make the substitution $1/r^{'}=\mu \left(1+e \cos \chi \right)$ and implement the method of partial fractions.
\item Then make the substitutions, $\cos \chi =2 \cos^2 \frac{\chi}{2}-1 $ and $\psi=\frac{\chi}{2}-\frac{\pi}{2}$.
\item Implement the variable transformation given by $\sin \alpha = \frac{\sqrt{1-m^2} \sin \psi}{\sqrt{1-m^2 \sin^2 \psi}}$, where $m^2$ is defined by Eq. \eqref{nsqr}.
\end{enumerate}
As a result the integrals of motion are expressed as functions of standard elliptic integrals, given by
\begin{subequations}
\begin{equation}
I_1 \left(  \alpha , e, \mu, a, Q \right)= -\left[ C_3
 I_3\left(  \alpha , e, \mu, a, Q \right)  + C_4 I_4 \left(  \alpha, e, \mu, a, Q \right) \right], \label{I1a}
\end{equation}
\begin{eqnarray}
I_2 \left(  \alpha , e, \mu, a, Q \right) &&=  \left[ C_5 I_5 \left(  \alpha , e, \mu, a, Q \right)+ C_6 I_6 \left(  \alpha , e, \mu, a, Q \right) +  C_7 I_3 \left(  \alpha , e, \mu, a, Q \right)+ \right. \nonumber \\
&& \left.  C_8 I_4 \left(  \alpha , e, \mu, a, Q \right) \right], \label{I2a}
\end{eqnarray}
  \begin{eqnarray}
 I_3\left(  \alpha , e, \mu, a, Q \right) =&& \frac{1}{\sqrt{1-m^2}\left( m^2 + p_{2}^2\right) } \left[m^2 F\left(\alpha, k^2 \right) + p_{2}^2 \Pi \left( \frac{-\left( p_{2}^2 +m^2\right) }{1-m^2}, \alpha, k^2\right)  \right], \label{I3} \nonumber \\
 \\
   I_4 \left(  \alpha , e, \mu, a, Q \right)=&& \frac{1}{\sqrt{1-m^2}\left( m^2 + p_{3}^2\right) } \left[m^2 F\left(\alpha, k^2 \right) + p_{3}^2 \Pi \left( \frac{-\left( p_{3}^2 +m^2\right) }{1-m^2}, \alpha, k^2\right)  \right], \label{I4} \nonumber \\
   \\
   I_5 \left(  \alpha , e, \mu, a, Q \right)&&= \frac{1}{\sqrt{1-m^2}\left( m^2 + p_{1}^2\right)^2} \left[ m^4 F\left(\alpha, k^2 \right) + 2 p_{1}^2 m^2 \Pi\left(s^2, \alpha, k^2\right)  \right. \nonumber \\ 
&&\left. + p_{1}^4 I_7 \left(  \alpha , e, \mu, a, Q \right)\right],\label{I5} \\
  I_6\left(  \alpha , e, \mu, a, Q \right)&&= \frac{1}{\sqrt{1-m^2}\left( m^2 + p_{1}^2\right) } \left[m^2 F\left(\alpha, k^2 \right) + p_{1}^2 \Pi\left(s^2, \alpha, k^2\right) \right], \label{I6}
   \end{eqnarray}
\begin{eqnarray}
   I_7\left(  \alpha , e, \mu, a, Q \right)&&= \frac{s^4 \sin \alpha \cos \alpha \sqrt{1-k^2 \sin^2 \alpha}}{2\left( 1-s^2\right) \left( k^2 -s^2\right) \left(1-s^2 \sin^2 \alpha \right) } -\frac{s^2}{2 \left(1-s^2 \right) \left( k^2 -s^2 \right) } K\left(\alpha, k^2\right)  \nonumber \\
 && -\frac{1}{2\left(1-s^2 \right) } F\left(\alpha, k^2 \right) + \frac{\left[ s^4 -2s^2 \left( 1+k^2\right) + 3k^2\right] }{2\left( 1-s^2\right) \left( k^2 -s^2\right)} \Pi\left( s^2 , \alpha, k^2\right), \label{I7} \\
 I_8\left(  \alpha ,  e, \mu, a, Q \right)=&&\frac{2\mu \left( 1-e^2\right)}{\sqrt{C-A+\sqrt{{B}^2-4 AC}}}F\left( \alpha, k^2\right), \label{integral1}
\end{eqnarray}
\label{Iintegrals}
\end{subequations}
where 
\begin{subequations}
\begin{eqnarray}
C_3=&&\frac{2\left( 1-e^2\right)\mu \left[L a^2  -2 x r_{+}  \right]  }{ \sqrt{ \left(A -B +C\right) \left(1-a^2 \right) } \left( a^2 \mu - a^2 \mu e -r_{+}\right) }, \label{C3}\\
C_4 =&& \frac{2\left( 1-e^2\right)\mu \left[-L a^2 +2 x r_{-}  \right]  }{ \sqrt{ \left(A-B+C \right) \left(1-a^2 \right)}\left( a^2 \mu - a^2 \mu e -r_{-}\right) }, \label{C4} \\
C_5=&& \frac{4 E \left(1+ e \right) }{\mu \sqrt{\left( A-B+C\right) } \left(1-e \right) }, \label{C5} \ \
C_6= \frac{8E \left(1+ e \right)}{\sqrt{ \left( A-B+C\right) }},\\
C_7=&& \frac{4  a^2 \mu \left( 1-e^2\right) \left(-L a + 2 E r_{-} \right) }{r_{-}\sqrt{\left( A-B+C\right) \left(1-a^2 \right) }\left( a^2 \mu -a^2 \mu e -r_{+}\right) }, \label{C7}\\
C_8=&& \frac{4  a \mu \left( 1-e^2\right) \left(-2 L r_{-}\sqrt{1-a^2} -2 E a r_{-} +L a^2 \right) }{r_{-}\sqrt{\left( A-B+C\right) \left(1-a^2 \right) }\left( a^2 \mu -a^2 \mu e -r_{-}\right) }, \label{C8} \\
A=&&Q a^2 e^2 \mu^4 \left( 1-e^2\right)^2, \label{A}\\
B=&& 2e \mu^3 \left( 1-e^2\right)^2 \left[ 2 Q a^2 \mu - x^2 -Q\right], \label{B}  \\
C=&&\mu^3 \left( 1-e^2\right)^2 \left[3 \mu Q a^2 -2 x^2 -2 Q \right]  + \left(1-E^2 \right) \left(1-e^2 \right), \label{C} \\
n^2 =&& \frac{4A}{2A-B - \sqrt{B^2 -4 A C}},  \ \ m^2 = \frac{4A}{2A-B + \sqrt{B^2 -4 A C}}, \label{nsqr}\\
 k^2=&&\frac{n^2 -m^2}{1-m^2},  \ \ s^2= \frac{-p_{1}^2 -m^2}{1-m^2},  \label{ksqr} \\
p_{1}^2=&& \frac{2e}{1-e}, \ \ p_{2}^2= \frac{2e a^2 \mu}{a^2 \mu -a^2 \mu e - r_{+}}, \ \
p_{3}^2= \frac{2e a^2 \mu }{a^2 \mu -a^2 \mu e- r_{-}}, \label{p2}
\end{eqnarray}
\begin{eqnarray}
x_{1,2}=&&\frac{-B\pm\sqrt{{B}^2-4 AC}}{2A}, \label{x12}
\end{eqnarray}
\label{constants}
\end{subequations}

and where the variables $E$, $L$ and $x$ can be written as functions of ($e$, $\mu$, $a$, $Q$) using Eqs. (\ref{Esqr}-\ref{Rvar}), which makes all the integrals to be only functions of ($e$, $\mu$, $a$, $Q$). The definition of the elliptic integrals involved, is given below \cite{Grad}:
\begin{subequations}
\begin{eqnarray}
F\left( \alpha, k^2 \right)=&&  \int_{0}^{\alpha}\frac{{\rm d}\alpha}{\sqrt{1-k^2 \sin^2 \alpha}}, \label{elliptica}\\
K\left( \alpha, k^2 \right)=&&  \int_{0}^{\alpha}\sqrt{1-k^2 \sin^2 \alpha}\cdot {\rm d}\alpha, \label{ellipticb}\\
\Pi \left(s^2, \alpha, k^2 \right)=&& \int_{0}^{\alpha}\frac{{\rm d}\alpha}{\left( 1- s^2 \sin^2 \alpha \right) \sqrt{1-k^2 \sin^2 \alpha}}. \label{ellipticc}
\end{eqnarray}
\end{subequations}

A complete derivation of these integrals is given in \ref{integrals}. Next, to solve the integrals $H_1$, $H_2$, and $H_3$ of Eqs. (\ref{phiintmtn}- \ref{Rthetaint}), we make the substitutions $z=\cos \theta^{'}$ and $z=z_{-}\sin \beta$ \citep{Fujita2009} which reduces these integrals to
\begin{subequations}
\begin{equation}
H_1\left( \theta , \theta_{0}, e, \mu , a, Q \right)=  \frac{2 L}{ z_{+} a \sqrt{1-E^2}} \left\lbrace F\left( \arcsin \left( \frac{\cos \theta_{0}}{z_{-}}\right) , \frac{z_{-}^2}{z_{+}^2}\right)-F\left( \arcsin \left( \frac{\cos \theta}{z_{-}}\right), \frac{z_{-}^2}{z_{+}^2}\right)  + \right. \nonumber
\end{equation}
\begin{equation}
\left. \Pi\left( z_{-}^2, \arcsin \left( \frac{\cos \theta}{z_{-}}\right) , \frac{z_{-}^2}{z_{+}^2}\right) -\Pi \left( z_{-}^2, \arcsin \left( \frac{\cos \theta_{0}}{z_{-}}\right) , \frac{z_{-}^2}{z_{+}^2}\right)    \right\rbrace , \label{H1}
\end{equation}
\begin{equation}
H_2\left(  \theta , \theta_{0}, e, \mu , a, Q \right)= \frac{2E a z_{+}}{\sqrt{1-E^2}}  \left\lbrace   K\left( \arcsin \left( \frac{\cos \theta}{z_{-}}\right)  , \frac{z_{-}^2}{z_{+}^2} \right) - F\left( \arcsin \left( \frac{\cos \theta}{z_{-}}\right)  , \frac{z_{-}^2}{z_{+}^2} \right) -  \right. \nonumber
\end{equation}
\begin{equation}
 \left. K\left(  \arcsin \left( \frac{\cos \theta_{0}}{z_{-}}\right), \frac{z_{-}^2}{z_{+}^2} \right)  +F\left( \arcsin \left( \frac{\cos \theta_{0}}{z_{-}}\right)  , \frac{z_{-}^2}{z_{+}^2} \right)   \right\rbrace , \label{H2}
\end{equation}
\begin{equation}
H_3\left(  \theta , \theta_{0}, e, \mu , a, Q \right) = \frac{1}{a\sqrt{1-E^2}z_{+}}\left\lbrace F\left( \arcsin \left( \frac{\cos \theta_{0}}{z_{-}}\right) ,\frac{z_{-}^{2}}{z_{+}^{2}}\right) -F\left( \arcsin \left( \frac{\cos \theta}{z_{-}}\right),\frac{z_{-}^{2}}{z_{+}^{2}}\right)  \right\rbrace,  \label{integral2}
\end{equation}
\label{Hequations}
where
\begin{equation}
z_{\pm}^{2}=\frac{-P^{'}\pm \sqrt{P^{'2}-4Q^{'}}}{2}, \ \ P^{'}=\frac{-L^2-Q-a^2\left(1-E^2 \right) }{a^2 \left( 1- E^2\right) }, \ \
Q^{'}=\frac{Q}{a^2\left( 1- E^2\right) }. \label{zpm}
\end{equation}
\end{subequations}

Hence, the equations of motion can now be written in short as
\begin{subequations}
\begin{eqnarray}
&&\phi-\phi_{0}=  \frac{1}{2} \left[ C_3 I_3\left(  \alpha ,  e, \mu, a, Q \right) + C_4 I_4\left(  \alpha ,  e, \mu, a, Q \right) -H_1 \left(  \theta , \theta_{0}, e, \mu , a, Q \right) \right] , \nonumber \\\label{phifinal}
 \\
 &&t-t_{0}= \frac{1}{2} \left[ C_5 I_5 \left(  \alpha ,  e, \mu, a, Q \right)+ C_6 I_6 \left(  \alpha ,  e, \mu, a, Q \right)+ C_7 I_3 \left(  \alpha ,  e, \mu, a, Q \right)  \right. \nonumber \\ 
  && \left. + C_8 I_4 \left(  \alpha ,  e, \mu, a, Q \right)+ H_2 \left(  \theta , \theta_{0}, e, \mu , a, Q \right) \right] \label{tfinal}
 , \\
 &&I_8 \left(  \alpha ,  e, \mu, a, Q \right) = H_3 \left(  \theta , \theta_{0}, e, \mu , a, Q \right), \label{rthetafinal}
\end{eqnarray}
\label{final}
\end{subequations}
where $I_3$, $I_4$, $I_5$, $I_6$, $I_8$, $H_1$, $H_2$ and $H_3$ are given by Eqs. (\ref{I3}-\ref{integral1}; \ref{Hequations}) respectively. Hence, all the integrals are written explicitly as functions of parameters ($e$, $\mu$, $a$, $Q$) through variables $\alpha\left( e, \mu, a, Q; \chi\right)$ and $\beta\left( e, \mu, a, Q; \theta\right)$ which are directly used to calculate ($r$,  $\theta$, $t$) through Eqs. (\ref{phifinal}-\ref{rthetafinal}). The radial motion, which varies with the $\alpha$, is assumed to have the starting point at the apastron distance, $r_a$ or $\alpha=0$, of the orbit and the starting point of the polar motion, $\beta_0$ or $\theta_0$, is an extra variable which can be chosen in the range $\{\theta_{-},\pi - \theta_{-}\}$. This is tantamount to shifting the starting point of the motion in time or adjusting the initial value of the observed time, $t_{0}$.

Once the initial points are fixed ($\alpha=0$, $\theta=\theta_{0}$), Eqs. \eqref{phifinal} and \eqref{tfinal} are used to calculate $\phi \left(r, \theta \right) $ and $t\left(r, \theta \right)$ respectively, whereas Eq. \eqref{rthetafinal} gives $r\left( \theta \right) $ or $\theta \left( r \right) $, which can be used to obtain $t\left( r \right)$ or $t\left( \theta \right)$ and $\phi \left( r \right)$ or $\phi \left( \theta \right)$. 

The elegant alternate forms presented here help us to write useful and simpler expressions of ($\phi$, $t$) for the equatorial eccentric trajectories, as shown later in \S \ref{equatorialtrajec}. Also, these results can be used to reduce the radial integrals for non-equatorial separatrix trajectories in the form of logarithmic and trigonometric functions, presented in \S \ref{separatrixtrajec}, which are useful in the study of gravitational waves from EMRIs.

\subsection{\underline{Bound orbit conditions in conic parameter space}}
 \label{bndcnd}
 The bound orbit regions have been studied and divided in the ($E$, $L$) space according to the different types of possible $r$ motion \cite{Hackmann2010}. The most relevant astrophysical bound orbit region corresponds to the case where $E<1$ and there are four real roots of $R(r)$, $r_1 > r_2 > r_3 > r_4>0$, such that the bound orbit either exists between $r_1$ and $r_2$ or $r_3$ and $r_4$, this has been defined as region III in the ($E$, $L$) plane by \cite{Hackmann2010}. Since $r_1$ and $r_2$ are the outer most turning points of the effective potential, the bound orbit should exist between these two in the astrophysical situations. We can implement this constraint in the ($e$, $\mu$, $a$, $Q$) space by imposing the condition $k^2<1$ on the parameter $k$ used in the radial integrals in \S \ref{analyticsoln}, where we have assumed that a bound orbit exists between $r_1$ and $r_2$, which requires $k^2<1$ as an essential condition for the elliptic integrals to have real values, Eqs. (\ref{elliptica}-\ref{ellipticc}). This further implies
 \begin{subequations}
 \begin{equation}
 n^2<1;
 \end{equation}
where the substitution of Eq. \eqref{nsqr} in the above expression yields
 \begin{equation}
 \left( A + B + C \right) >0,
 \end{equation}
 and by using Eqs. (\ref{A}-\ref{C}) and (\ref{Esqr}-\ref{Rvar}) this implies
\begin{equation}
\left[ \mu^3 a^2 Q \left(1+e \right)^2 +\mu^2 \left( \mu a^2 Q -x^2 -Q\right) \left(3-e \right) \left(1+e \right) +1 \right]>0. \label{boundcondreg3}
\end{equation} 
 Another necessary condition is that the periastron of the orbit $r_2=1/ \left[ \mu \left( 1+e \right)\right] $ is outside the horizon, which gives
 \begin{equation}
 \left[\mu \left( 1+e\right) \left( 1+ \sqrt{1-a^2}\right) \right] <1.
 \end{equation}
\end{subequations}
Hence, the necessary and independent conditions for this region can be collectively given as
\begin{subequations}
\begin{eqnarray}
\mu^3 a^2 Q \left(1+e \right)^2 +\mu^2 \left( \mu a^2 Q -x^2 -Q\right) \left(3-e \right) \left(1+e \right) +1  > 0, \label{deltaregiona}\\
\mu \left( 1+e\right) \left( 1+ \sqrt{1-a^2}\right)  <1, \label{deltaregionb} \\
E(e, \mu, a, Q) < 1. \label{deltaregionc}
\end{eqnarray}
\label{deltaregion}
\end{subequations}

There exists unstable bound orbits for $E>1$ specified as region IV in the ($E$, $L$) plane by \cite{Hackmann2010}, where the bound orbit exists between $r_2$ and $r_3$. Such a situation is not important from the astrophysical point of view, because the particle will follow a bound trajectory between the outermost turning points, i.e. $r_1$ and $r_2$, and hence the above conditions, Eq. \eqref{deltaregion}, together represent a necessary and sufficient condition for the existence of bound orbits.

\subsection{\underline{Equatorial bound orbits}}
\label{equatorialtrajec}
In this section, we apply the integrals of motion, Eqs. (\ref{phifinal}, \ref{tfinal}), to the eccentric equatorial trajectories, where $Q=0$ ($\theta=\pi/2$). We show that the forms derived in \S \ref{analyticsoln} reduce to very compact expressions of ($\phi$, $t$) involving trigonometric functions and elliptic integrals for the equatorial eccentric orbits. We implement the limit, $Q \rightarrow 0$ which leads to $A\rightarrow 0$, Eq. \eqref{A}, and reduces the factors $\left(1+x_1\right)$, $A\left( 1+x_2\right) $, using Eq. \eqref{x12}, to
\begin{subequations}
\begin{equation}
\left(1+x_1\right) \rightarrow \left(1-\frac{C}{B}\right) , \ \mathrm{and} \
A\left(1+x_2 \right) \rightarrow -B,
\end{equation}
which gives
\begin{equation}
A\left( 1+x_1 \right) \left(1+x_2 \right)=A-B+C= \mu \left(1-e^2 \right)^2 \left[ 1- \mu^2 x^2 \left( 3 -e^2 -2e \right) \right], \label{reduced}
\end{equation}
where the translation equation given by Eq. \eqref{Esqr} for $Q=0$ is used to substitute for $E^2$.
Also, $m^2$ and $n^2$ reduce to
\begin{equation}
m^2=\frac{2B}{B-C}=\frac{4 \mu^2 e x^2}{\left[ 1- \mu^2 x^2 \left( 3 -e^2 -2e \right) \right]}, \ \ \
 n^2= \frac{4A {B}^2 }{2{B}^2 \left( A-B \right)+ 2 A C }=0. 
\end{equation}
\end{subequations}
The substitution of these reduced expressions of $m^2$ and $n^2$ further simplifies the integrals $I_3$,  $I_4$, $I_5$, and $I_6$, as shown in \ref{Qeq0}, which finally yields the expressions for azimuthal angle and time coordinate for equatorial trajectories to be given by
\begin{subequations}
\begin{eqnarray}
\phi-\phi_{0}=&&-\frac{1}{2} I_1= a_1  \Pi  \left( -p_{2}^2, \psi, m^2\right) + b_1 \Pi\left( -p_{3}^2, \psi, m^2\right) \label{eqphi}, \\
t-t_{0}=&&\frac{1}{2} I_2=a_2 I_5 + b_2 I_6 +c_2 I_3 + d_2 I_4, \nonumber \\
&&=a_2 \left[ \frac{p_{1}^4 \sin \psi \cos \psi \sqrt{1-m^2 \sin^2 \psi}}{2\left( 1+p_{1}^2\right) \left( m^2 +p_{1}^2\right) \left(1+p_{1}^2 \sin^2 \psi \right) } - \frac{F\left( \psi, m^2\right) }{2\left(1+p_{1}^2 \right) } +  \frac{p_{1}^2K\left( \psi, m^2\right) }{2\left(1+p_{1}^2 \right) \left( m^2 + p_{1}^2\right) } \right] \nonumber \\
&&+ \Pi\left( -p_{1}^2 , \psi , m^2\right)  \left\lbrace a_2  \frac{\left[ p_{1}^4 +2p_{1}^2 \left( 1+m^2\right) + 3m^2\right] }{2\left( 1+p_{1}^2\right) \left( m^2 +p_{1}^2\right)} +b_2\right\rbrace + c_2 \Pi\left( -p_{2}^2 , \psi , m^2\right)  \nonumber \\
&& + d_2 \Pi\left( -p_{3}^2 , \psi , m^2\right), \label{eqtime}
\end{eqnarray}
where the substitution of Eq. \eqref{reduced} into Eqs. (\ref{C3}-\ref{C8}) yields the reduced forms of the constants given by
\begin{eqnarray}
a_1=&&\frac{C_3}{2}=\frac{\mu^{1/2}\left[ L a^2 -2 x r_{+}\right] }{\sqrt{1-a^2}\left(a^2 \mu- a^2 \mu e - r_{+} \right) \sqrt{1-\mu^2 x^2 \left(3 -e^2 -2 e \right) } }, \label{a1}\\
b_1=&&\frac{C_4}{2}=\frac{\mu^{1/2}\left[ -L a^2 +2 x r_{-}\right] }{\sqrt{1-a^2}\left(a^2 \mu- a^2 \mu e - r_{-} \right) \sqrt{1-\mu^2 x^2 \left(3 -e^2 -2 e \right) } }, \label{b1} \\
a_2=&& \frac{C_5}{2}=\frac{2  E}{\mu^{3/2}\left( 1-e\right)^2 \sqrt{1-\mu^2 x^2 \left(3 -e^2 -2 e \right) }}, \label{a2} \\
b_2=&& \frac{C_6}{2}=\frac{4E}{\mu^{1/2}\left( 1-e\right) \sqrt{1-\mu^2 x^2 \left(3 -e^2 -2 e \right) } }, \label{b2} 
\end{eqnarray}
\begin{eqnarray}
c_2=&& \frac{C_7}{2}=\frac{2a^2 \mu^{1/2}\left(-La + 2 E r_{-}\right)}{r_{-}\sqrt{\left[ 1-\mu^2 x^2 \left(3 -e^2 -2 e \right)\right] \left(1-a^2 \right)  }  \left( a^2 \mu -a^2 \mu e -r_{+}\right) }, \label{c2} 
\end{eqnarray}
\begin{eqnarray}
d_2=&& \frac{C_8}{2}= \frac{2a \mu^{1/2}\left(-2 L r_{-}\sqrt{1-a^2}-2E r_{-}a +La^2\right)}{r_{-}\sqrt{\left[ 1-\mu^2 x^2 \left(3 -e^2 -2 e \right)\right]  \left(1-a^2 \right)}  \left( a^2 \mu -a^2 \mu e -r_{-}\right)}. \label{d2}
\end{eqnarray}
\label{finaleq}
\end{subequations}
The brackets of the factor $\left[ 1-\mu^2 x^2 \left(3 -e^2 -2 e \right)\right] $ in the expressions of $c_2$ and $d_2$ above were missing in the journal version \citep{RMCQG2019}, and has been corrected here. 
The corresponding fundamental frequency formulae for the equatorial trajectories are 
\begin{equation}
\nu_{\phi}=\frac{c^{3} \cdot \left[ \phi\left(\psi=\pi/2 \right) - \phi_{0}\right] }{ 2 \pi G M \cdot  \left[ t\left(\psi=\pi/2 \right)-t_{0}\right]  },  \ \ \
\nu_{r}=\frac{c^{3}}{ G M \cdot t_{r} }=\frac{c^{3}}{2 G M \cdot \left[ t\left(\psi=\pi/2 \right)-t_0 \right]  }. \label{eqradialfreq}
\end{equation}

These compact expressions, Eqs. (\ref{finaleq}, \ref{eqradialfreq}), for the equatorial eccentric trajectories have their importance in various astrophysical studies, in addition to, gyroscope precession and phase space studies.

\begin{table}
\begin{center}
\caption{This table summarizes all the integrals solved in \S \ref{analyticsoln}, \ref{equatorialtrajec} to calculate the integrals of motion in the Kerr geometry, where all the constants are defined in the text.}
\scalebox{0.75}{
\begin{tabular}{|c|c|}
\hline
\hline
 & Analytic solution of $\phi$  and $t$ for $Q\neq0$\\
\hline
&\\
 & $ \phi-\phi_{0}= \frac{1}{2} \left( C_3 I_3 + C_4 I_4 -H_1  \right)$; $ t-t_{0}=\frac{1}{2} \left( C_5 I_5 + C_6 I_6 + C_7 I_3 + C_8 I_4 + H_2  \right)$ \\
\hline
&\\
$I_1$ & $\displaystyle{\int_{r_{a}}^{r}} \frac{1}{\Delta \sqrt{R}}\frac{\partial R}{\partial L} {\rm d} r^{'} = - \left[ C_3 I_3 + C_4 I_4\right]$\\
\hline
&\\
$I_2$ & 
$\displaystyle{\int_{r_{a}}^{r}} \frac{1}{\Delta \sqrt{R}}\frac{\partial R}{\partial E} {\rm d} r^{'}=\left[ C_5 I_5 + C_6 I_6  +  C_7 I_3 +  C_8 I_4 \right],$ \\
\hline
& \\
$I_3$ & $\displaystyle{\int_{0}^{\psi}  \frac{{\rm d} \psi}{\left( 1+ p_{2}^2 \sin^2 \psi \right) \sqrt{1-m^2 \sin^2 \psi} \sqrt{1-n^2 \sin^2 \psi}}=\frac{1}{\sqrt{1-m^2}\left( m^2 + p_{2}^2\right) } \left[m^2 F\left(\alpha, k^2 \right) + p_{2}^2 \Pi\left( \frac{-p_{2}^2 -m^2}{1-m^2}, \alpha, k^2\right)  \right]}$\\
& \\
& where $\displaystyle{\sin \alpha = \frac{\sqrt{1-m^2}\sin \psi}{\sqrt{1-m^2 \sin^2 \psi}}}$, $\displaystyle{\psi=\frac{\chi}{2}-\frac{\pi}{2}}$ and $\displaystyle{1/r=\mu\left(1+ e \cos \chi \right) }$ \\
\hline
& \\
$I_4$ &  $\displaystyle{\int_{0}^{\psi}  \frac{{\rm d} \psi}{\left( 1+ p_{3}^2 \sin^2 \psi \right) \sqrt{1-m^2 \sin^2 \psi} \sqrt{1-n^2 \sin^2 \psi}}=\frac{1}{\sqrt{1-m^2}\left( m^2 + p_{3}^2\right) } \left[m^2 F\left(\alpha, k^2 \right) + p_{3}^2 \Pi\left( \frac{-p_{3}^2 -m^2}{1-m^2}, \alpha, k^2\right)  \right]}$\\
\hline
& \\
$I_5$ & $\displaystyle{\int_{0}^{\psi} \frac{{\rm d} \psi}{\left( 1+ p_{1}^2 \sin^2 \psi \right) ^2 \sqrt{1-m^2 \sin^2 \psi} \sqrt{1-n^2 \sin^2 \psi}}}$ \\
& $\displaystyle{=\frac{1}{\sqrt{1-m^2}\left( m^2 + p_{1}^2\right)^2} \left[ m^4 F\left(\alpha, k^2 \right) + 2 p_{1}^2 m^2 \Pi\left(s^2, \alpha, k^2\right) + p_{1}^4 I_7 \left(  \alpha , e, \mu, a, Q \right)\right],}$\\
\hline
& \\
$I_6$ & $\displaystyle{\int_{0}^{\psi} \frac{{\rm d} \psi}{\left( 1+ p_{1}^2 \sin^2 \psi \right) \sqrt{1-m^2 \sin^2 \psi} \sqrt{1-n^2 \sin^2 \psi}}=\frac{1}{\sqrt{1-m^2}\left( m^2 + p_{1}^2\right) } \left[m^2 F\left(\alpha, k^2 \right) + p_{1}^2 \Pi\left(s^2, \alpha, k^2\right) \right]}$\\
\hline
& \\
$I_7$  & $\displaystyle{\int_{0}^{\psi} \frac{{\rm d} \psi}{\left( 1+ s^2 \sin^2 \psi\right)^2 \sqrt{1-k^2 \sin^2 \psi} }=\frac{s^4 \sin \alpha \cos \alpha \sqrt{1-k^2 \sin^2 \alpha}}{2\left( 1-s^2\right) \left( k^2 -s^2\right) \left(1-s^2 \sin^2 \alpha \right) } + \frac{\left[ s^4 -2s^2 \left( 1+k^2\right) + 3k^2\right] }{2\left( 1-s^2\right) \left( k^2 -s^2\right)} \Pi\left( s^2 , \alpha, k^2\right)}$ \\
& $\displaystyle{-\frac{1}{2\left(1-s^2 \right) } F\left(\alpha, k^2 \right) -\frac{s^2}{2 \left(1-s^2 \right) \left( k^2 -s^2 \right) } K\left(\alpha, k^2\right)}$\\
\hline
& \\
$H_1$ & $\displaystyle{\int_{\theta_{0}}^{\theta} \frac{1}{\sqrt{\Theta}}\frac{\partial \Theta}{\partial L}{\rm d}\theta^{'}=\frac{2 L}{ z_{+} a \sqrt{1-E^2}} \left\lbrace \Pi\left( z_{-}^2, \arcsin \left(  \frac{\cos \theta}{z_{-}} \right), \frac{z_{-}^2}{z_{+}^2}\right) -\Pi\left( z_{-}^2, \arcsin \left(  \frac{\cos \theta_{0}}{z_{-}} \right) , \frac{z_{-}^2}{z_{+}^2}\right) -  \right. }$\\
& $ \displaystyle{\left. F\left( \arcsin \left(  \frac{\cos \theta}{z_{-}} \right), \frac{z_{-}^2}{z_{+}^2}\right) +F\left( \arcsin \left(  \frac{\cos \theta_{0}}{z_{-}} \right) , \frac{z_{-}^2}{z_{+}^2}\right) \right\rbrace} $\\
\hline
& \\
$H_2$ & $\displaystyle{\int_{\theta_{0}}^{\theta} \frac{1}{\sqrt{\Theta}}\frac{\partial \Theta}{\partial E}{\rm d}\theta^{'}=\frac{2E a z_{+}}{\sqrt{1-E^2}} \left\lbrace  K\left( \arcsin \left(  \frac{\cos \theta}{z_{-}} \right) , \frac{z_{-}^2}{z_{+}^2} \right) -K\left( \arcsin \left(  \frac{\cos \theta_{0}}{z_{-}} \right) , \frac{z_{-}^2}{z_{+}^2}\right)   -\right. }$ \\
& $ \displaystyle{\left.  F\left( \arcsin \left(  \frac{\cos \theta}{z_{-}} \right) , \frac{z_{-}^2}{z_{+}^2} \right) +  F\left( \arcsin \left(  \frac{\cos \theta_{0}}{z_{-}} \right) , \frac{z_{-}^2}{z_{+}^2} \right)   \right\rbrace.}$\\
\hline
\hline
 & Analytic solution of $\phi$  and $t$ for $Q=0$\\
\hline
& \\
& $\phi-\phi_{0}=\displaystyle{ a_1  \Pi  \left( -p_{2}^2, \psi, m^2\right) + b_1 \Pi\left( -p_{3}^2, \psi, m^2\right)}$ \\
\hline 
& \\
& $t-t_{0}=\displaystyle{a_2 \left[ \frac{p_{1}^4 \sin \psi \cos \psi \sqrt{1-m^2 \sin^2 \psi}}{2\left( 1+p_{1}^2\right) \left( m^2 +p_{1}^2\right) \left(1+p_{1}^2 \sin^2 \psi \right) } - \frac{F\left( \psi, m^2\right) }{2\left(1+p_{1}^2 \right) } +  \frac{p_{1}^2K\left( \psi, m^2\right) }{2\left(1+p_{1}^2 \right) \left( m^2 + p_{1}^2\right) } \right]+ d_2 \Pi\left( -p_{3}^2 , \psi , m^2\right) }$ \\
& $\displaystyle{+ c_2 \Pi\left( -p_{2}^2 , \psi , m^2\right) + \Pi\left( -p_{1}^2 , \psi , m^2\right)  \left\lbrace a_2  \frac{\left[ p_{1}^4 +2p_{1}^2 \left( 1+m^2\right) + 3m^2\right] }{2\left( 1+p_{1}^2\right) \left( m^2 +p_{1}^2\right)} +b_2\right\rbrace }$ \\
\hline
\end{tabular}
\label{integraltable}
}

\end{center}

\end{table}

\section{Non-equatorial separatrix trajectories}
\label{Qseparatrix}
The separatrix orbits have been studied for the equatorial plane around a Kerr black hole \cite{Perez-Giz2009,Levin2009}. They have been shown as homoclinic orbits which asymptote to an energetically bound and unstable circular orbit. Here, we discuss the non-equatorial counterpart of these separatrix trajectories where these orbits asymptote to an energetically bound, unstable spherical orbit. These non-equatorial homoclinic trajectories are critical in calculating the evolution of test objects transiting from inspiral to plunge, which is not always confined to the equatorial plane, as in EMRIs emitting gravitational radiation.

In this section, we first deduce the expressions of $E$ and $L$ for the spherical orbits as functions of the radius $r_s$, and ($a$, $Q$). We then derive the exact expressions for the conic parameters ($e$, $\mu$) for non-equatorial separatrix orbits as a function of the radius of the corresponding spherical orbit, $r_s$, and ($a$, $Q$). We also show that these formulae reduce to the equatorial case, previously derived in \cite{Levin2009}, when $Q\rightarrow 0$ is applied. Next, we derive the exact analytic expressions for the non-equatorial separatrix trajectories by reducing it from the general trajectory formulae, Eqs. (\ref{phifinal}-\ref{rthetafinal}). We find that in this case, the radial part of the solutions can be reduced to a form that involves only logarithmic and trigonometric functions.

\subsection{\underline{Energy and angular momentum of spherical orbits}}
\label{ELforsphorbits}
Spherical orbits are the non-equatorial counterparts of circular orbits and set a crucial signpost in the dynamical study of non-equatorial and separatrix trajectories. The exact expressions for energy and angular momentum  for the spherical orbits can be derived by substituting $e=0$ and $\mu=1/r_s$, where $r_s$ is the radius of the orbit, in the expressions for $E$, $L$, and $x$ given by Eqs. (\ref{Esqr}-\ref{Rvar}), which yields
\begin{subequations}
\begin{equation}
 E= \frac{\left\lbrace \splitfrac{2a^4 Q + \left( r_s-3\right) \left( r_s-2 \right)^2 r_{s}^4 -a^2 r_s \left[r_{s}^{2}\left(3r_s -5 \right) + Q \left(r_s\left(r_s -4 \right) +5  \right)   \right]}{ -2 a \left[r_s \left(r_s -2 \right) +a^2  \right] \sqrt{a^2 Q^2 -r_s^3 Q \left(r_s -3 \right) + r_s^5} } \right\rbrace^{1/2}  }{r_{s}^2 \left[r_{s} \left( r_s-3\right)^2 -4a^2 \right]^{1/2} }, 
 \label{Ensph} 
\end{equation}
\begin{equation}
 x=\frac{\left\lbrace \splitfrac{-2a^4Q +r_{s}^2 \left(r_s -3 \right) \left[r_{s}^2 -\left(r_s -3 \right) Q \right] +a^2 r_s \left( r_s^3 +r_s^2 -2 Q r_s +8Q \right)}{ -2a  \left[r_s \left(r_s -2 \right) +a^2  \right] \sqrt{a^2 Q^2 -r_s^3 Q \left(r_s -3 \right) + r_s^5} }\right\rbrace^{1/2} }{r_s^{1/2} \left[r_{s} \left( r_s-3\right)^2 -4a^2 \right]^{1/2}}, \label{xsph} 
\end{equation}
 and
 \begin{equation}
 L=x+aE. \label{Lsph}
\end{equation}
\label{ELsphfinal}
\end{subequations} 
Similar formulae were derived in terms of inclination angle using an approximation in \citep{Grossman2012}, whereas we have written the exact form in terms of the fundamental parameters and constant of motion $Q$. These formulae reduce to the energy and angular momentum formulae for circular orbits when $Q=0$ is substituted \cite{Bardeen1972}:  
\begin{eqnarray}
E=\frac{r_{c}^{2}-2r_{c} + a\sqrt{r_{c}}}{r_{c}\left( r_{c}^{2}- 3r_{c} + 2a\sqrt{r_{c}}\right) ^{1/2}}, \ \ L=\frac{\sqrt{r_{c}}\left( r_{c}^{2}-2a\sqrt{r_{c}}+a^{2}\right) }{r_{c}\left( r_{c}^{2}- 3r_{c} + 2 a \sqrt{r_{c}}\right)^{1/2}}. \label{ELcircular}
\end{eqnarray}

\subsection{\underline{Exact expressions of conic variables for non-equatorial separatrix orbits}}
\label{emuseparatrix}
\begin{figure}
 \mbox{ \subfigure[]{
\includegraphics[scale=0.46]{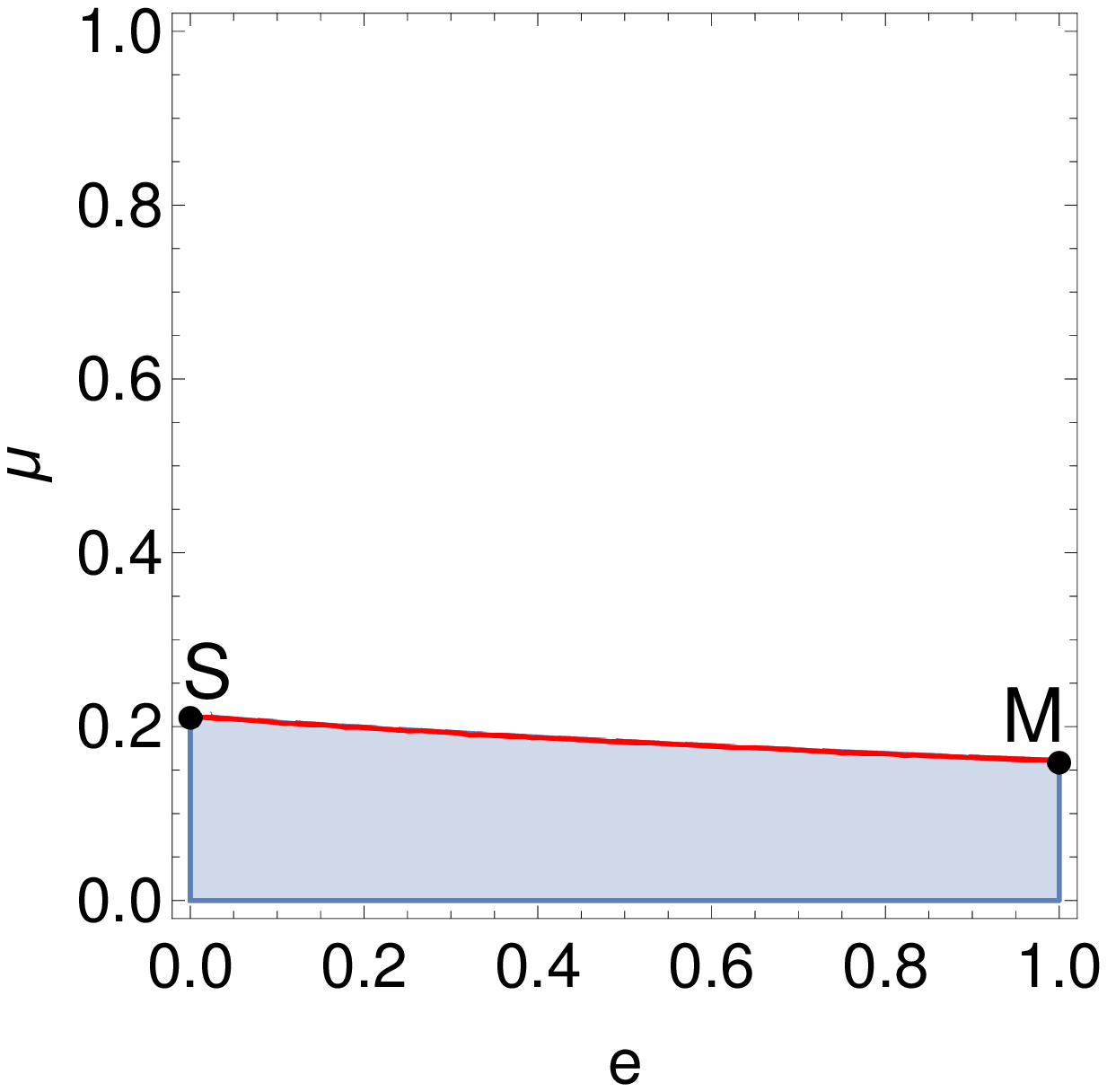} \label{seporbitsa}}
\hspace{0.7cm}
\subfigure[]{
\includegraphics[scale=0.7]{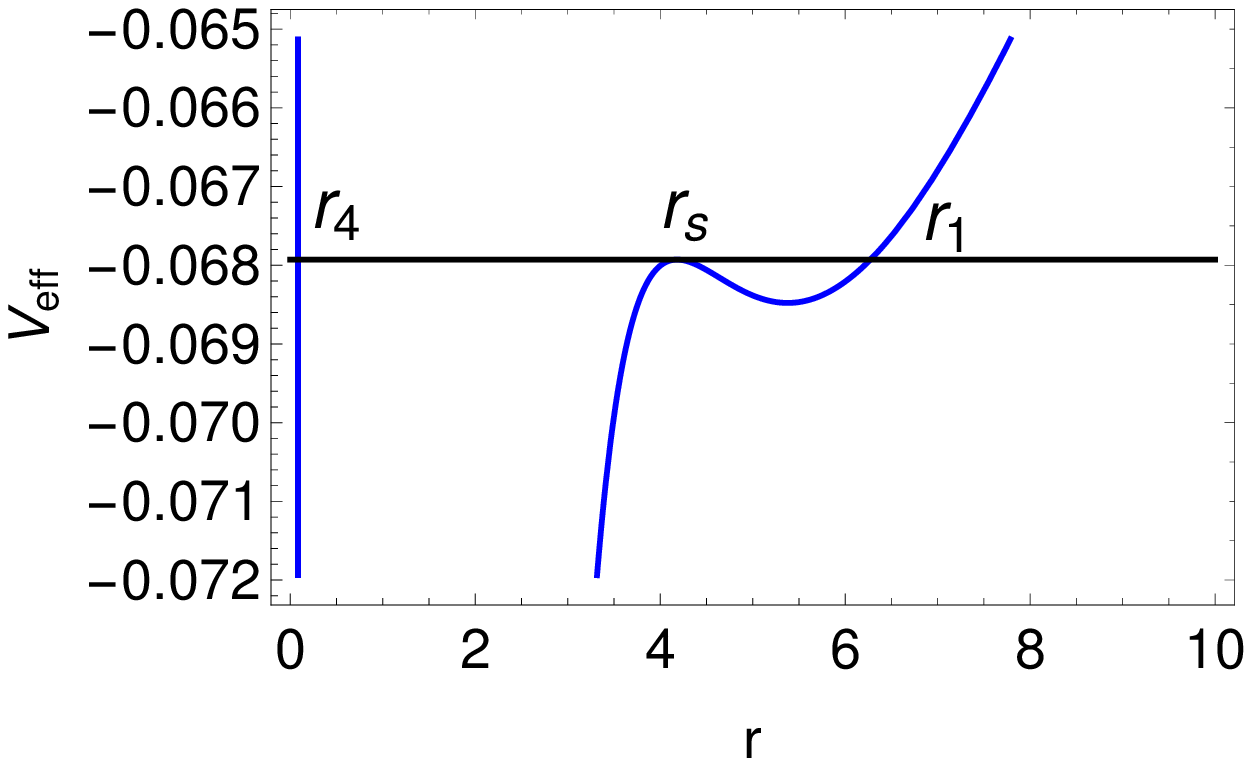}\label{seporbitsb}} }
\caption{\label{seporbits}(a) The shaded region depicts bound orbit region in the ($e$, $\mu$) plane determined by Eqs. \eqref{deltaregion} for $a=0.5$ and $Q=5$. The red boundary of the region represents non-equatorial separatrix orbits with eccentricity of the orbit varying along the curve. The black dot represented by S corresponds to the ISSO with ($e=0$, $\mu=1/r_s$), whereas M represents the MBSO with ($e=1$, $\mu=1/2r_s$); (b) The effective potential, Eq. \eqref{Qeffpot}, is shown for a non-equatorial separatrix orbit with $E=0.92959309$ , $L=2.15349738$, $a=0.5$, and $Q=5$, where the horizontal line represents the total energy given by $\left(E^2-1\right)/2$.}
\end{figure}

Similar to the case of equatorial plane, the non-equatorial separatrix trajectories can be parametrized by the radius of unstable spherical orbits, $r_s$, for a given combination of $a$ and $Q$, where $r_s$ varies from MBSO to ISSO. The energy and angular momentum of the separatrix orbits can be determined by Eqs. (\ref{Ensph}-\ref{Lsph}) by varying $r_s$ between the extrema MBSO and ISSO radii.
In the ($e$, $\mu$) plane, these homoclinic orbits forms the boundary (other than $e=0$ and $e=1$ curves) of the allowed bound orbit region defined by Eq. \eqref{deltaregion} for a fixed $a$ and $Q$; see red curve in Fig. \ref{seporbitsa}. The locus of this boundary in the ($e$, $\mu$) plane is obtained when equality is applied to the inequality Eq. \eqref{deltaregiona}, which results in
\begin{equation}
\left[ \mu^3 a^2 Q \left(1+e \right)^2 +\mu^2 \left( \mu a^2 Q -x^2 -Q\right) \left(3-e \right) \left(1+e \right) +1 \right] = 0. \label{sepeq}
\end{equation} 

 ISSO is a homoclinic orbit with $e=0$ and MBSO is a homoclinic orbit with $e=1$; hence the endpoints of the separatrix curve (red curve in Fig. \ref{seporbitsa}) represents the ISSO and MBSO radii. At these endpoints, the parameter $\mu$ takes values as described below:
\begin{subequations}
\begin{equation}
\mathrm{For \ ISSO,} \ e=0 \ \mathrm{for \ the \ homoclinic \ orbit \ gives} \ \mu=\frac{ 2r_a }{2 r_p r_a}=\frac{1}{r_p}=\frac{1}{r_s}. \nonumber 
\end{equation}
\begin{equation}
\mathrm{For \ MBSO,} \ e=1 \ \mathrm{for \ the \ homoclinic \ orbit \ gives} \ \mu=\frac{1 + r_p /r_a}{2 r_p}=\frac{1}{2 r_p}=\frac{1}{2 r_s}. \nonumber
\end{equation}
\end{subequations} 

\begin{figure}
 \mbox{ \subfigure[]{
\includegraphics[scale=0.6]{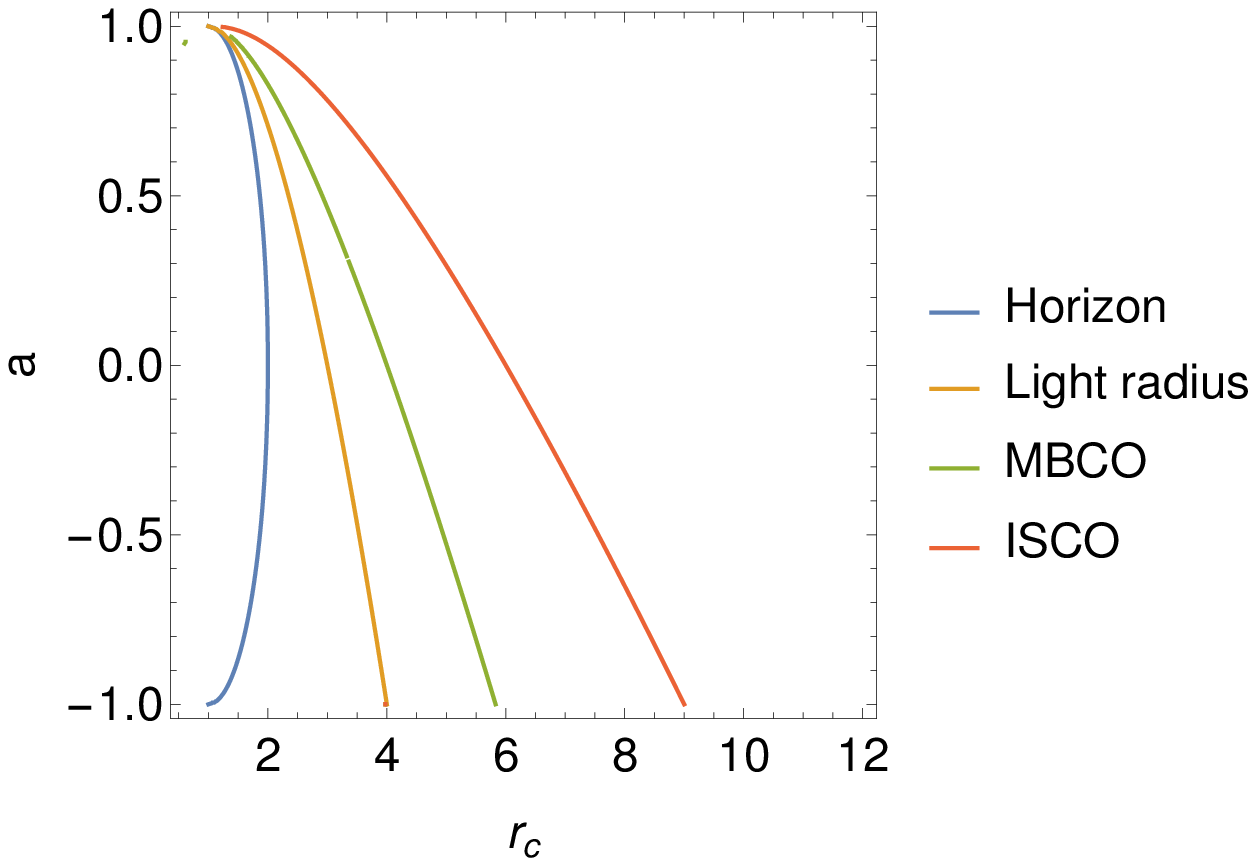} \label{plotQ0}}
\hspace{0.7cm}
\subfigure[]{
\includegraphics[scale=0.6]{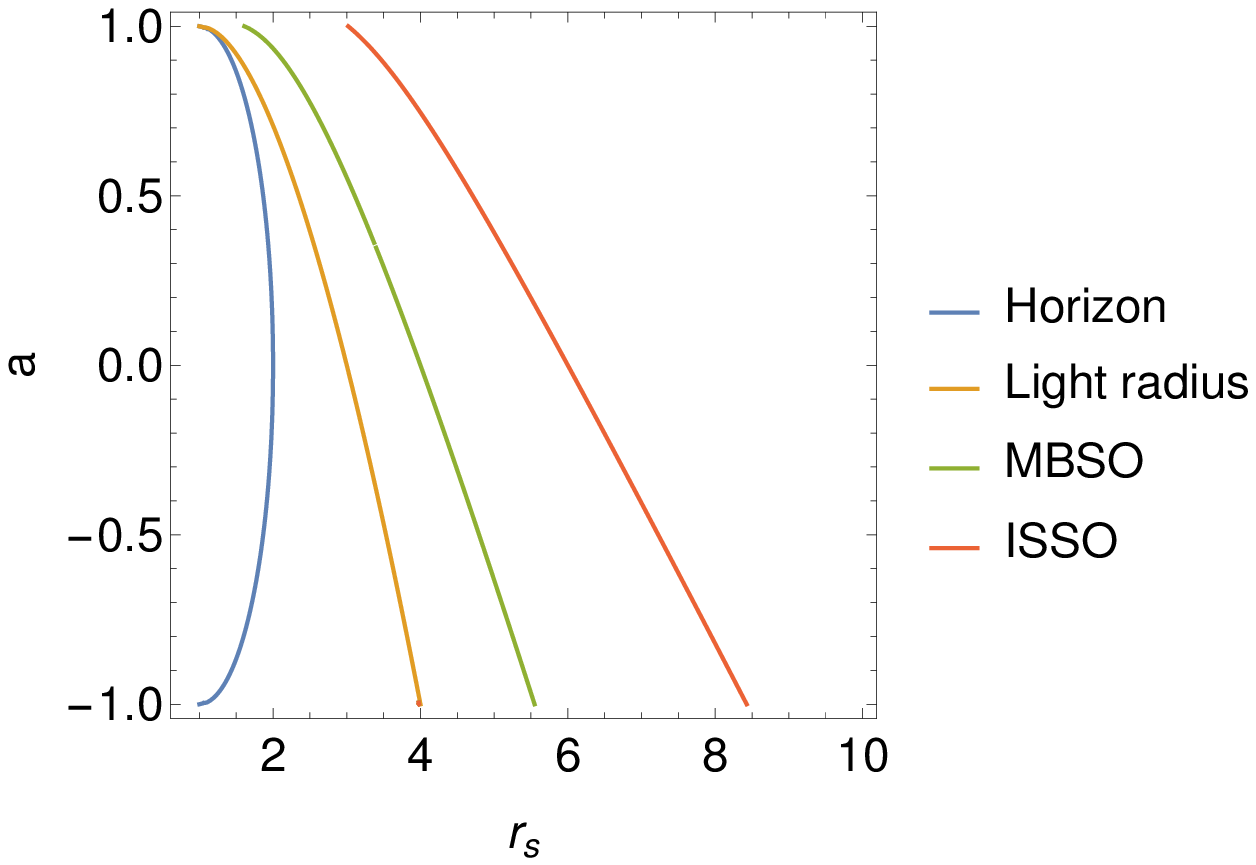}\label{plotQ5}} }
 \mbox{ \subfigure[]{
\includegraphics[scale=0.6]{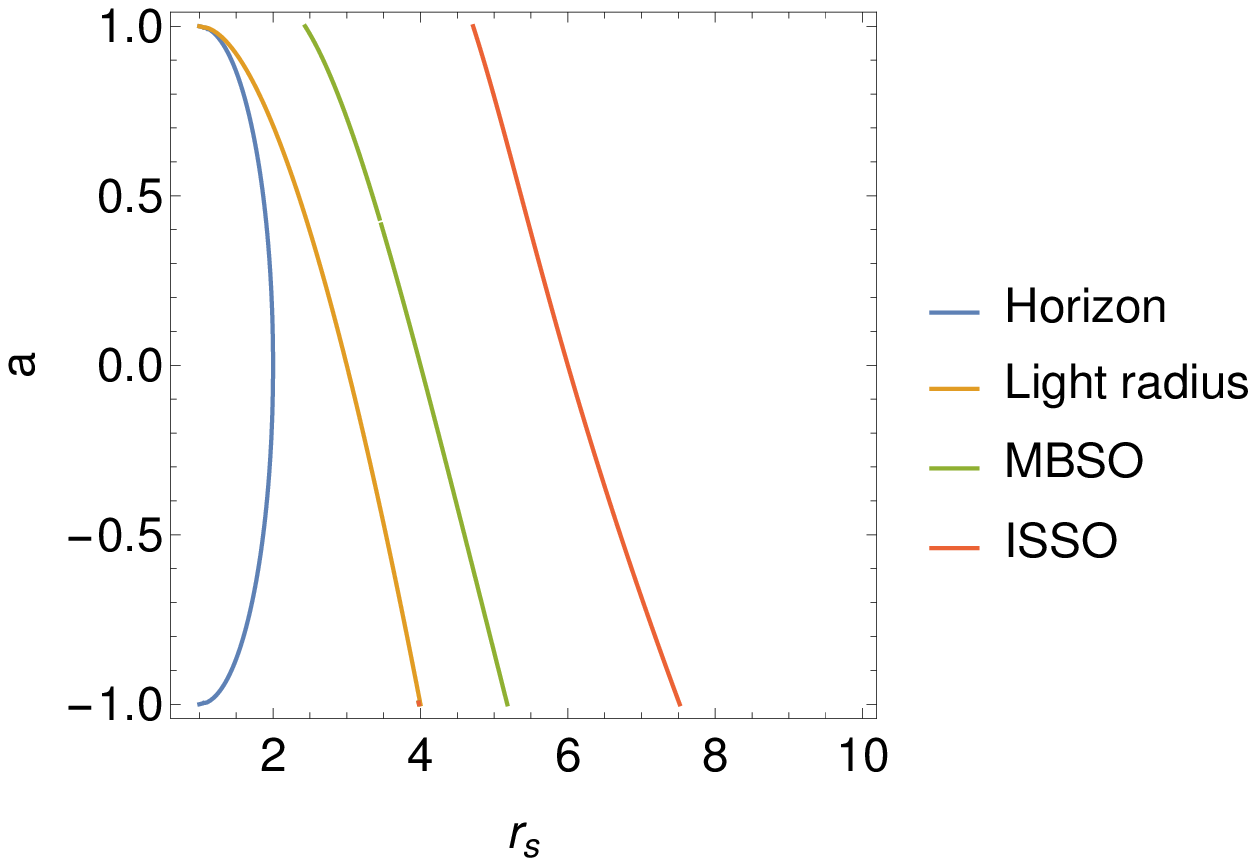} \label{plotQ10}}
\hspace{0.7cm}
\subfigure[]{
\includegraphics[scale=0.6]{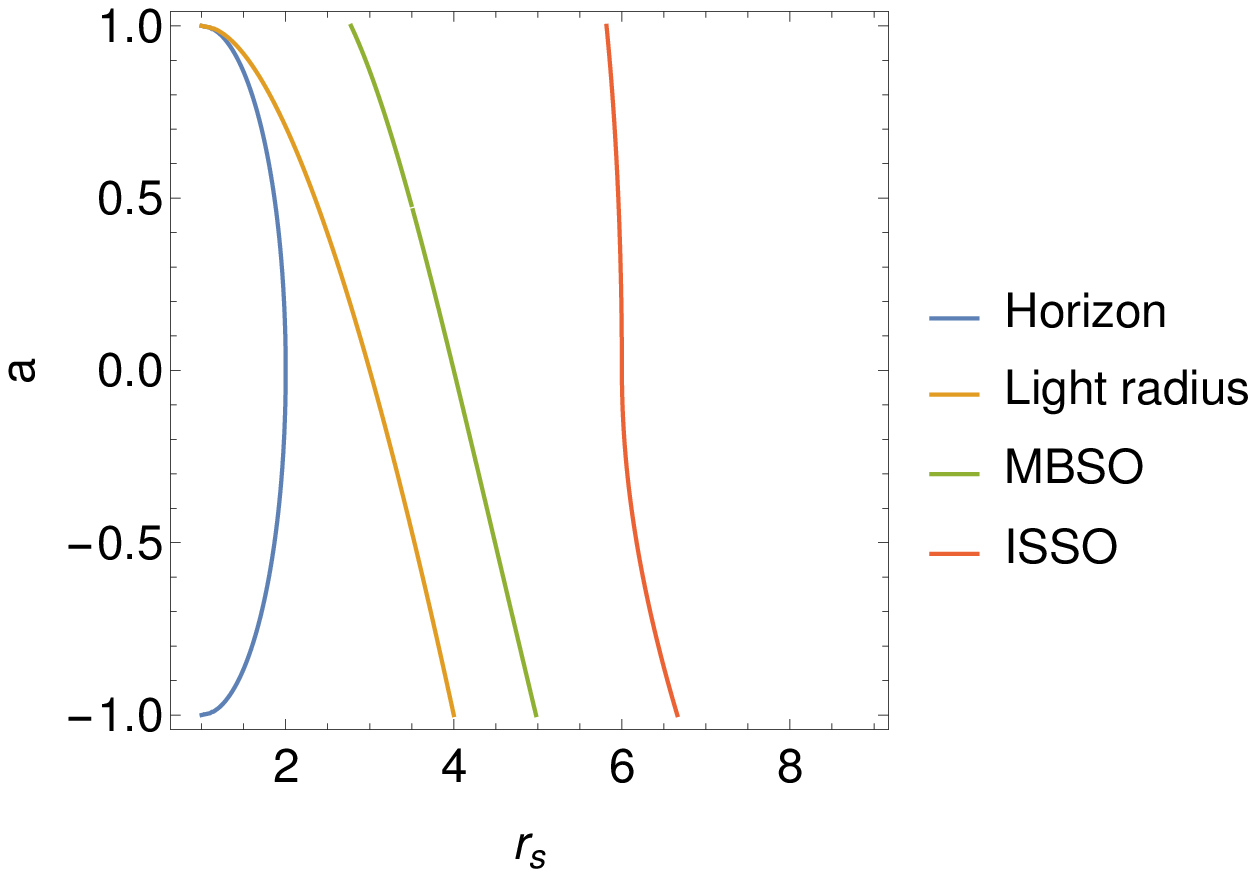}\label{plotQ12}} }
\caption{\label{plotQ} The contours of different important radii around the Kerr black hole in the ($r_s$, $a$) plane for (a) $Q=0$, (b) $Q=5$, (c) $Q=10$, and (d) $Q=12$. }
\end{figure}

The equations for ISSO and MBSO radii can be obtained using the equation of separatrix curve, Eq. \eqref{sepeq}, by plugging in ($e=0$, $\mu=1/r_s$) and ($e=1$, $\mu=1/2 r_s$) to derive ISSO and MBSO respectively (as shown in \ref{sphapp}). Hence, the equations for these radii are given by
\begin{eqnarray}
&&r_{s}^9 -12 r_{s}^8 - 6a^2 r_{s}^7 + 36 r_{s}^7 + 8 a^2 Q r_{s}^6 -28 a^2 r_{s}^6 -24 a^2 Q r_{s}^5 + 9 a^4 r_{s}^5 -24 a^4 Q r_{s}^4 + \nonumber \\
&&48 a^2 Q r_{s}^4 + 16 a^4 Q^2 r_{s}^3 -8a^4 Q r_{s}^3 -48 a^4 Q^2 r_{s}^2 + 48 a^4 Q^2 r_{s} -16 a^6 Q^2=0, \label{ISSOequation}
\end{eqnarray}
for ISSO and 
\begin{eqnarray}
&&r_{s}^8 -8 r_{s}^7 - 2a^2 r_{s}^6 + 16 r_{s}^6 + 2 a^2 Q r_{s}^5 -8 a^2 r_{s}^5 -6 a^2 Q r_{s}^4 + a^4 r_{s}^4 -2 a^4 Q r_{s}^3 + \nonumber \\
&& 8 a^2 Q r_{s}^3 +  a^4 Q^2 r_{s}^2 -2a^4 Q r_{s}^2 -2 a^4 Q^2 r_{s} + a^4 Q^2 =0. \label{MBSOequation}
\end{eqnarray}
for MBSO. The light radius for the spherical orbits can be obtained by equating the denominator of Eq. \eqref{Ensph} to zero, so that $E\rightarrow \infty$, which has the well known form for the equatorial light radius \citep{Bardeen1972} given by 
\begin{equation}
X= 2 \left\lbrace 1+ \cos \left[ \frac{2}{3} \arccos \left( -a\right) \right]  \right\rbrace. \label{ISOrad}
\end{equation}
Fig. \ref{plotQ} shows the contours of these radii in the ($r_s$, $a$) plane for various $Q$ values. 

The effective potential diagram for the non-equatorial separatrix orbits shows double roots ($r_2=r_3$) of $R(r)$ at the periastron of the eccentric orbit and it also represents the spherical orbit radius, $r_s$ (see Fig \ref{seporbitsb}). One of the remaining two roots of $R(r)$ represents the apastron ($=r_1> r_s$) of the eccentric orbit and the other inner root ($=r_4< r_s$) is not the part of bound trajectory.

 Now, following a similar method used in \cite{Levin2009}, we derive the expressions for $e$ and $\mu$ for separatrix orbits with $Q\neq0$. We write $R(r)=0$ in the form
\begin{subequations}
\begin{equation}
u^4 + a^{'} u^3 +b^{'}u^2 +c^{'}u +d^{'}=0, \label{quartic}
\end{equation}
where $u=1/r$ and
\begin{equation}
a^{'}=  - \frac{2 \left[ x^2 +Q \right] }{a^2 Q}, \ \ \
b^{'}=\frac{\left(x^2 + 2a E x + a^2 +Q \right) }{a^2 Q }, \ \ \
c^{'}= - \frac{2}{a^2 Q}, \ \ \
d^{'}= \frac{1-E^2}{a^2 Q}. \label{adash} 
\end{equation}

\end{subequations}
For the separatrix orbits, Eq. \eqref{quartic} can be written as 
\begin{equation}
\left(u - u_s \right)^2 \cdot \left[u^2 -\left( u_1 + u_4\right) u + u_1 u_4 \right]=0, \label{reg4quartic}
\end{equation}
where $u_s=1/r_s$, $u_1=1/r_1$ apastron of the orbit, and $u_4=1/r_4$ corresponds to the inner most root of $R(r)$. 
The comparison of $u^3$ and constant term of the above equation with those of Eq. \eqref{quartic} further gives the expression
\begin{equation}
u_1=\frac{1}{2}\left[-\left( a^{'}+ 2 u_s\right) - \sqrt{\left( a^{'}+ 2 u_s\right)^2 -\frac{4d^{'}}{u_{s}^2}}  \right] . \label{u1sep}
\end{equation}
The conic parameters for such an orbit are given by
\begin{eqnarray}
e_s=&&\frac{u_s -u_1}{u_s+u_1}, \ \ \ \ \ \
\mu_s= \frac{u_s+u_1}{2}, \label{emusepQ}
\end{eqnarray}
where the substitution of $u_1$ and $u_s=1/r_s$ yields
\begin{subequations}
\begin{eqnarray}
e_s=&&\frac{4 +a^{'}r_s+\sqrt{\left(r_s a^{'}+ 2 \right)^2 -4d^{'}r_{s}^{4}}}{-a^{'}r_s-\sqrt{\left( r_s a^{'}+ 2\right)^2 -4d^{'}r_{s}^{4}}}, \label{esepgen}\\
\mu_{s}=&&\frac{1}{4r_s}\left[-a^{'}r_s - \sqrt{\left(r_s a^{'}+ 2 \right)^2 -4d^{'}r_{s}^{4}} \right]; \label{musepgen}
\end{eqnarray}
since a homoclinic orbit has same energy and angular momentum of the unstable spherical orbit, as shown in Fig. \ref{seporbitsb}; hence $a^{'}$ and $d^{'}$ can be rewritten using the formulae of $E$ and $L$ for the spherical orbits, Eqs. (\ref{Ensph}-\ref{Lsph}), to be
\begin{equation}
 a^{'}= \frac{2\left\lbrace  \splitfrac{2a^4Q -r_{s}^2 \left(r_s -3 \right) \left[r_{s}^2 -\left(r_s -3 \right) Q \right] -a^2 r_s \left( r_s^3 +r_s^2 -2 Q r_s +8Q \right)}{ +2a  \left[r_s \left(r_s -2 \right) +a^2  \right] \sqrt{a^2 Q^2 -r_s^3 Q \left(r_s -3 \right) + r_s^5} - Q r_s \left[ r_s\left(r_s -3 \right)^2 -4 a^2 \right] } \right\rbrace   }{a^2 Q r_s \left[r_{s} \left( r_s-3\right)^2 -4a^2 \right]}, 
\label{adashsep}
\end{equation}
\begin{equation}
 d^{'}=\frac{\left\lbrace \splitfrac{-2a^4 Q - \left( r_s-3\right) \left( r_s-2 \right)^2 r_{s}^4 +a^2 r_s \left[r_{s}^{2}\left(3r_s -5 \right) + Q \left(r_s\left(r_s -4 \right) +5  \right)   \right]}{ +r_{s}^4 \left[r_{s} \left( r_s-3\right)^2 -4a^2 \right]+2 a \left[r_s \left(r_s -2 \right) +a^2  \right] \sqrt{a^2 Q^2 -r_s^3 Q \left(r_s -3 \right) + r_s^5} } \right\rbrace  }{a^2 Q r_{s}^4 \left[r_{s} \left( r_s-3\right)^2 -4a^2 \right] } \label{ddashsep}.
\end{equation}
\label{emusepfinal}
\end{subequations}
These expressions reduce to the ($e$, $\mu$) formulae for the equatorial separatrix orbits (see \ref{separatrixapp} for the details) when the limit $Q\rightarrow 0$ is implemented, to the forms previously derived by \cite{Levin2009}:

\begin{eqnarray}
e_s=-\frac{r_{c}^{2} - 6 r_{c} - 3 a^{2} + 8 a \sqrt{r_{c}}}{r_{c}^{2}+a^{2}-2r_{c}},
 \label{esepfor}\ \
\mu_s=\frac{r_{c}^{2}+a^{2}-2r_{c}}{4 r_{c}\left( \sqrt{r_{c}} - a \right) ^{2}}.
\end{eqnarray}

\subsection{\underline{Exact forms for the non-equatorial separatrix trajectories}}
\label{separatrixtrajec}
In this section, we show the reduction of our general trajectory solutions, Eqs. \ref{final}, for the case of separatrix orbits with $Q\neq0$ to simple expressions. The separatrix or homoclinic orbits represent a curve in the ($e$, $\mu$) plane for a fixed $a$ and $Q$ combination, Fig. \ref{seporbits}, which is also the boundary of the bound orbit region defined by Eqs. \eqref{deltaregion}. This separatrix curve is defined by Eq. \eqref{sepeq}, which gives us the relation
\begin{equation}
x^2+Q = \frac{1+ 4 \mu^3 a^2 Q \left( 1+ e\right)}{\mu^2 \left( 3-e\right) \left(1+ e \right)};
\end{equation} 
this further reduces the expressions of $A$, $B$, $C$ (Eqs. (\ref{A}-\ref{C})) and correspondingly the expressions of $n^2$ and $m^2$ to 
\begin{subequations}
\begin{equation}
n^2=1 \ \ \mathrm{or} \ \ k^2=1, \label{ksep1}
\end{equation}
\begin{equation}
m^2= \frac{a^2 Q \mu^3 e \left(1+ e \right) \left( 3-e\right)  }{\left[ 1+ 2 a^2 \left( -1+e^2\right) Q \mu^3 \right] }.
\end{equation}
\end{subequations}
The integrals governing the vertical motion ($\theta$ integrals) given by Eqs. (\ref{H1}, \ref{H2}, \ref{integral2}) retain their same form as they do not involve $k^2=1$, whereas, the radial integrals given by Eqs. (\ref{I1a}-\ref{integral1}) reduce further, when $k^2=1$ is substituted. The elliptic integrals reduce to forms involving trigonometric and logarithmic functions using the following identities given here
\begin{subequations}
\begin{equation}
\Pi \left( q^2 , \alpha , 1 \right) =  \frac{1}{1-q^2} \left[ \mathrm{ln} \left( \tan \alpha + \sec \alpha \right) - q \ \mathrm{ln} \sqrt{\frac{1+ q \sin \alpha}{1- q \sin \alpha}}  \right], \mathrm{where}  \  q^2>0,  \ q^2 \neq 1,  \nonumber
\end{equation}
\begin{equation}
= \frac{1}{1-q^2} \left[ \mathrm{ln} \left( \tan \alpha + \sec \alpha \right) + \mid q \mid \tan^{-1} \left(  \mid q \mid \sin \alpha \right)  \right], \ \mathrm{where} \ q^2 <0, 
\end{equation}
\begin{eqnarray}
F \left( \alpha, 1 \right) &&=\mathrm{ln} \left( \tan \alpha + \sec \alpha \right), \\
K \left( \alpha, 1 \right) &&= \sin \alpha.
\end{eqnarray}
\label{sepred}
\end{subequations}
\begin{table}
\begin{center}
\caption{This table summarizes the trajectory solution derived in \S \ref{separatrixtrajec} for the non-equatorial separatrix orbits.}
\scalebox{0.73}{
\begin{tabular}{|c|c|}
\hline
\hline
 & Analytic solutions\\
\hline
& \\
$S_3$ & $\displaystyle{=\frac{1}{\left(1+ p_2^{2} \right)} \left[ \frac{p_2^{2}}{\sqrt{-\left(p_2^{2} + m^2 \right) }  } \mathrm{ln} \sqrt{\frac{\sqrt{1-m^2} + \sqrt{- \left(p_2^{2} + m^2 \right)} \sin \alpha}{\sqrt{1-m^2} - \sqrt{- \left(p_2^{2} + m^2 \right)} \sin \alpha}}  +\frac{\mathrm{ln} \left( \tan \alpha + \sec \alpha \right)}{\sqrt{1-m^2}  }\right]}$ \\
& \\
\hline
& \\
$S_4$ & $\displaystyle{=\frac{1}{\left(1+ p_3^{2} \right)} \left[  \frac{p_3^{2}}{\sqrt{-\left(p_3^{2} + m^2 \right) } } \mathrm{ln} \sqrt{\frac{\sqrt{1-m^2} + \sqrt{- \left(p_3^{2} + m^2 \right)} \sin \alpha}{\sqrt{1-m^2} - \sqrt{- \left(p_3^{2} + m^2 \right)} \sin \alpha}}  + \frac{\mathrm{ln} \left( \tan \alpha + \sec \alpha \right)}{\sqrt{1-m^2}  } \right] }$ \\
& \\
\hline
 & \\
 $S_5$ & $\displaystyle{= \frac{1}{\sqrt{1-m^2} \left( m^2 + p_{1}^2\right)^2} \left[ \frac{m^2 \left( m^2 - p_{1}^2 m^2 +2 p_1^2\right) }{\left( 1+p_1^2\right) }\mathrm{ln} \left( \tan \alpha + \sec \alpha \right) + p_{1}^4 S_7 + \frac{2 p_{1}^2 m^2 \left(1-m^2 \right) }{\left( 1+p_1^2\right)} \mid s \mid \tan^{-1} \left[ \mid s \mid \sin \alpha \right] \right]}$ \\
\hline
& \\
$S_6$ & $ \displaystyle{ =\frac{\mathrm{ln} \left( \tan \alpha + \sec \alpha \right)}{\sqrt{1-m^2} \left(1+ p_1^{2} \right) } + \frac{p_{1}^2 \sqrt{1-m^2 } }{ \left( m^2 + p_{1}^2\right) \left( 1+ p_{1}^2\right) } \mid s \mid \tan^{-1} \left[ \mid s \mid \sin \alpha \right]}$ \\
\hline & \\
$S_7$ & $\displaystyle{=\frac{1}{2 \left(1- s^2 \right)^2}\left[ \frac{s^4 \sin \alpha \cos^2 \alpha}{ \left( 1- s^2 \sin^2 \alpha \right)  } + 2 \ \mathrm{ln} \left( \tan \alpha + \sec \alpha \right) - s^2 \sin \alpha  + \left( 3-s^2\right) \mid s \mid \tan^{-1} \left( \mid s \mid \sin \alpha \right)  \right]}$ \\
\hline
\hline
\end{tabular}
\label{septable}
}

\end{center}

\end{table}
 The final and simple expressions for the azimuthal angle, $\left( \phi-\phi_{0}\right) $, $\left( t-t_{0}\right) $, and the equation relating $r-\theta$ motion for the non-equatorial separatrix trajectories  (see \ref{Qsepintegrals} for the derivation) are given by 
\begin{subequations}
\begin{equation}
\phi-\phi_{0}=  \frac{1}{2} \left\lbrace \frac{\sqrt{\mu\left( 1+e \right) \left( 3-e \right)}}{\sqrt{ e  \left[ 1+ 2 a^2 \left( -1 + e^2 \right) Q \mu^3 \right]  \left(1-a^2 \right) }} \left[\frac{\left[L a^2  -2 x  r_{+}  \right]}{\left( a^2 \mu - a^2 \mu e -r_{+}\right)} S_3+  \frac{\left[-L a^2  +2 x r_{-}  \right]}{\left( a^2 \mu - a^2 \mu e -r_{-}\right) } S_4  \right] -H_1  \right\rbrace , \label{phisep}
\end{equation} 
\begin{equation}
t-t_{0}= \frac{\sqrt{\left( 1+e \right) \left( 3-e \right)}}{\sqrt{e \mu \left[ 1+ 2 a^2 \left( -1 + e^2 \right) Q \mu^3 \right] }}\left\lbrace \frac{ E  }{\mu \left( 1- e \right)^2   } S_5 + \frac{  a^2 \mu \left(-L a + 2 E r_{-} \right) }{r_{-}\sqrt{ \left(1-a^2 \right) }\left( a^2 \mu -a^2 \mu e -r_{+}\right) } S_3+ \frac{2 E }{\left( 1- e \right) }S_6 \right. \nonumber 
\end{equation}
\begin{equation}
 \left.   + \frac{  a \mu \left(-2 L r_{-}\sqrt{1-a^2} -2 E a r_{-} +L a^2 \right) }{r_{-}\sqrt{\left(1-a^2 \right) }\left( a^2 \mu -a^2 \mu e -r_{-}\right) } S_4 \right\rbrace+ \frac{1}{2} H_2 , \label{tsep}
\end{equation}
\begin{equation}
\frac{2\mu \left( 1-e^2\right)a z_{+}\sqrt{1-E^2}}{\sqrt{C-A+\sqrt{{B}^2-4 AC}}} \mathrm{ln}\left( \tan \alpha + \sec \alpha \right) =\left\lbrace F\left( \arcsin \left( \frac{\cos \theta_{0}}{z_{-}}\right) ,\frac{z_{-}^{2}}{z_{+}^{2}}\right)-  F\left( \arcsin \left( \frac{\cos \theta}{z_{-}}\right),\frac{z_{-}^{2}}{z_{+}^{2}}\right) \right\rbrace. \label{rthetasep}
\end{equation}
\label{Qsepfinal}
\end{subequations}
where integrals $S_3-S_7$ are summarized in Table \ref{septable}, and $H_1\left(  \theta , \theta_{0}, e, \mu , a, Q \right)$, $H_2\left(  \theta , \theta_{0}, e, \mu , a, Q \right)$ are given by Eq. \eqref{H1}, \eqref{H2} respectively.

These expressions have their utility in evaluating the trajectory evolution of inspiralling objects near the separatrix, and just before plunging, for extreme mass ratio inspirals (EMRIs) in gravitational wave astronomy \cite{Glampedakisetal2002,Drascoetal2005,Drasco2006}.

\section{Trajectories}
\label{trajec}
 The analytic solution  of the integrals of motion presented in this paper in \S \ref{analyticsoln} provides a direct and exact recipe to study bound trajectories without involving numerical integrations. These expressions have their utility in calculating extreme mass ratio inspirals (EMRIs) in gravitational wave astronomy, where numerical models consider an adiabatic progression through series of geodesics around a Kerr black hole \cite{Glampedakisetal2002,Drascoetal2005,Drasco2006}. We now discuss various kinds of bound geodesics around Kerr black hole using our analytic solution for the integrals of motion. We use the translation formulae, Eqs. (\ref{Esqr}-\ref{Rvar}), to obtain the integrals of motion only in terms of ($e$, $\mu$, $a$, $Q$) parameters. To sketch the trajectories, we have chosen the starting point for the trajectories to be ($\beta_{0}=\pi/2$, $\alpha=0$) as it follows from Eq. \eqref{Rthetaint}. We use Eq. \eqref{rthetafinal} to calculate corresponding small change in $\theta$ or $\beta$ with the small change in $r$ or $\alpha$ and substitute corresponding ($r$, $\theta$) or ($\alpha$, $\beta$) values in Eqs. \eqref{phifinal} and \eqref{tfinal} to calculate ($\phi$, $t$).
 
\begin{figure}
 \mbox{ \subfigure[]{
\includegraphics[scale=0.31]{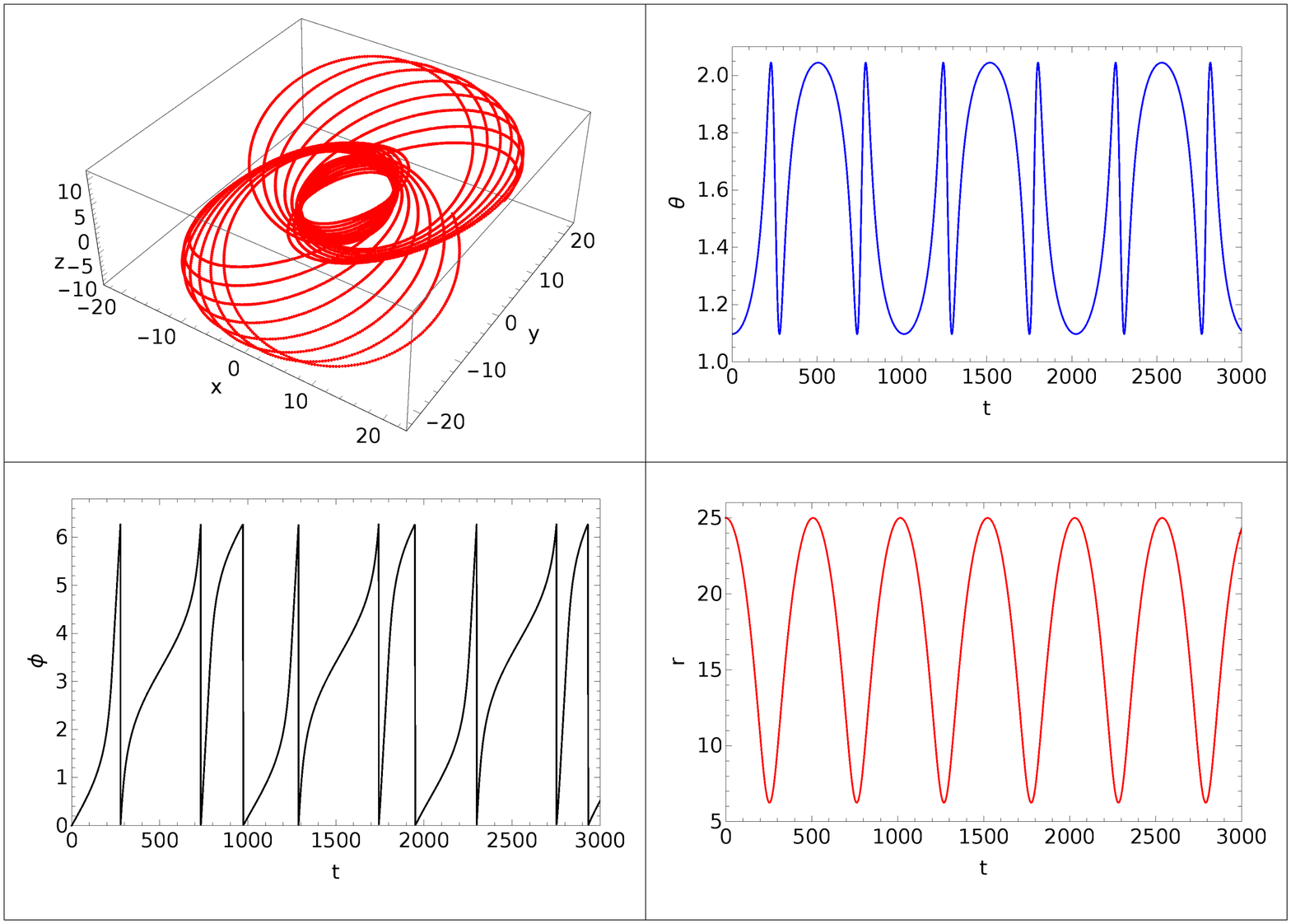} }
\hspace{-0.5cm}
\subfigure[]{
\includegraphics[scale=0.31]{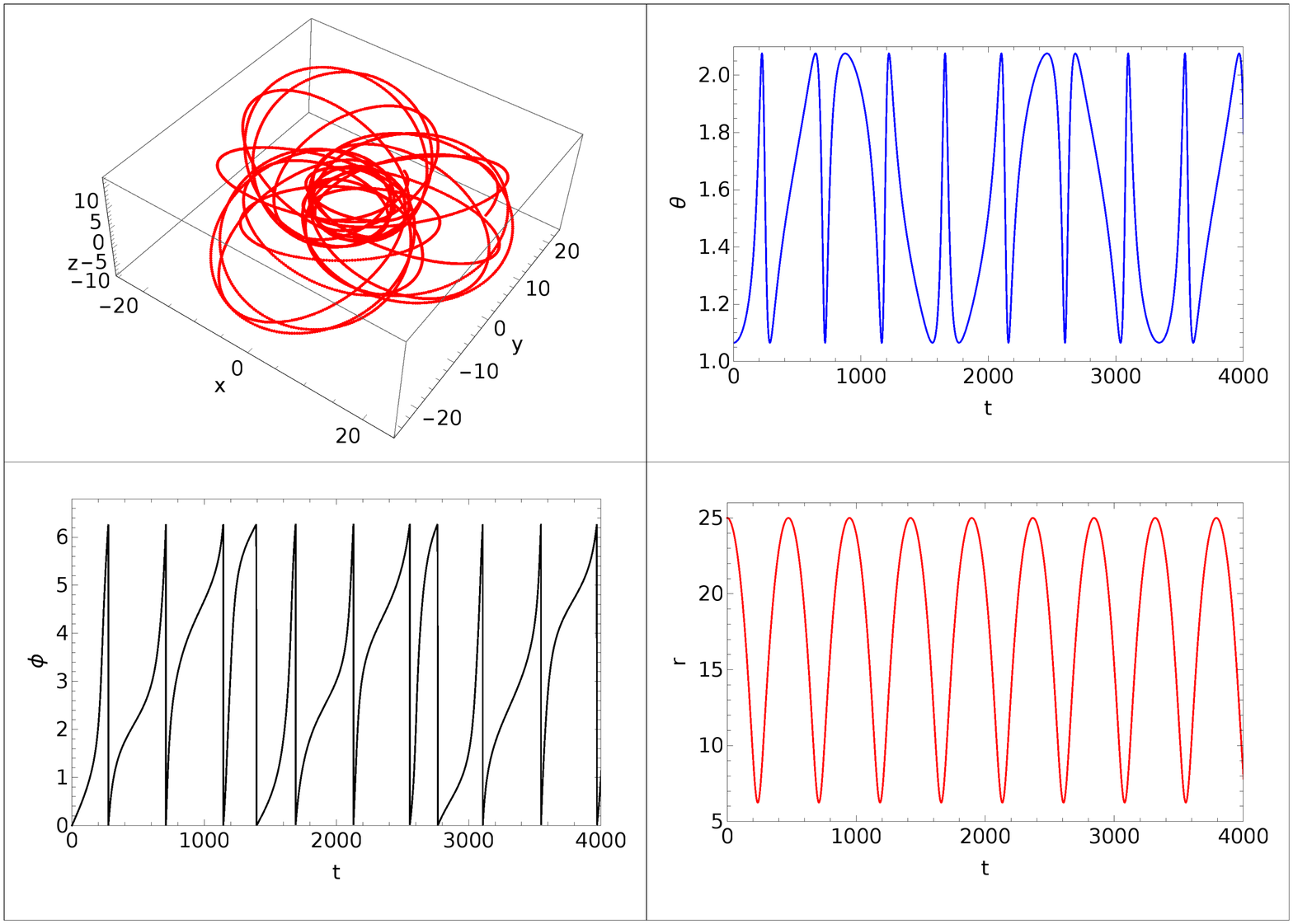}}} 
\mbox{ \subfigure[]{
\includegraphics[scale=0.31]{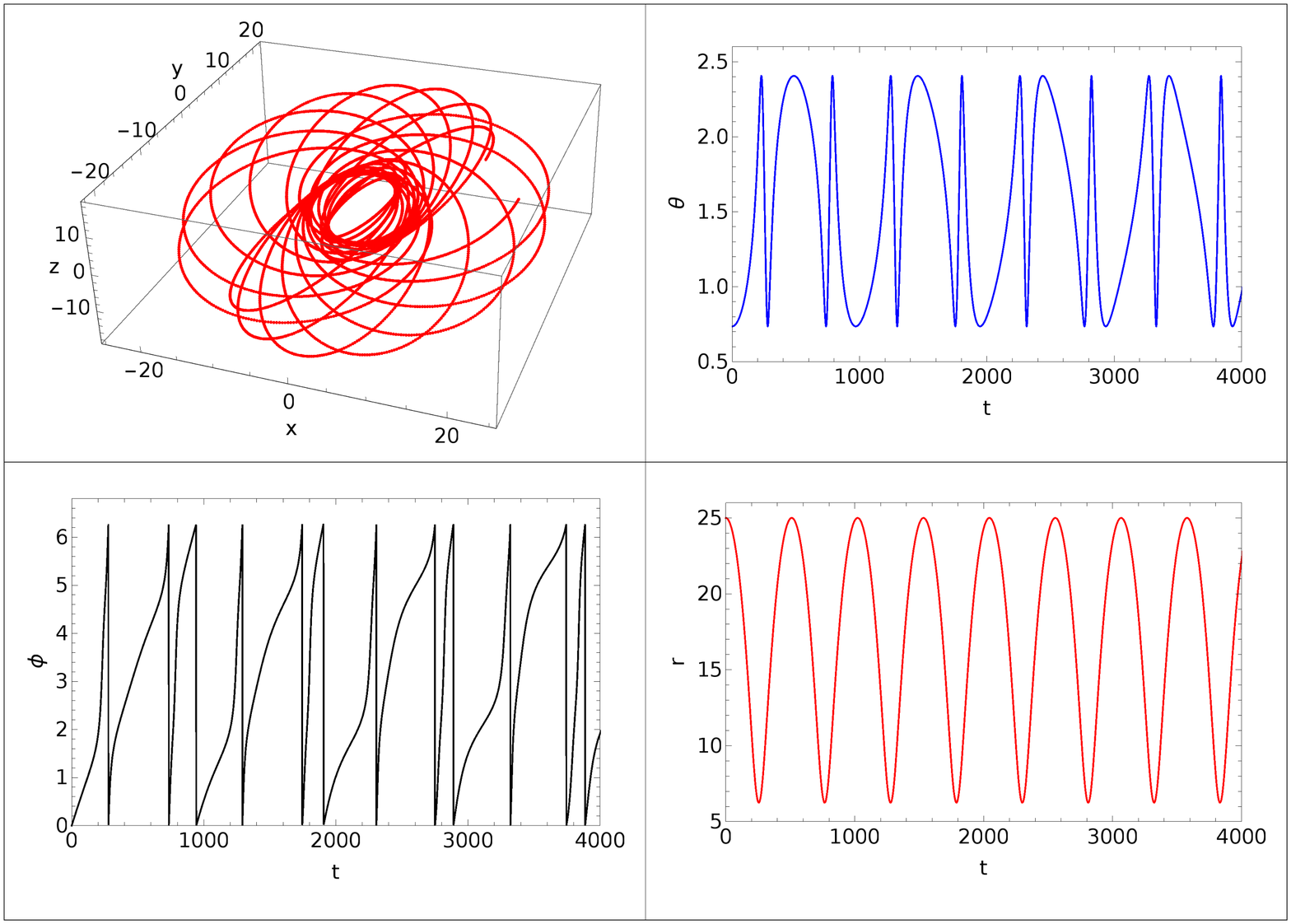}}
\hspace{-0.4cm}
\subfigure[]{
\includegraphics[scale=0.31]{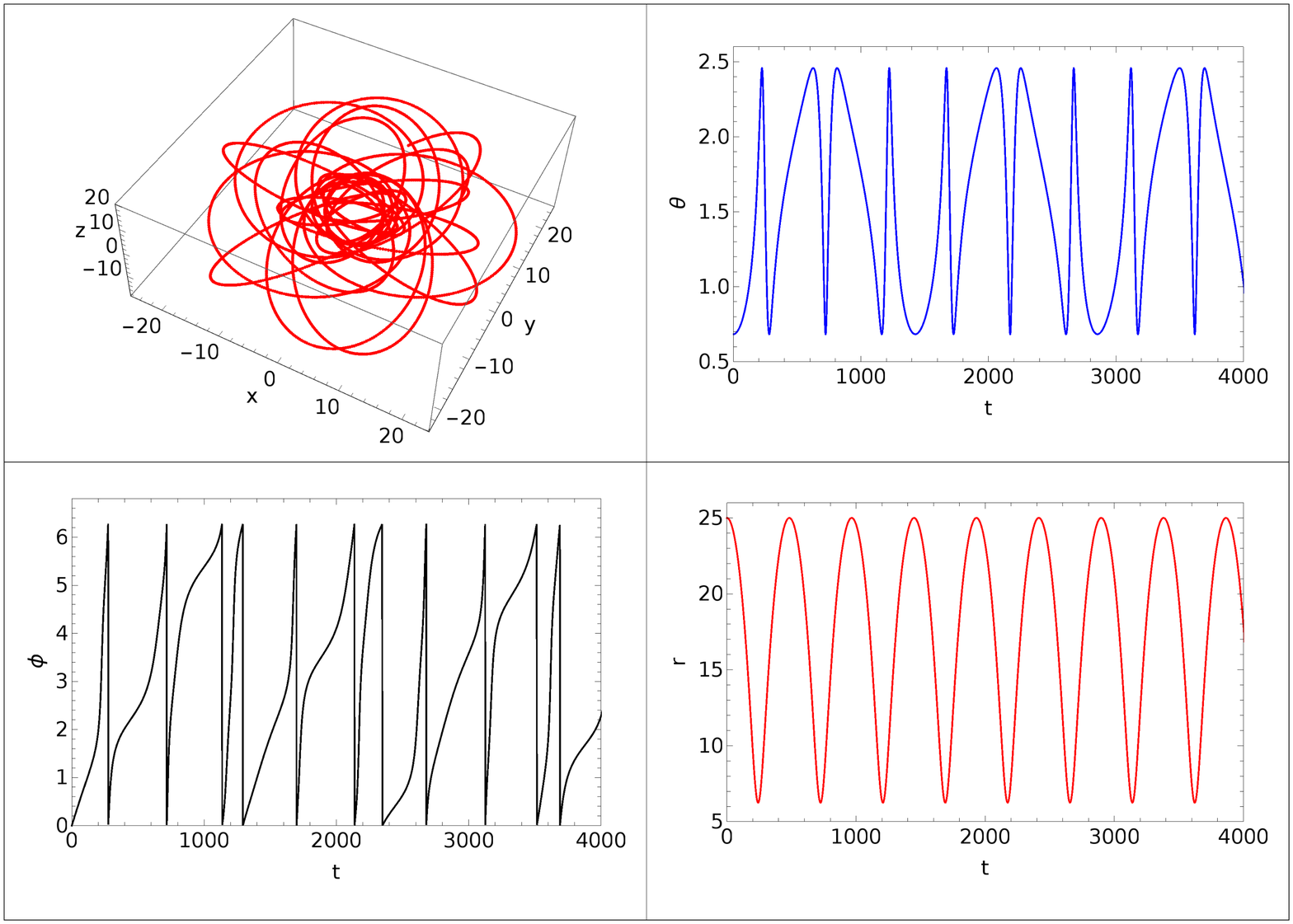}}}
 \caption{\label{eccpro}The figure shows prograde eccentric bound orbits (a) E1, (b) E2, (c) E3, and (d) E4  in the table \ref{parameters}, for various combinations of ($e$, $\mu$, $a$, $Q$) satisfying Eq. \eqref{boundcondreg3} and also presents the evolution of corresponding $\theta$, $\phi$ and $r$ with coordinate time, $t$.}
 \end{figure}

\begin{figure}
 \mbox{ \subfigure[]{
\includegraphics[scale=0.31]{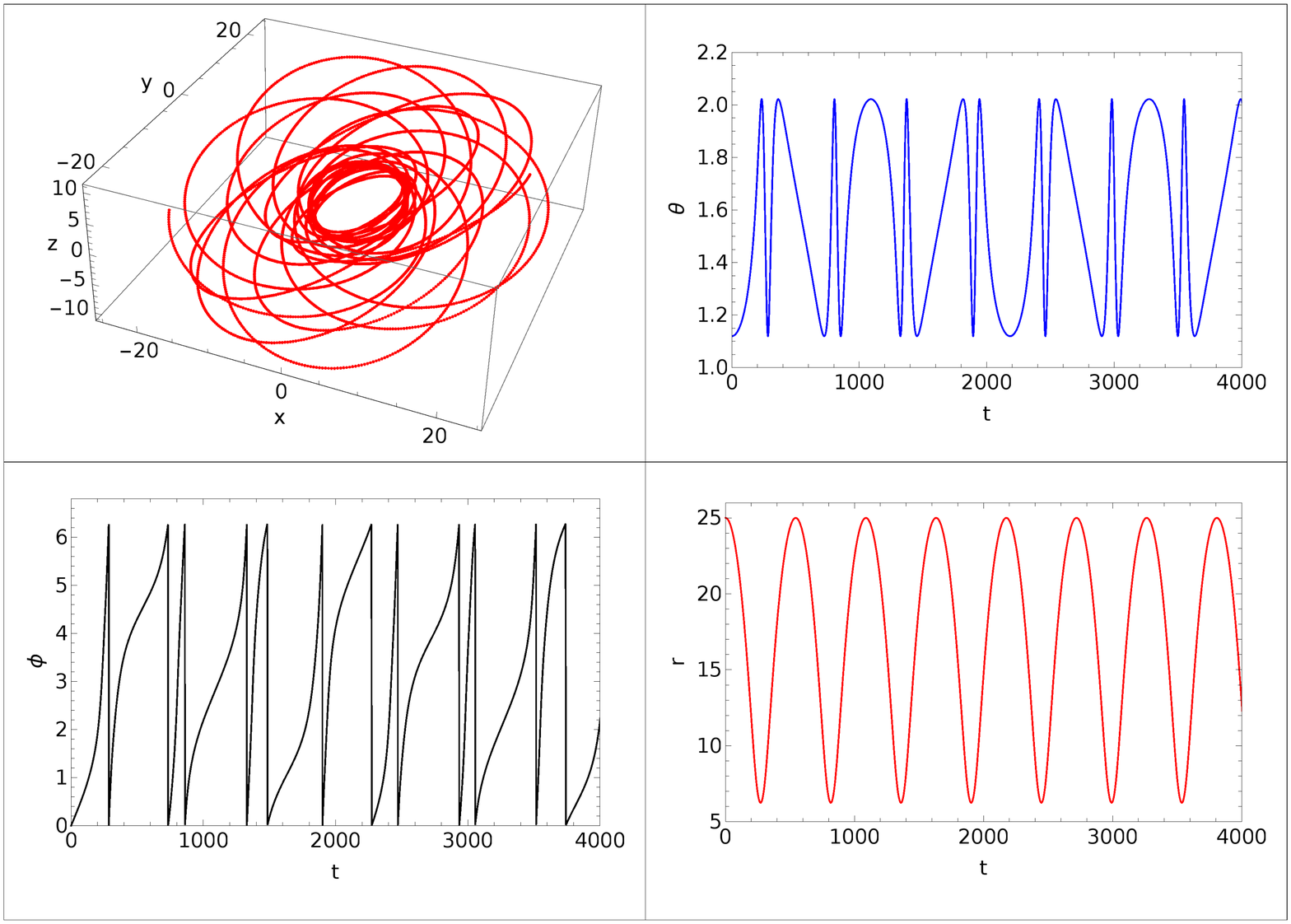} }
\hspace{-0.5cm}
\subfigure[]{
\includegraphics[scale=0.31]{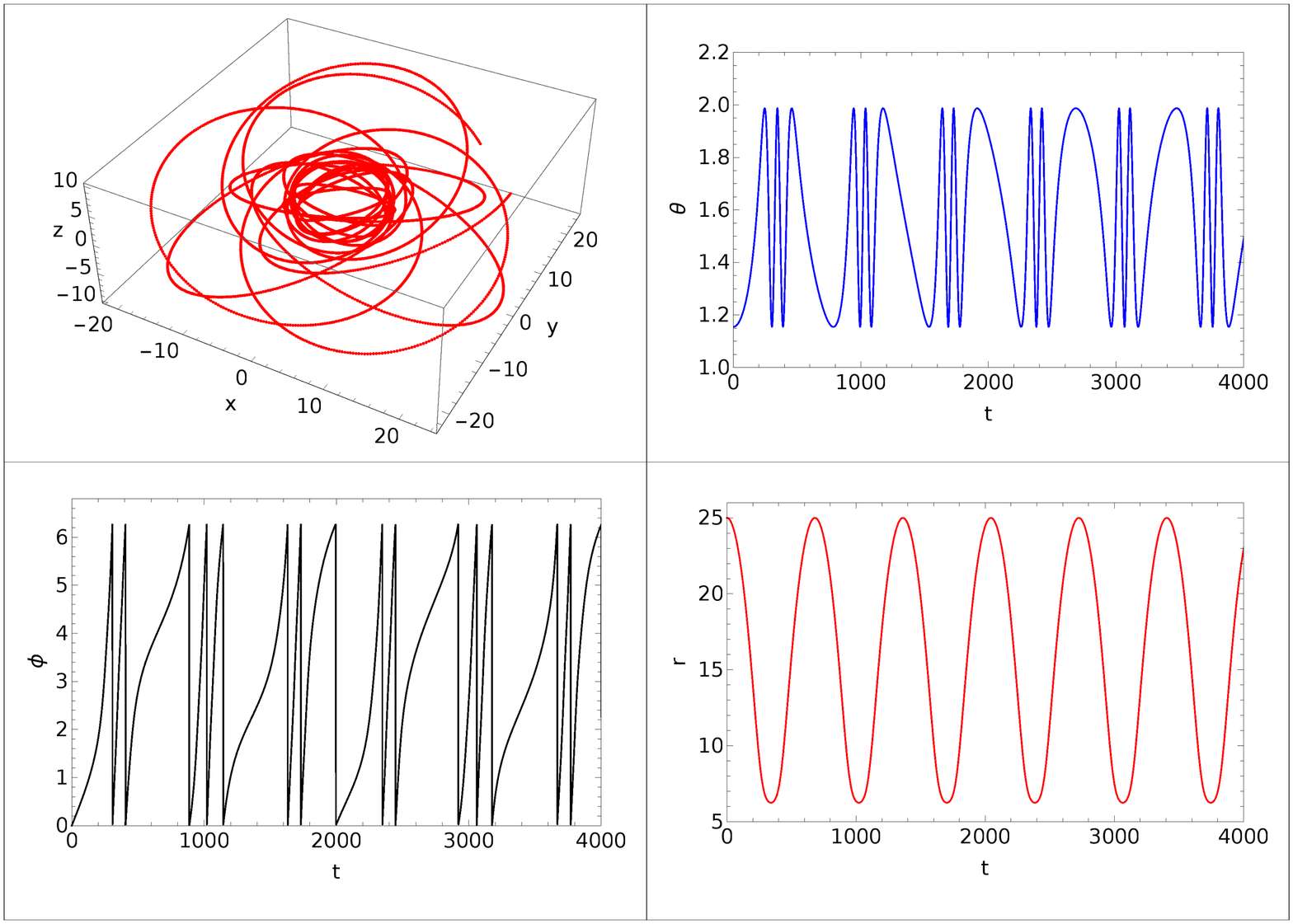}}} 
\mbox{ \subfigure[]{
\includegraphics[scale=0.31]{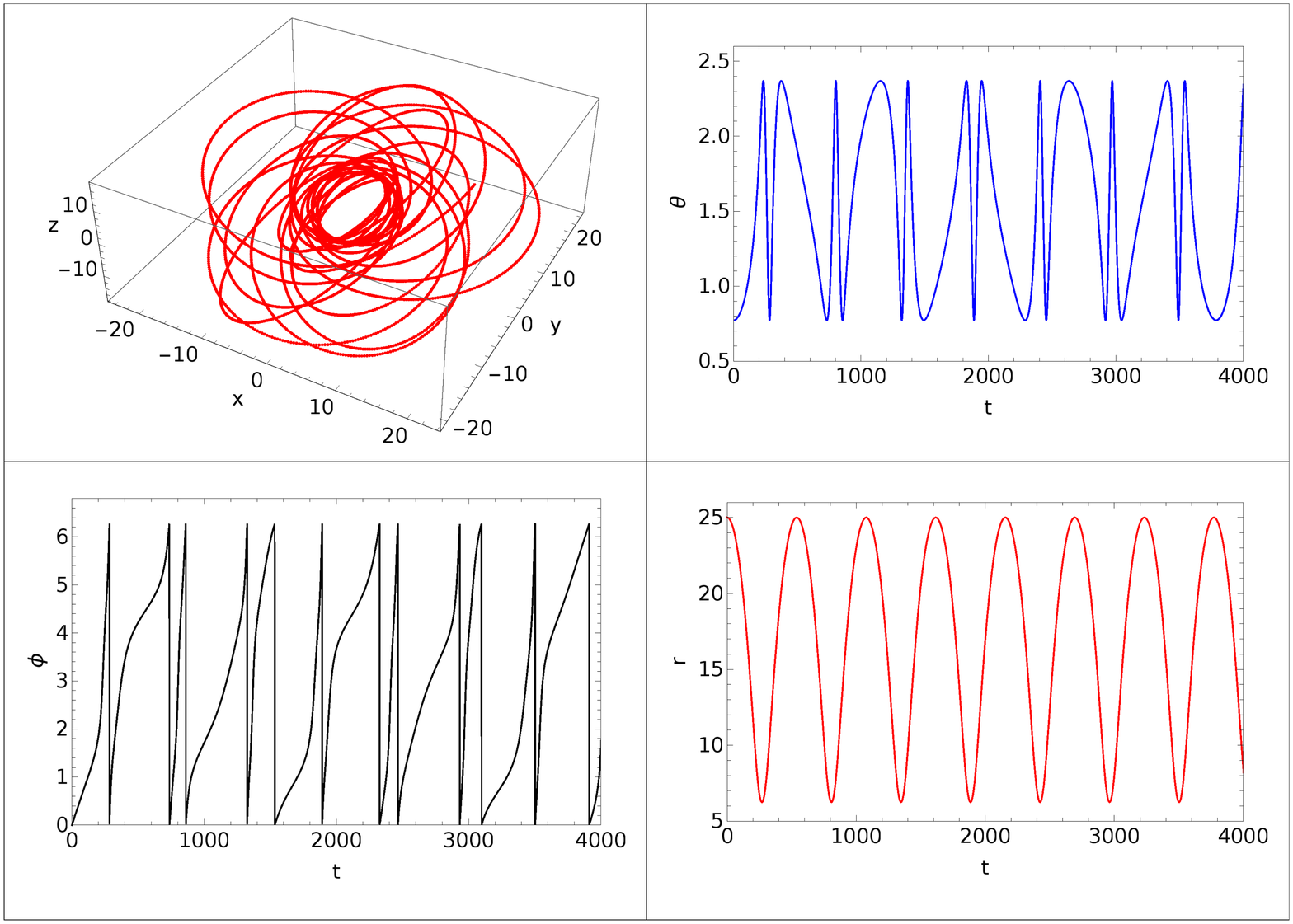}}
\hspace{-0.4cm}
\subfigure[]{
\includegraphics[scale=0.31]{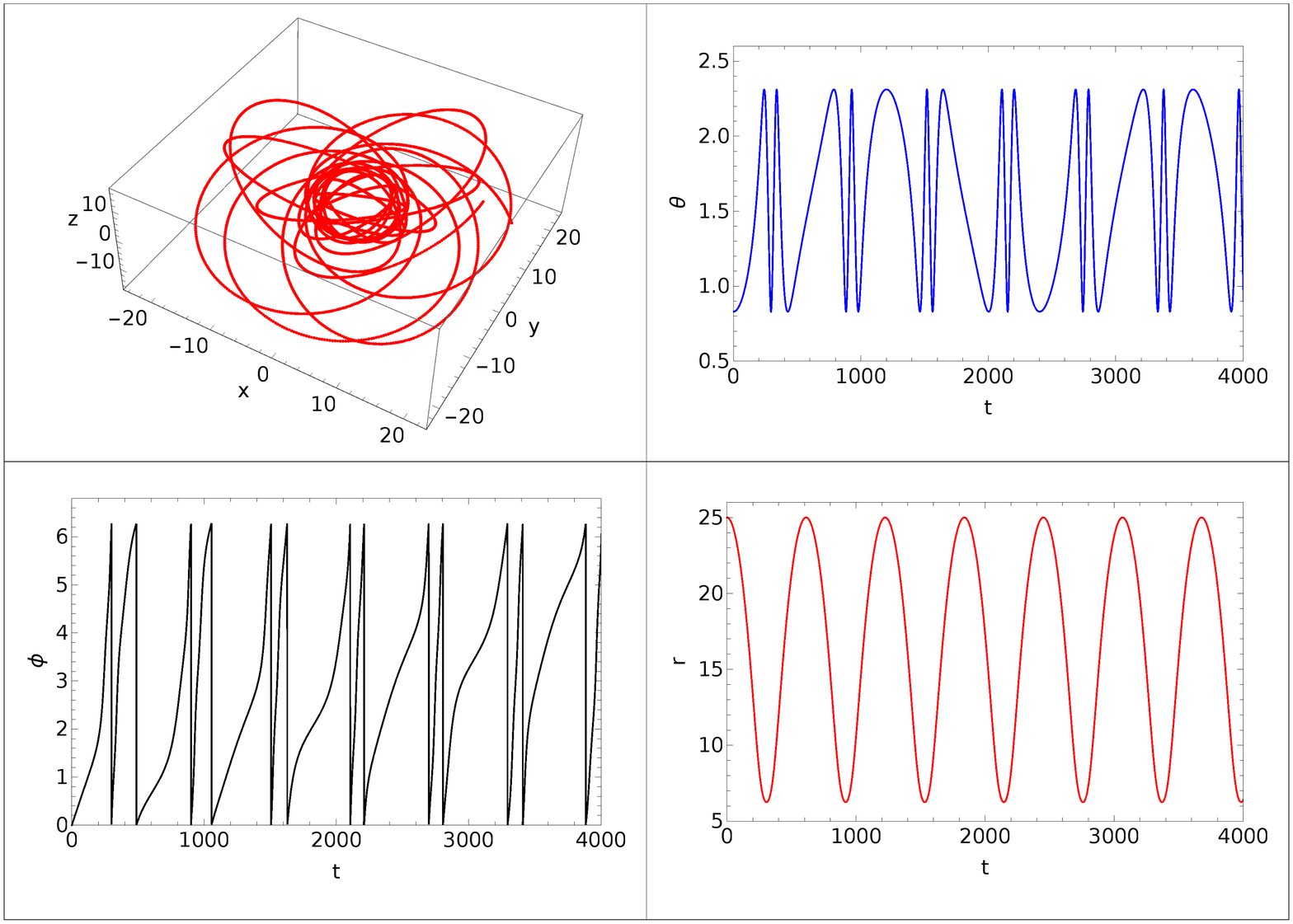}}}
 \caption{\label{eccret}The figure shows retrograde eccentric bound orbits (a) E5, (b) E6, (c) E7, and (d) E8 in the table \ref{parameters}, for various combinations of ($e$, $\mu$, $a$, $Q$) satisfying Eq. \eqref{boundcondreg3} and also presents the evolution of corresponding $\theta$, $\phi$ and $r$ with coordinate time, $t$.}
 \end{figure}

\begin{figure}
 \mbox{ \subfigure[]{
\includegraphics[scale=0.318]{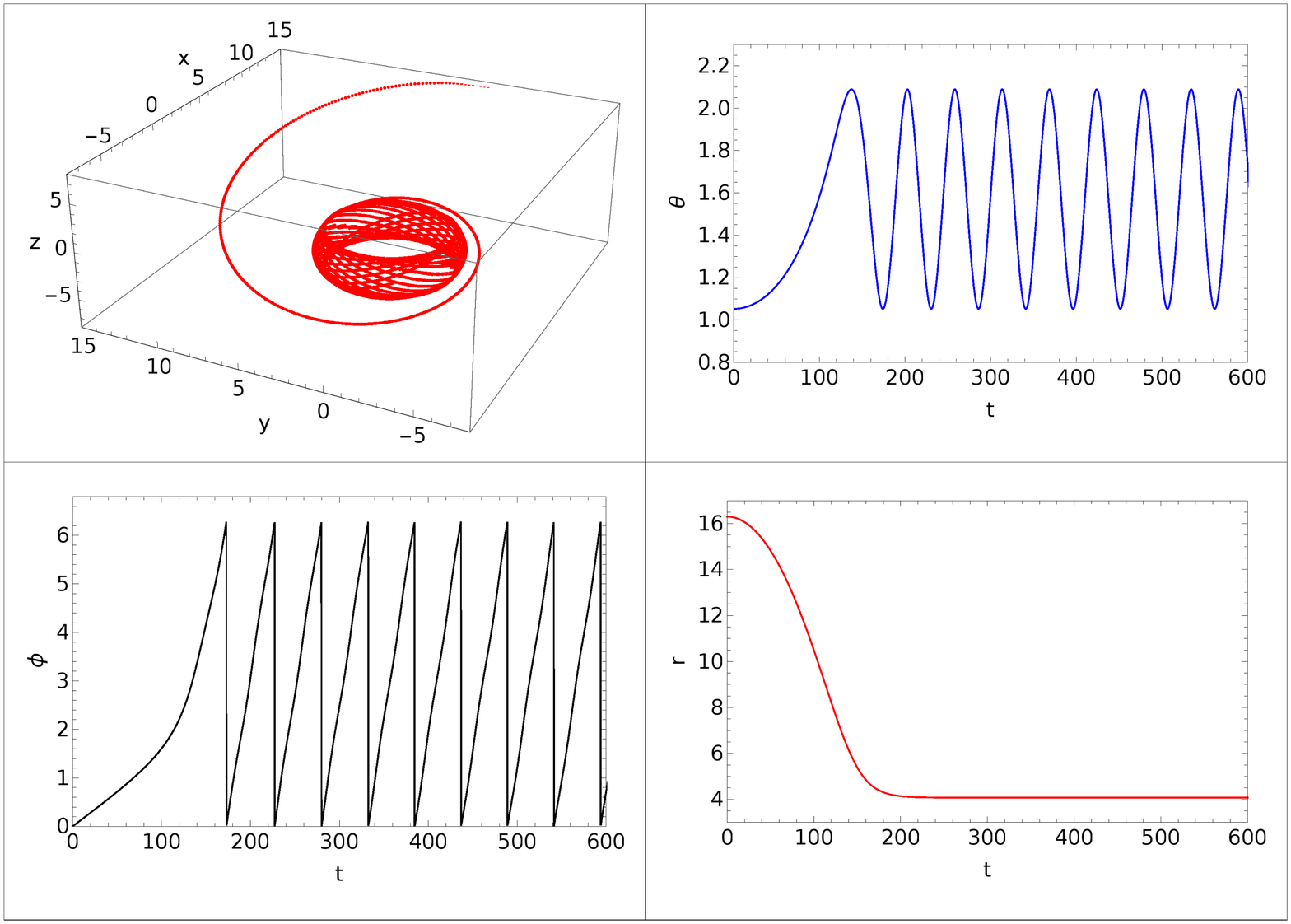} }
\hspace{-0.5cm}
\subfigure[]{
\includegraphics[scale=0.31]{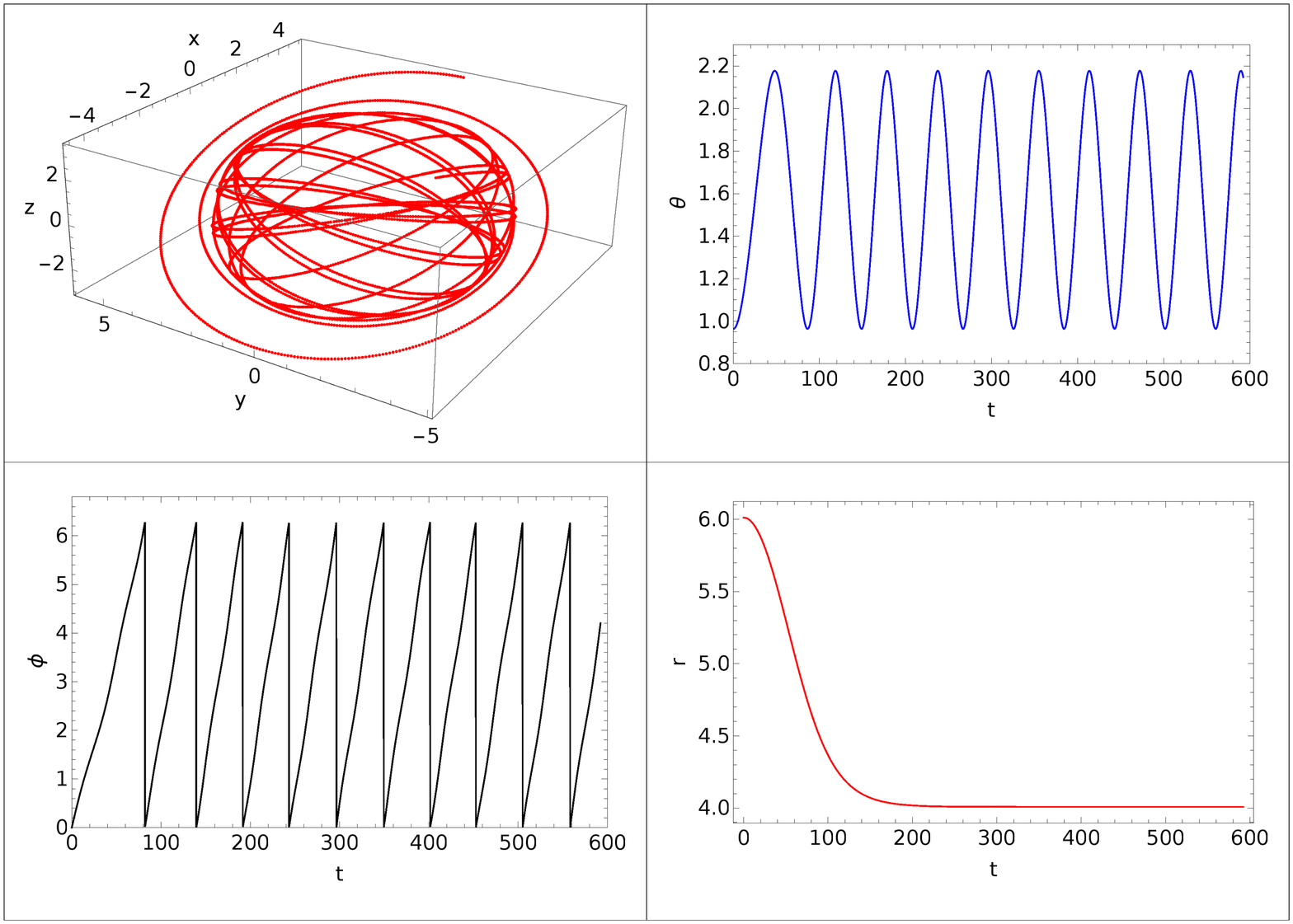}}} 
\mbox{ \subfigure[]{
\includegraphics[scale=0.313]{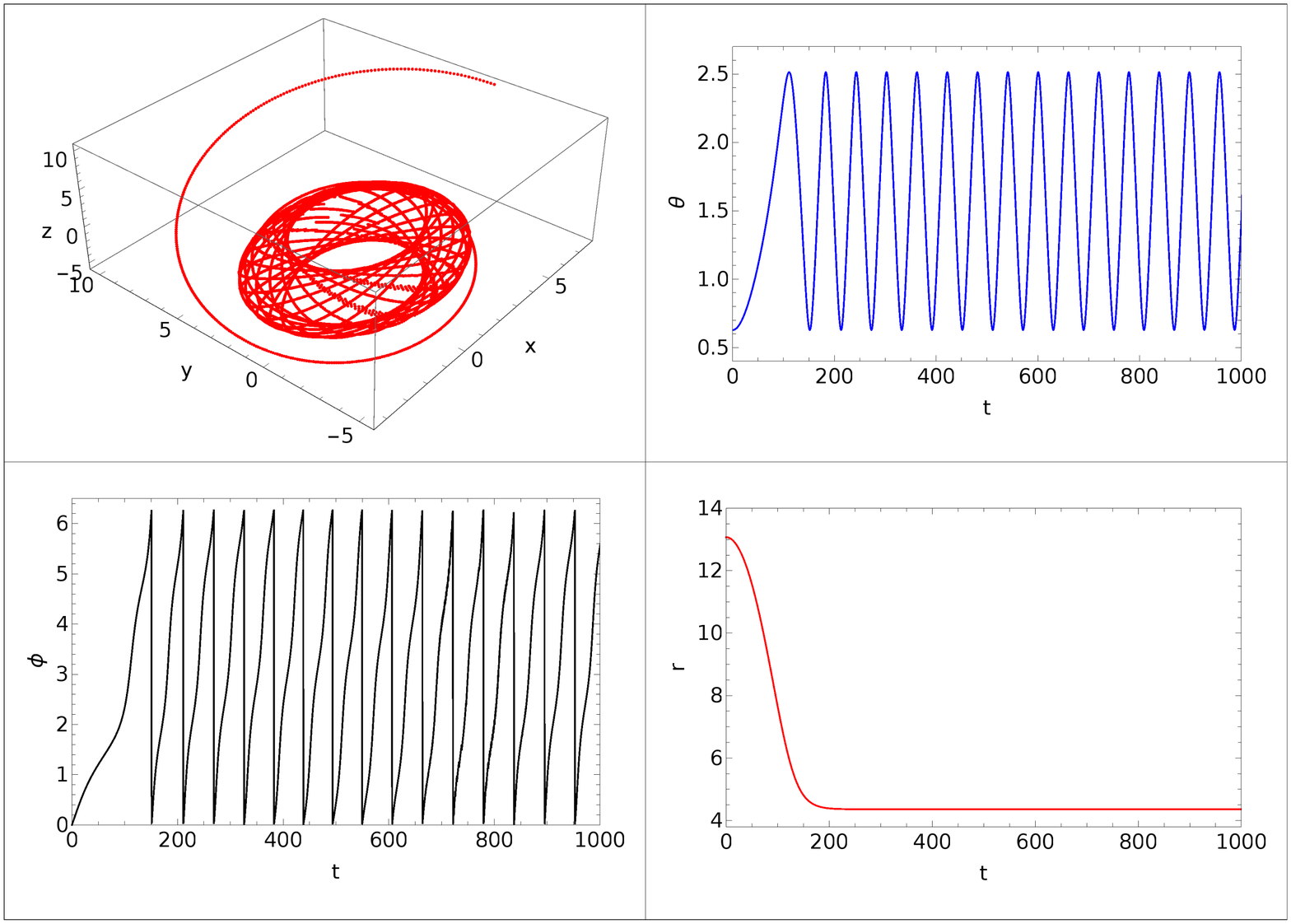}}
\hspace{-0.4cm}
\subfigure[]{
\includegraphics[scale=0.313]{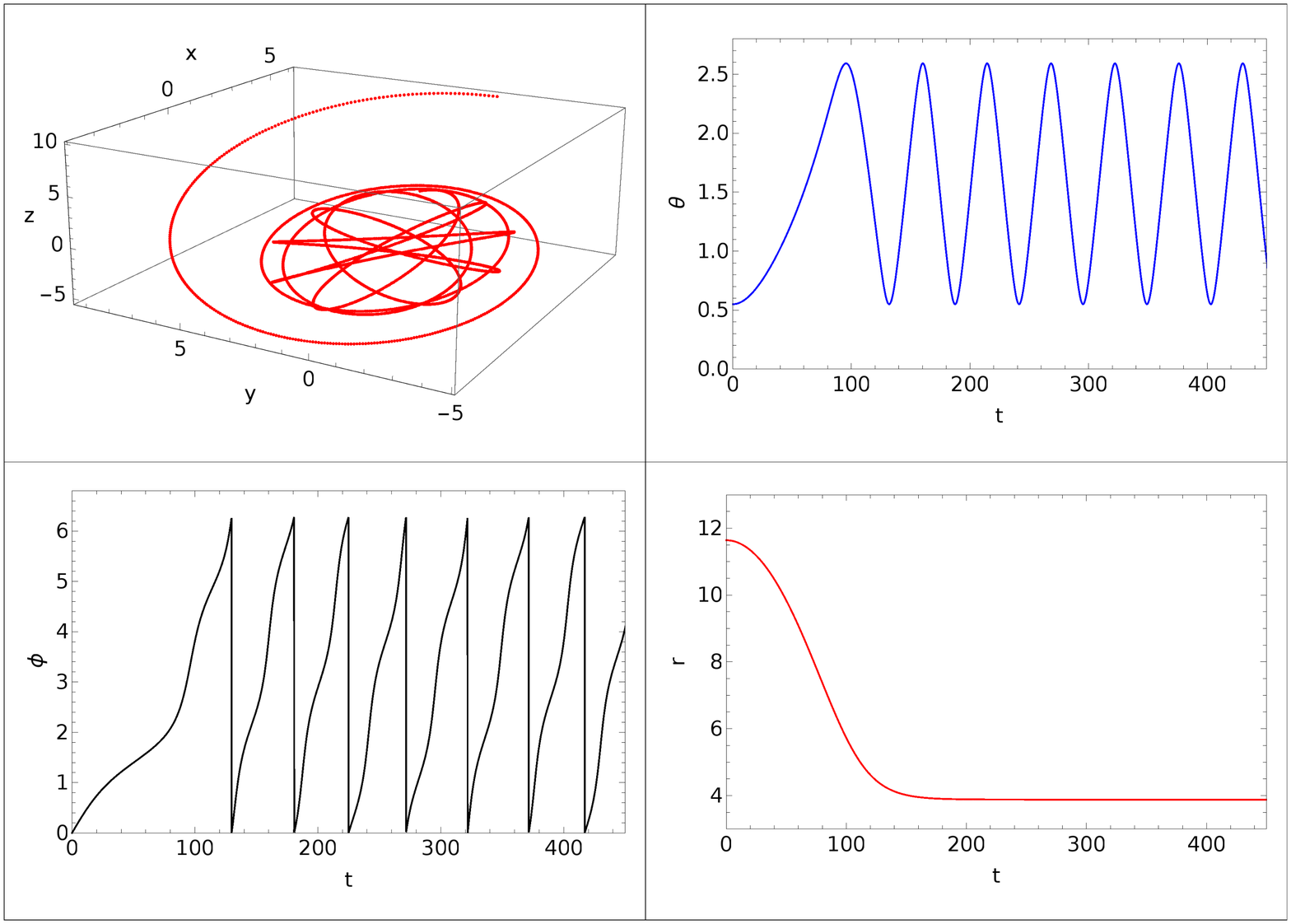}}}
 \caption{\label{seppro}The figure shows the prograde homoclinic orbits (a) H1, (b) H2, (c) H3, and (d) H4 in the table \ref{parameters}, for various combinations of ($e$, $\mu$, $a$, $Q$) and also presents the evolution of corresponding $\theta$, $\phi$ and $r$ with coordinate time, $t$.}
 \end{figure}

\begin{figure}
 \mbox{ \subfigure[]{
\includegraphics[scale=0.31]{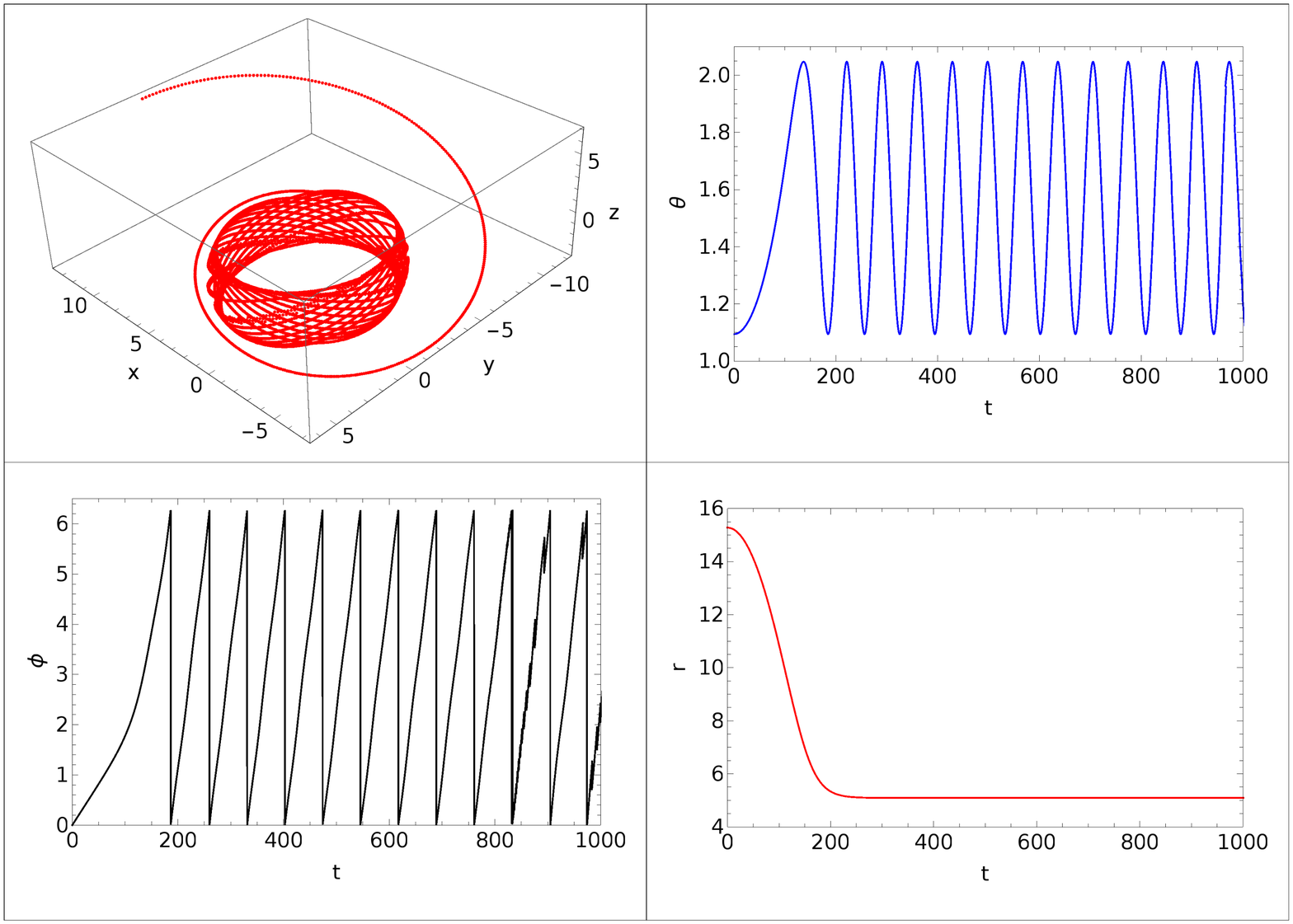} }
\hspace{-0.5cm}
\subfigure[]{
\includegraphics[scale=0.311]{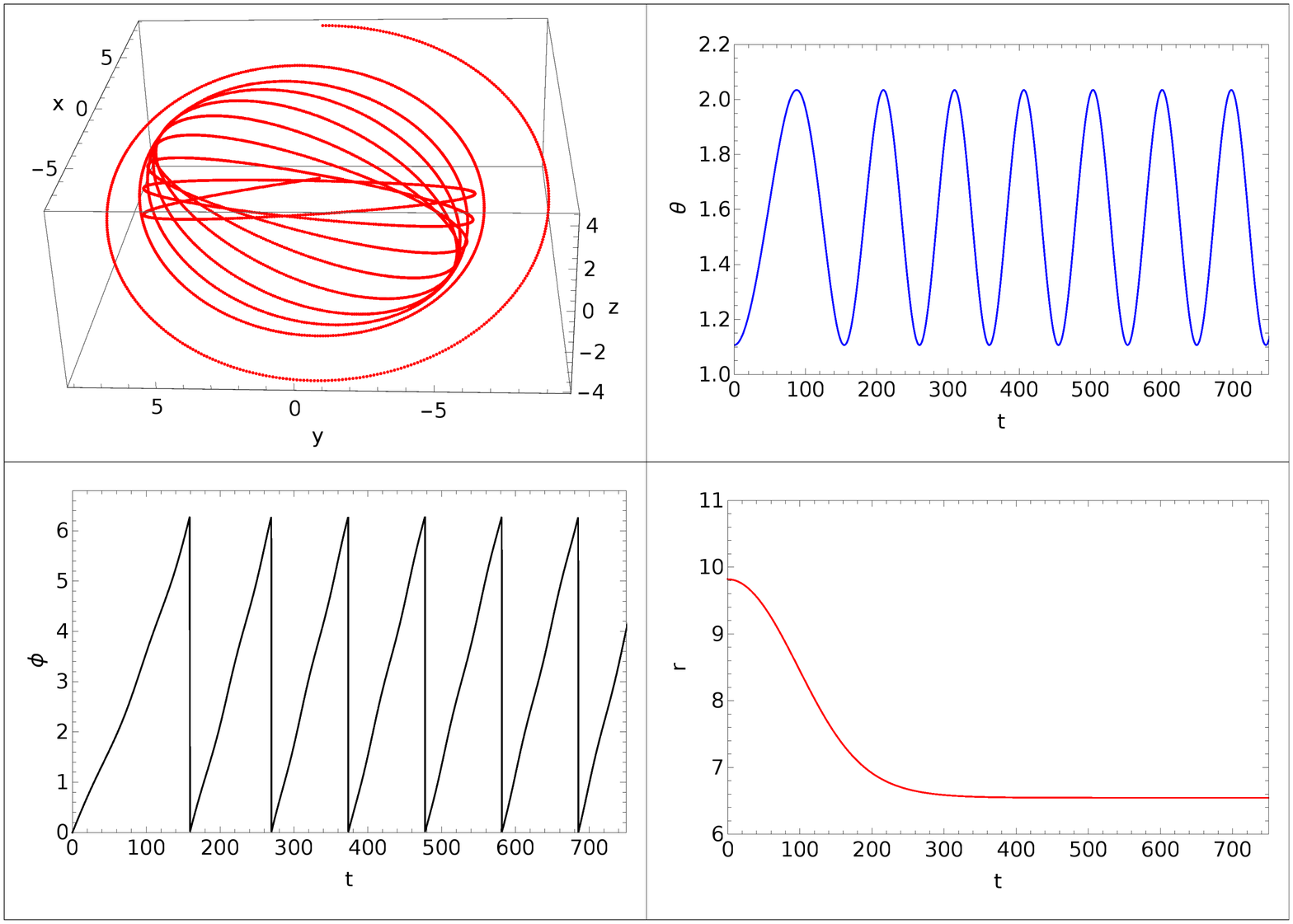}}} 
\mbox{ 
\subfigure[]{
\includegraphics[scale=0.31]{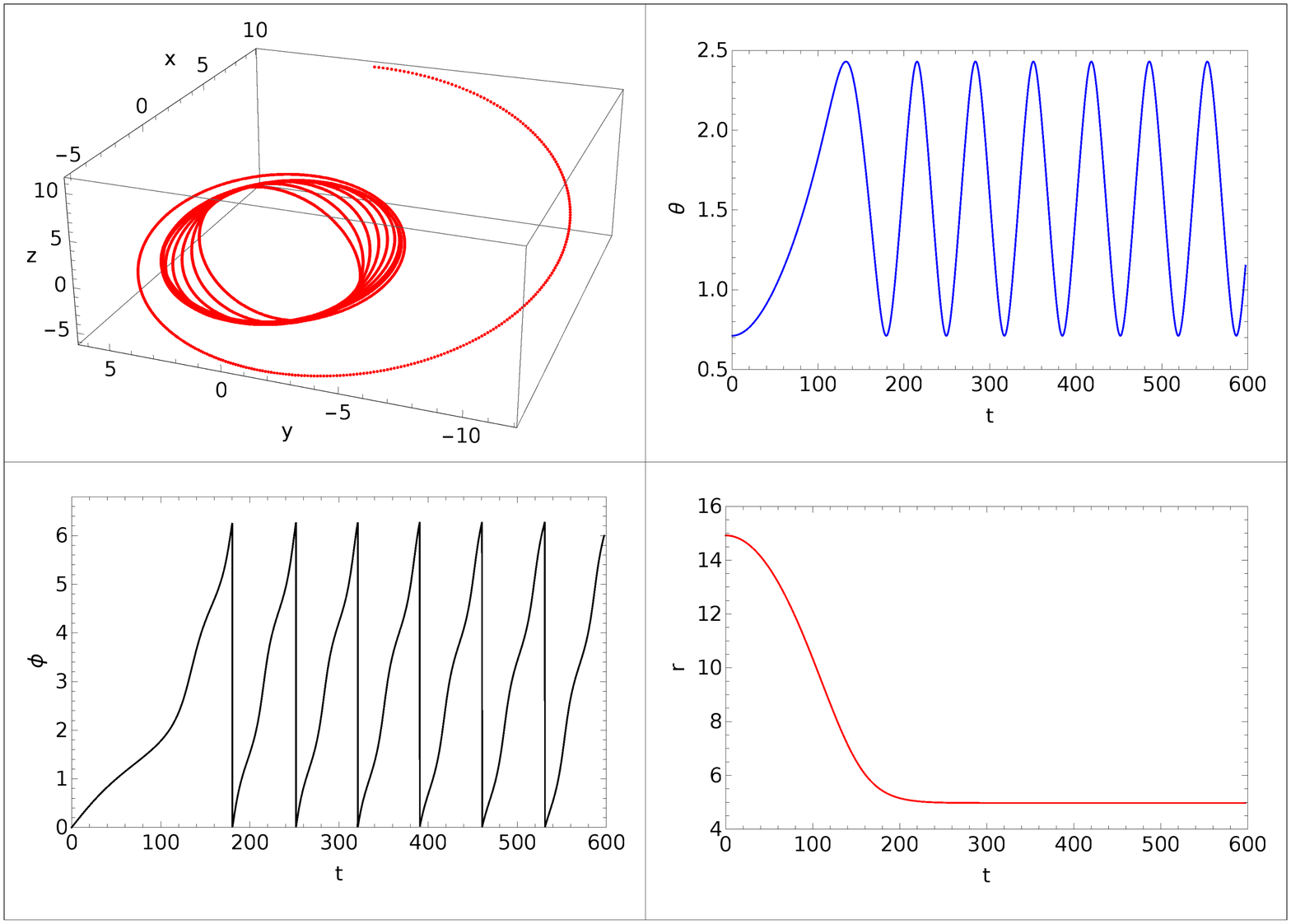}}
\hspace{-0.4cm}
\subfigure[]{
\includegraphics[scale=0.309]{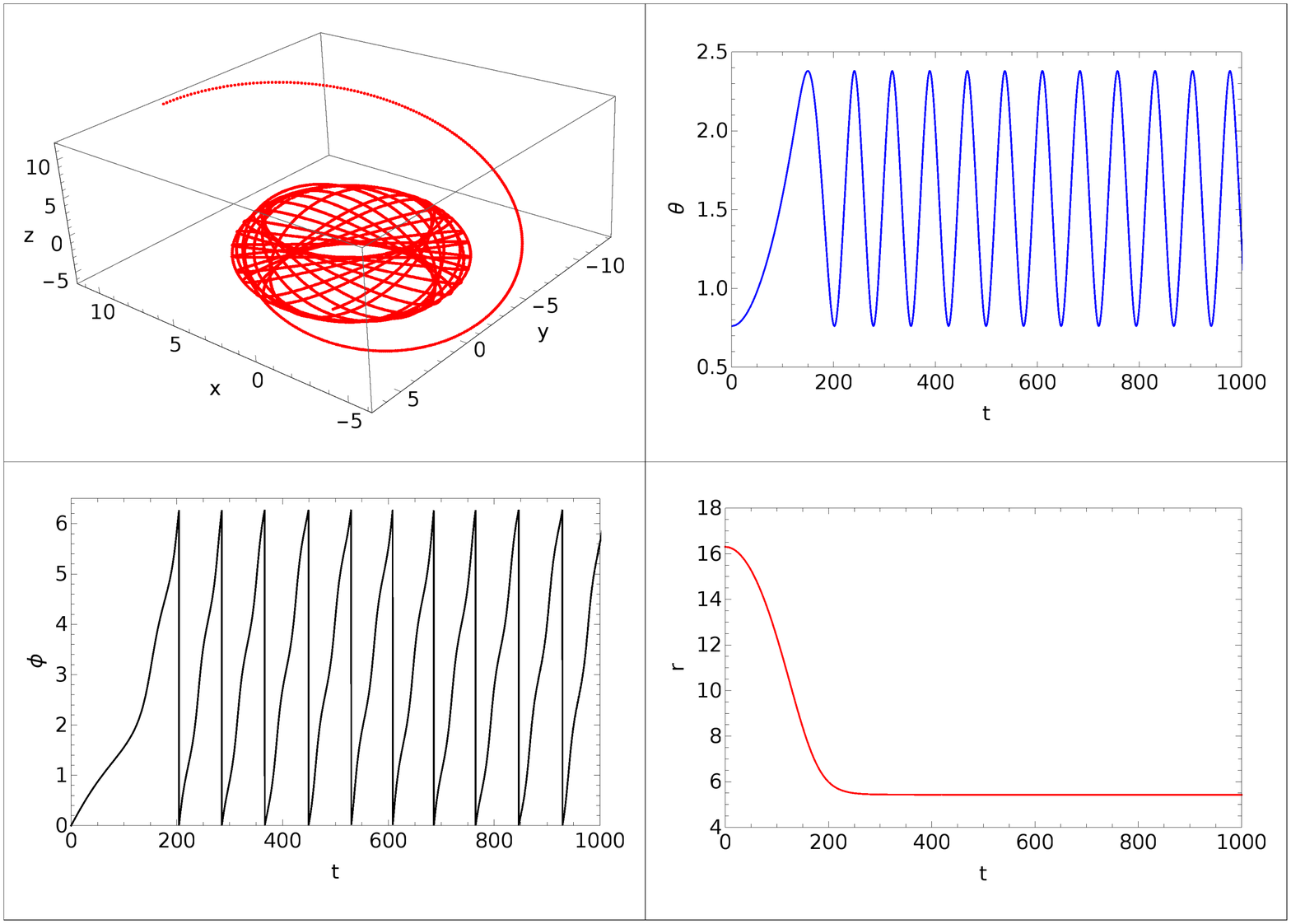}}}
 \caption{\label{sepret}The figure shows the retrograde homoclinic orbits (a) H5, (b) H6, (c) H7, and (d) H8 in the Table \ref{parameters}, for various combinations of ($e$, $\mu$, $a$, $Q$) and also presents the evolution of corresponding $\theta$, $\phi$ and $r$ with coordinate time, $t$.}
 \end{figure}

\begin{figure}
 \mbox{ \subfigure[]{
\includegraphics[scale=0.45]{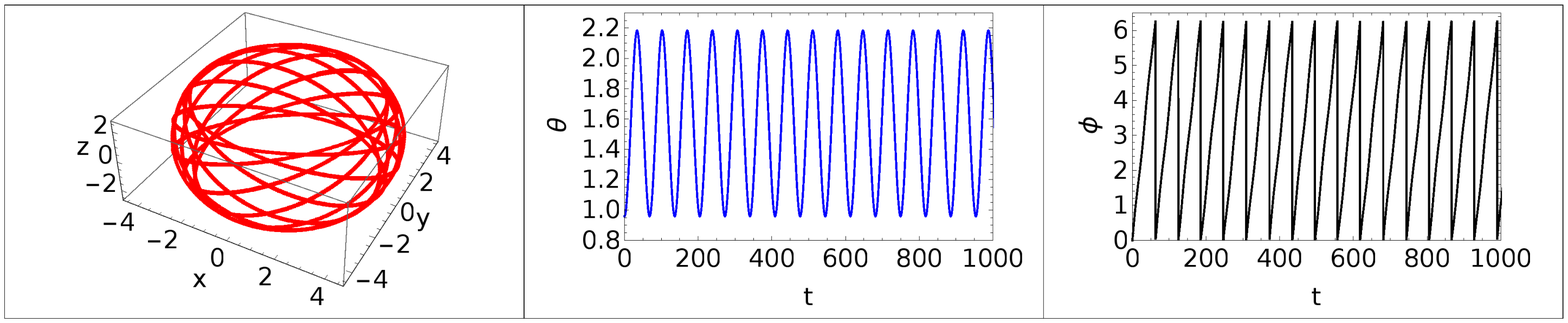} }} 
\mbox{ \subfigure[]{
\includegraphics[scale=0.46]{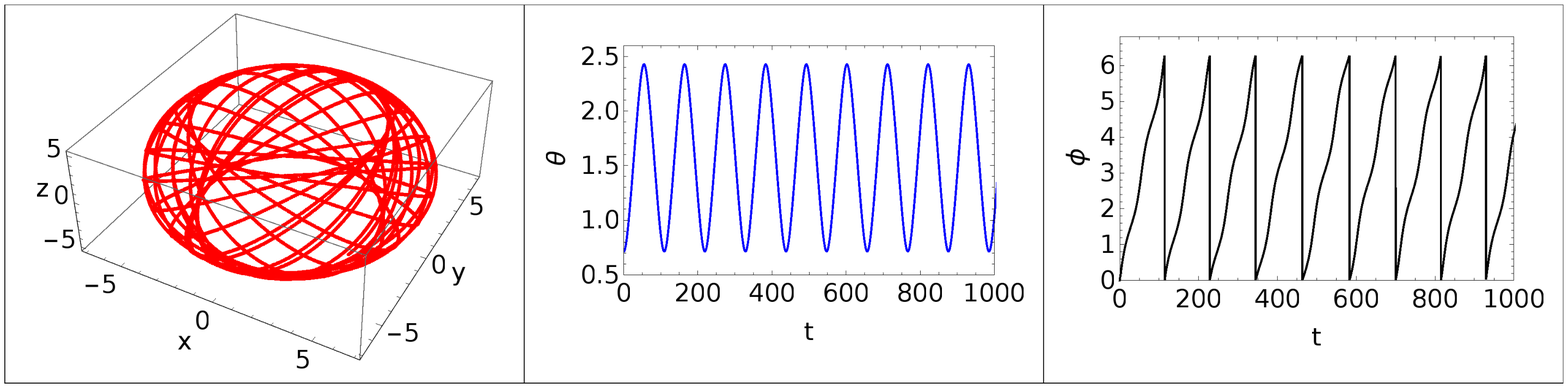}}}
 \caption{\label{sph}The figure shows the spherical orbits for (a) prograde, S1, (b) retrograde, S2, in the Table \ref{parameters} along with the corresponding evolution of $\theta$, $\phi$ and $r$ with coordinate time, $t$.}
 \end{figure}

\begin{figure}
 \mbox{ \subfigure[]{
\includegraphics[scale=0.29]{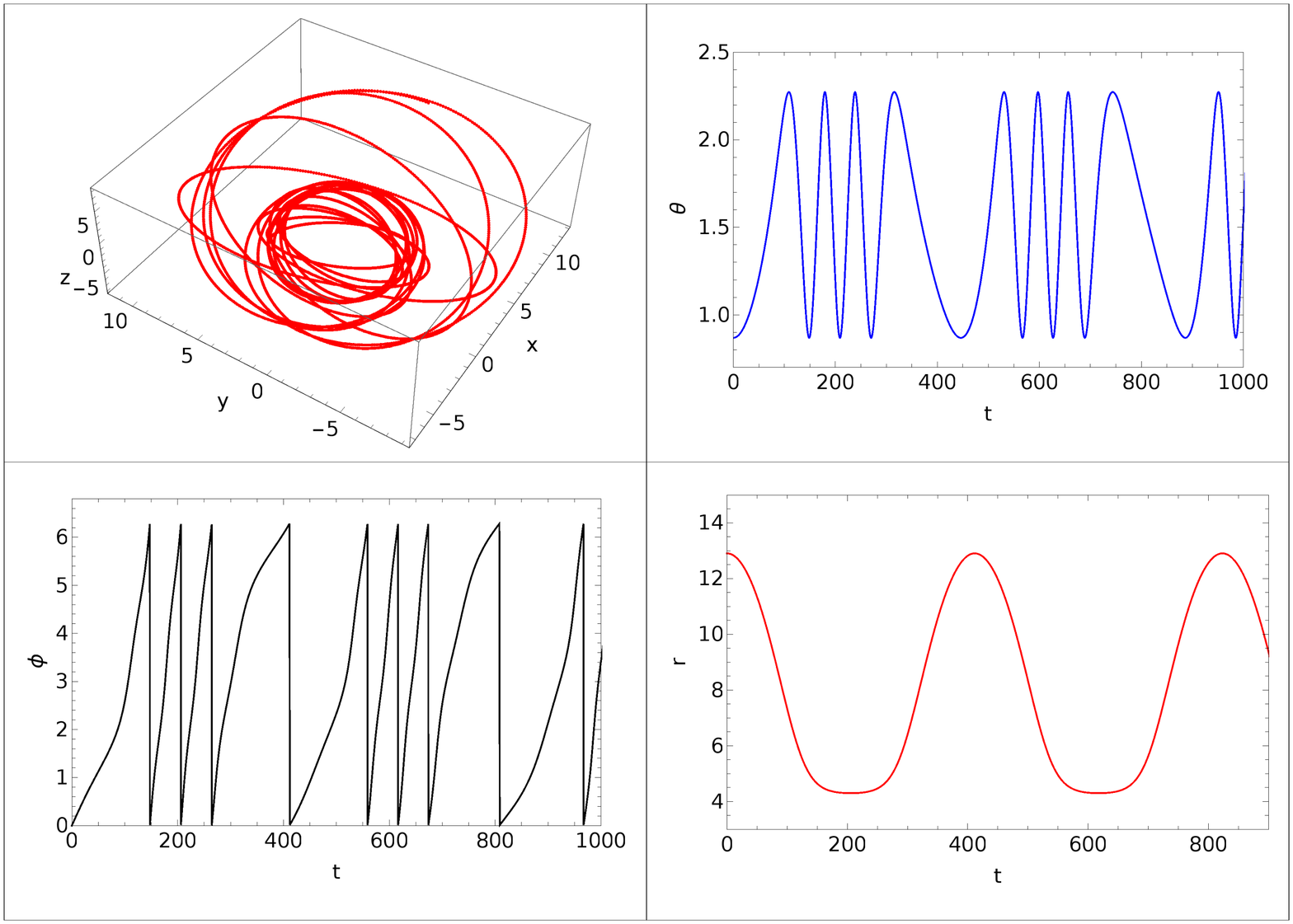} }
\hspace{-0.55cm}
\subfigure[]{
\includegraphics[scale=0.29]{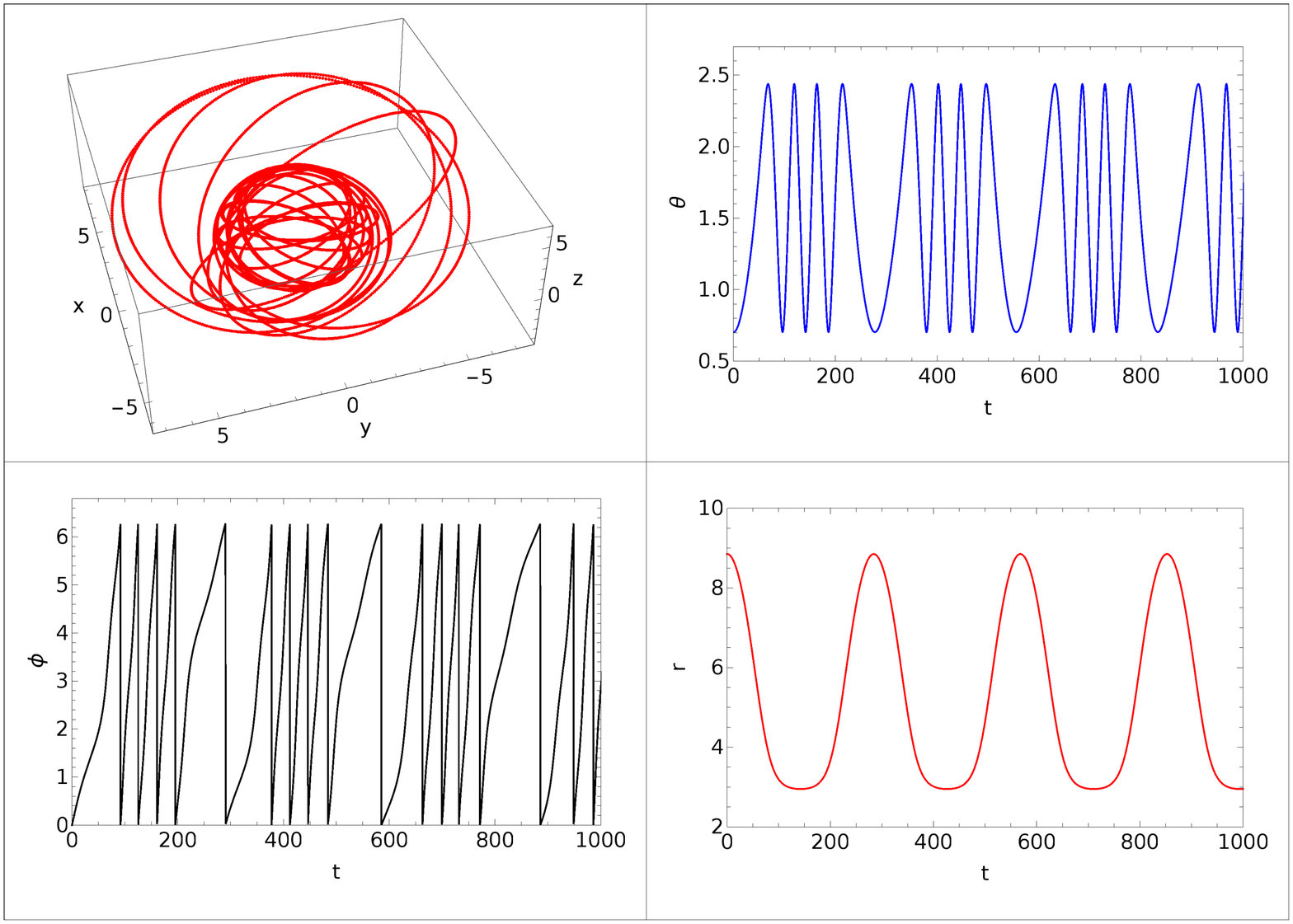}}} 
\mbox{ \subfigure[]{
\includegraphics[scale=0.29]{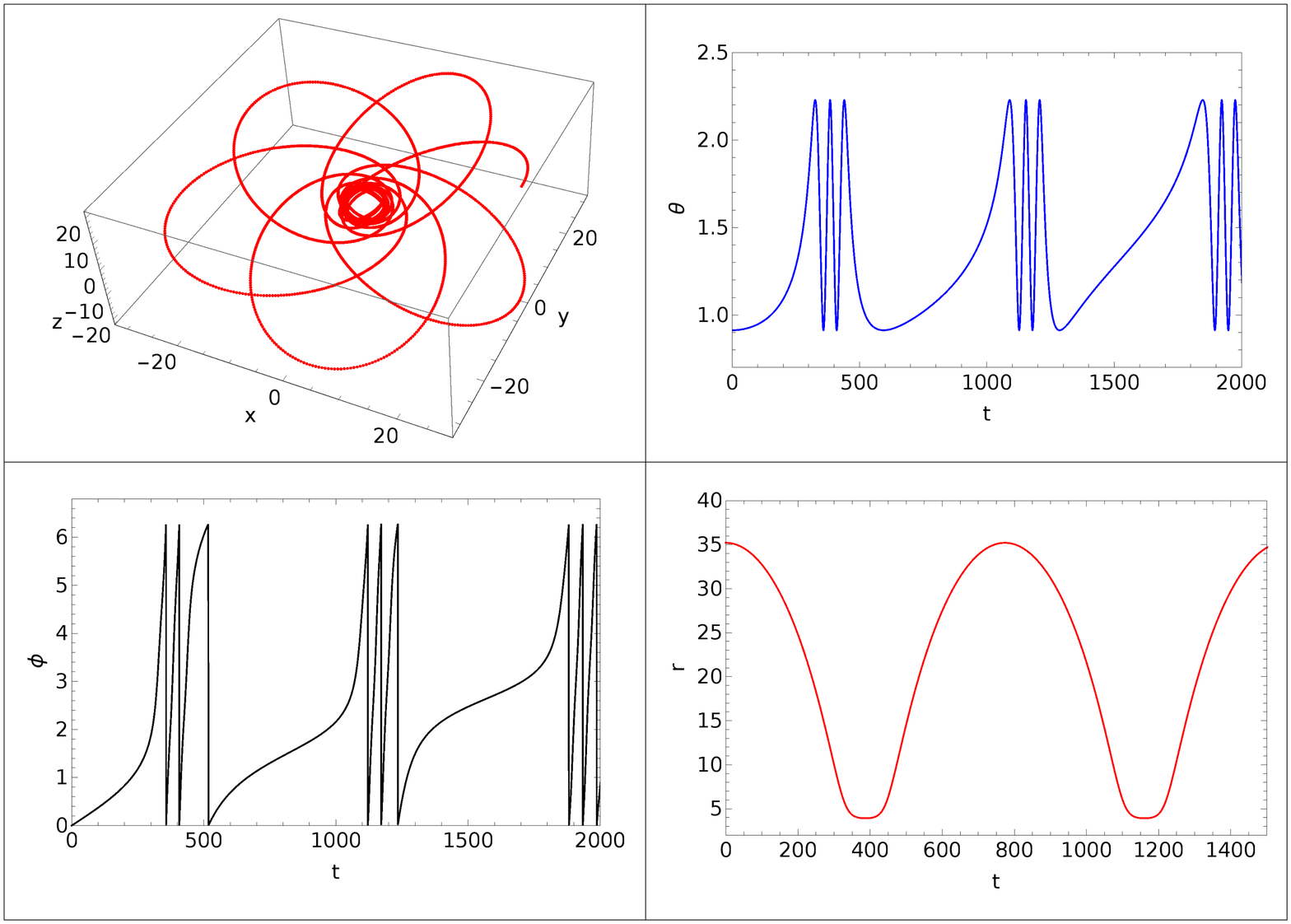}}
\hspace{-0.4cm}
\subfigure[]{
\includegraphics[scale=0.2901]{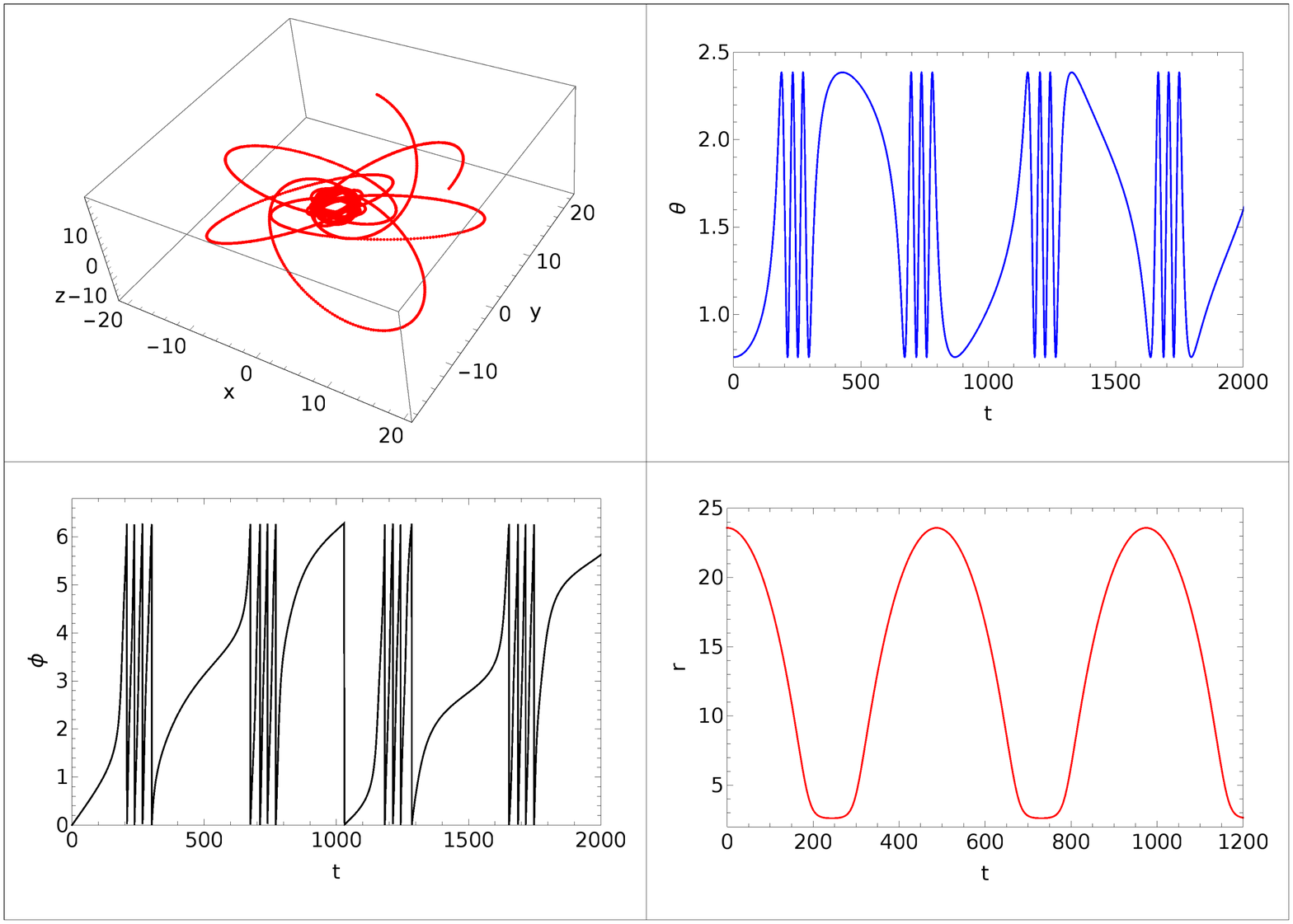}}}
\mbox{ \subfigure[]{
\includegraphics[scale=0.29]{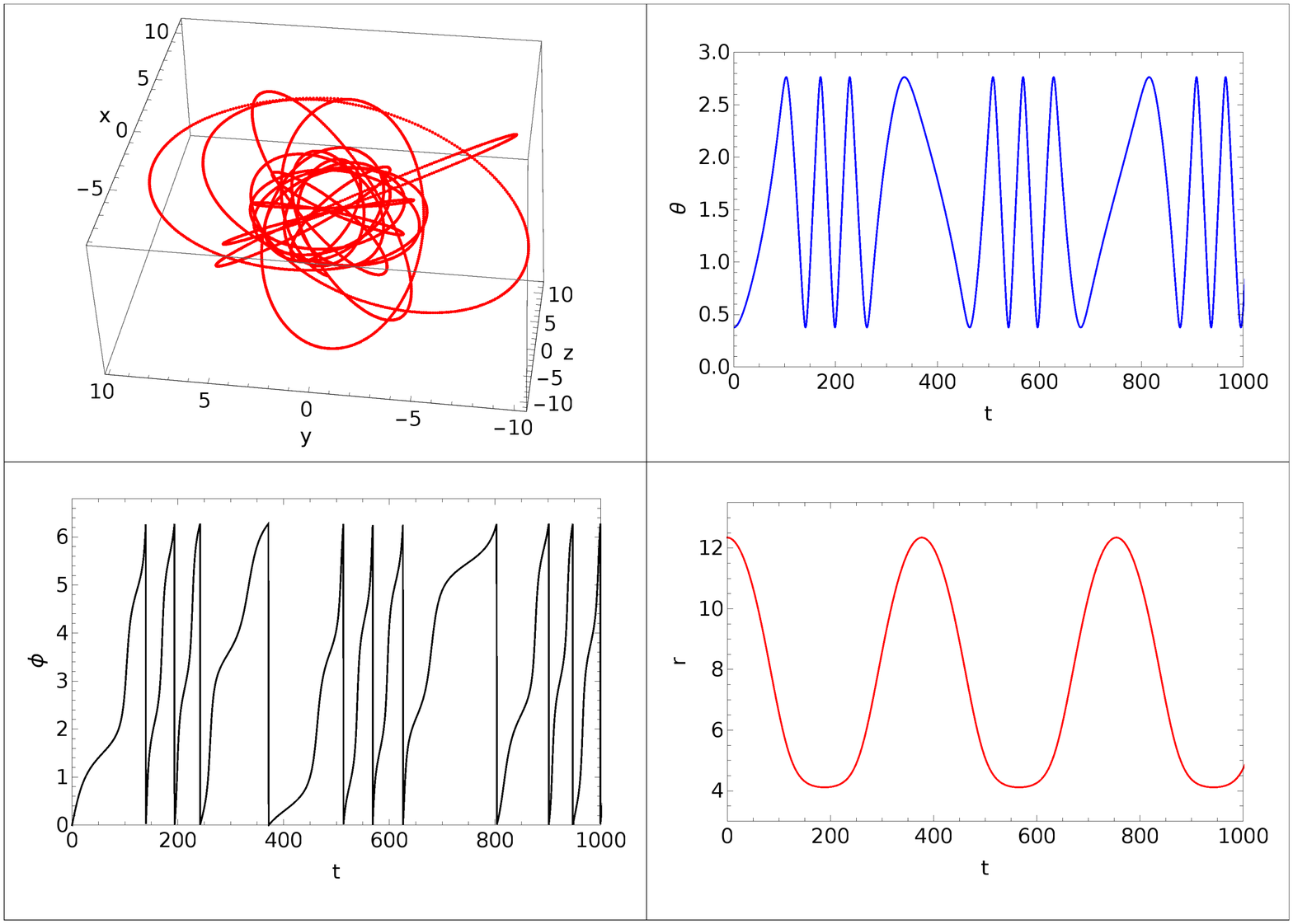}}
\hspace{-0.4cm}
\subfigure[]{
\includegraphics[scale=0.29]{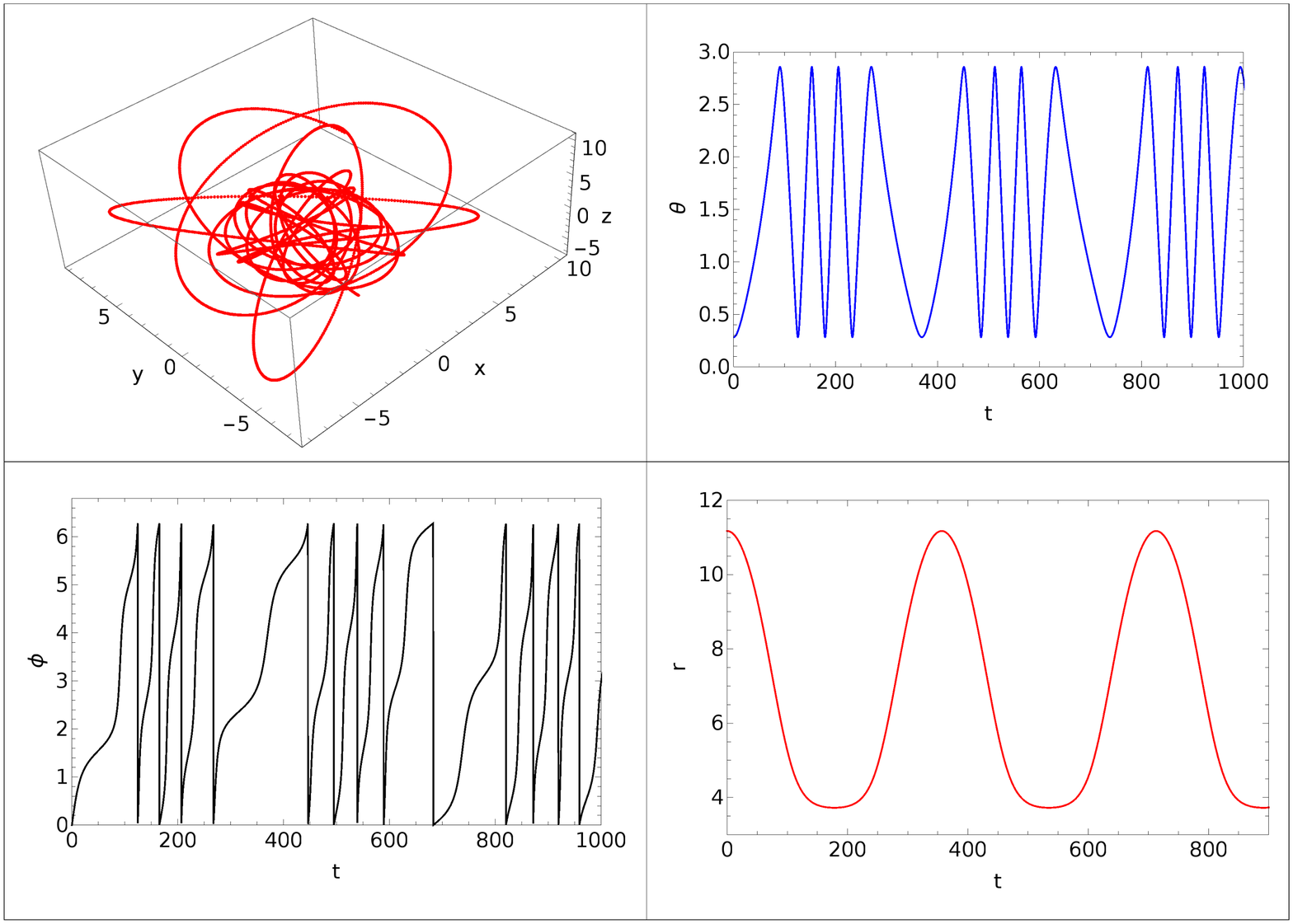}}}
 \caption{\label{zoomwhirl}The figure shows zoom whirl orbits (a) Z1, (b) Z2, (c) Z3, (d) Z4, (e) Z5, and (f) Z6 in the table \ref{parameters}, for various combinations of ($e$, $\mu$, $a$, $Q$) satisfying Eq. \eqref{boundcondreg3} and also presents the evolution of corresponding $\theta$, $\phi$ and $r$ with coordinate time, $t$.}
 \end{figure}

 There are various possible kinds of bound orbits. Here, we take up the each case and sketch these trajectories for different combinations of ($a$, $Q$), where the parameters values are tabulated in the Table \ref{parameters}. We take up slow rotating ($a=0.2$) and fast rotating black hole situations ($a=0.5$ or $a=0.8$), with both prograde and retrograde cases, for various $Q$ values. The various features of these orbits are enumerated below: 
\begin{enumerate}
\item \underline{\textit{Eccentric orbits}}: Figs. \ref{eccpro} and \ref{eccret} represent eccentric bound prograde and retrograde trajectories respectively, where the parameter values are depicted in the Table \ref{parameters}. The particle periodically oscillates between the periastron and the apastron, and is also bound between $\theta = \arccos \left( z_{-} \right)$ and $\theta = \arccos \left( -z_{-} \right) $ as shown in ($t$-$r$) and ($t$-$\theta$) plots in Figs. \ref{eccpro} and \ref{eccret}, whereas ($t$-$\phi$) plots depict that $\phi$ varies between 0 to $2\pi$. We have fixed ($e$, $\mu$) of the plotted trajectories and show the variation with change in $a$ and $Q$ parameters. The motion of the trajectory increases in the vertical direction with increase in $Q$ parameter.

\begin{table}
\begin{center}
\caption{This following table summarizes the values of conic parameters ($e$, $\mu$) chosen in the listed orbit simulations to study eccentric, homoclinic and spherical orbits for different ($a$, $Q$) combinations for both prograde and retrograde cases constructed using Eqs. \eqref{final}.}
\scalebox{0.6}{
\begin{tabular}{|c|c|c|c|c|c|c|}
\hline
Type of& Orbit $\#$ &Inverse latus-& Eccentricity & Spin of & Carter's constant& Varying \\
orbit & & rectum of the orbit  & of the orbit  & the black hole  & &parameter \\
 & & $\mu$ &$e$ & $a$ &$Q$ &\\
\hline
\hline
& & & & & &\\
 Eccentric orbits  & E1& 0.1 & 0.6 & 0.2 & 3 & $a$\\
& E2& 0.1 & 0.6 & 0.8 & 3 &\\
\hline
& & & & & &\\
& E3 & 0.1 &  0.6 & 0.2& 8 & $a$\\
& E4 & 0.1 & 0.6 & 0.8 & 8 & \\
\hline
& & & & & &\\
& E5 & 0.1 & 0.6 & -0.2 & 3 & $a$\\
& E6 & 0.1 & 0.6 & -0.8& 3 &  \\
\hline
& & & & & &\\
& E7 & 0.1 & 0.6 & -0.2& 8 & $a$\\
& E8 & 0.1 & 0.6 & -0.8& 8 &\\
\hline
\hline
& & & & & &\\
Homoclinic orbits& H1 &0.153 & 0.6 & 0.2 & 3 & $a$ and $e$\\
& H2& 0.208& 0.2& 0.5& 3 &\\
\hline
& & & & & &\\
& H3& 0.153 & 0.5 & 0.2& 8 & $a$\\
& H4& 0.172& 0.5& 0.5 & 8 &\\
\hline
& & & & & &\\
& H5& 0.127 & 0.5& -0.2& 3 & $a$ and $e$\\
& H6& 0.127 & 0.2 & -0.5 & 3&\\
\hline
& & & & & &\\
& H7& 0.134 & 0.5& -0.2& 8& $a$\\
& H8& 0.123 & 0.5 & -0.5 & 8 &\\
\hline
\hline
& & & & & &\\
Spherical orbits& S1&0.222 & 0 & 0.5 &3 & $a$ and $Q$\\
& S2&0.144 & 0 & -0.5 & 8 &\\
\hline
\hline
& & & & & &\\
Zoom-whirl & Z1 & 0.155& 0.5 & 0.2& 5 & $a$\\
& Z2 & 0.226& 0.5 & 0.8& 5 &\\
\hline
& & & & & &\\
& Z3 & 0.142& 0.8 & 0.2& 5 & $a$\\
& Z4 & 0.212& 0.8 & 0.8& 5 &\\
\hline
& & & & & &\\
& Z5 & 0.162& 0.5 & 0.5& 10 &$a$\\
& Z6 & 0.179& 0.5 & 0.8& 10 &\\
\hline
\end{tabular}
\label{parameters}
}

\end{center}
\end{table}

\item \underline{\textit{Homoclinic/Separatrix orbits}}: Homoclinic orbits are the separatrices between eccentric bound and plunge orbits, where the particle asymptotically approaches the unstable spherical/circular orbit in both the distant past and the distant future.  The energy and angular momentum of the orbiting particle simultaneously correspond to a stable eccentric bound orbit and an unstable spherical/circular orbit. Separatrix orbits in the equatorial plane of a Kerr black hole are well studied, \cite{Levin2009,Perez-Giz2009,Shaughnessy2003}. The homoclinic orbits form an important group in Kerr dynamics as they represent the transition between inspiral and plunge orbits and hence, have their significance in the study of gravitational wave spectrum under the adiabatic approximation. The homoclinic or separatrix orbits correspond to the boundary of the region in ($e, \mu, a, Q$) space, defined by Eq. \eqref{sepeq}.
Separatrix orbits with $Q\neq0$ also have similar features as the equatorial separatrix orbits, where the particle asymptotically approaches the unstable spherical orbit. Figs. \ref{seppro} and \ref{sepret} show prograde and retrograde non-equatorial homoclinic/separatrix orbits respectively (see Table \ref{parameters} for parameter values). We see from H3 and H4 trajectories that an increase in spin parameter, $a$, increases the range of $\theta$. The orbit initially follows an eccentric path and  asymptotically approaches the periastron radius which also corresponds to the unstable spherical orbit radius as shown in ($t$-$r$) plots of Figs. \ref{seppro} and \ref{sepret}. 

\item \underline{\textit{Spherical orbits}}: Fig. \ref{sph} shows prograde and retrograde innermost stable spherical orbits (ISSO), which are also the homoclinic orbits with $e=0$. All the spherical stable orbits exist outside ISSO, whereas unstable spherical orbits are found between ISSO and MBSO.

\item \underline{\textit{Zoom-whirl orbits}}: Zoom whirl orbits are orbits where the particle takes a finite number of revolutions at the periastron before going back to the apastron, which is an extreme form of the periastron precession. Their significance in gravitational astronomy has been studied for the case of equatorial Kerr orbits \cite{Glampedakis2002}. Here, we discuss zoom-whirl orbits with $Q\neq0$ as shown in Fig. \ref{zoomwhirl}, where the particle takes finite revolutions with varying $\theta$ at the periastron before turning back to the apastron. We have chosen the value of $\mu$ very near to the separatrix, where usually the zoom whirl behavior is seen, for different values of ($e$, $a$, $Q$) combinations. As expected, the particle spends more time at the periastron, compared to the time taken at apastron, to take a finite number of revolutions which is making the $t-r$ plots appear flatter near the periastron, (see Fig. \ref{zoomwhirl}). We again see that the increase in $a$ increases the range of vertical motion of the orbit like for the eccentric orbits case. Homoclinic/Separatrix orbit family is the limiting case of the zoom-whirl orbit family where the particle takes infinite revolutions as it asymptotes to the unstable spherical orbit. 
\end{enumerate}
 
 \begin{figure}
   \begin{center}
\includegraphics[scale=0.62]{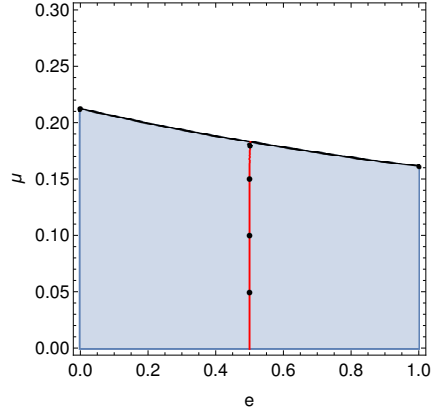} 
 \caption{\label{a5Q5}The shaded region depicts the bound orbit region defined by Eq. \eqref{boundcondreg3} in the ($e$, $\mu$) plane for $a=0.5$ and $Q=5$. The black curve represents the homoclinic orbits where the end points depict $e=0$ and $e=1$ homoclinic orbits corresponding to the ISSO and MBSO respectively. The red curve represents $e=0.5$ and we study orbits with different $\mu$ values as depicted by the dots on this curve. }
   \end{center}
 \end{figure}
 
 
Now, we discuss how different kinds of orbits are distributed in the bound orbit region in the ($e$, $\mu$) plane defined by the Eq. \eqref{deltaregion} for a fixed combination of ($a$, $Q$). We fix $a=0.5$ and $Q=5$ and show the shaded bound orbit region in Fig. \ref{a5Q5}, that represents the eccentric orbits allowed. The black curve which is the boundary of the shaded region represents homoclinic or separatrix orbits. The curve defined by $e=0$ represents all the spherical orbits with its end point at ISSO, which intersects with the separatrix line. We fix $e=0.5$ depicted by the red curve in Fig. \ref{a5Q5} and take different values of $\mu$, as depicted by the black dots on the red curve, and plot the corresponding trajectories and study their corresponding behavior.

 \begin{figure}
 \mbox{ \subfigure[]{
\includegraphics[scale=0.3]{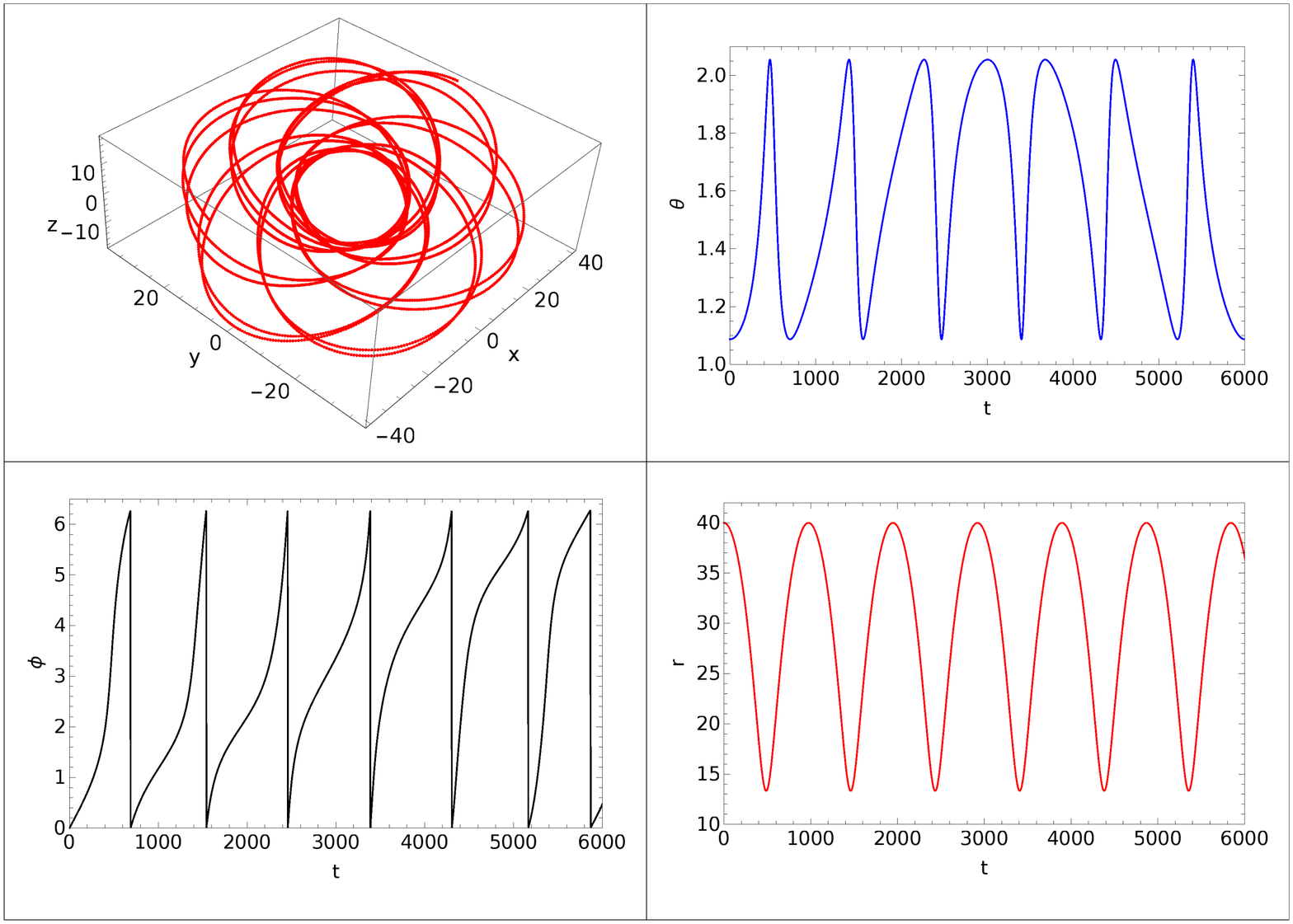} }
\hspace{-0.7cm}
\subfigure[]{
\includegraphics[scale=0.303]{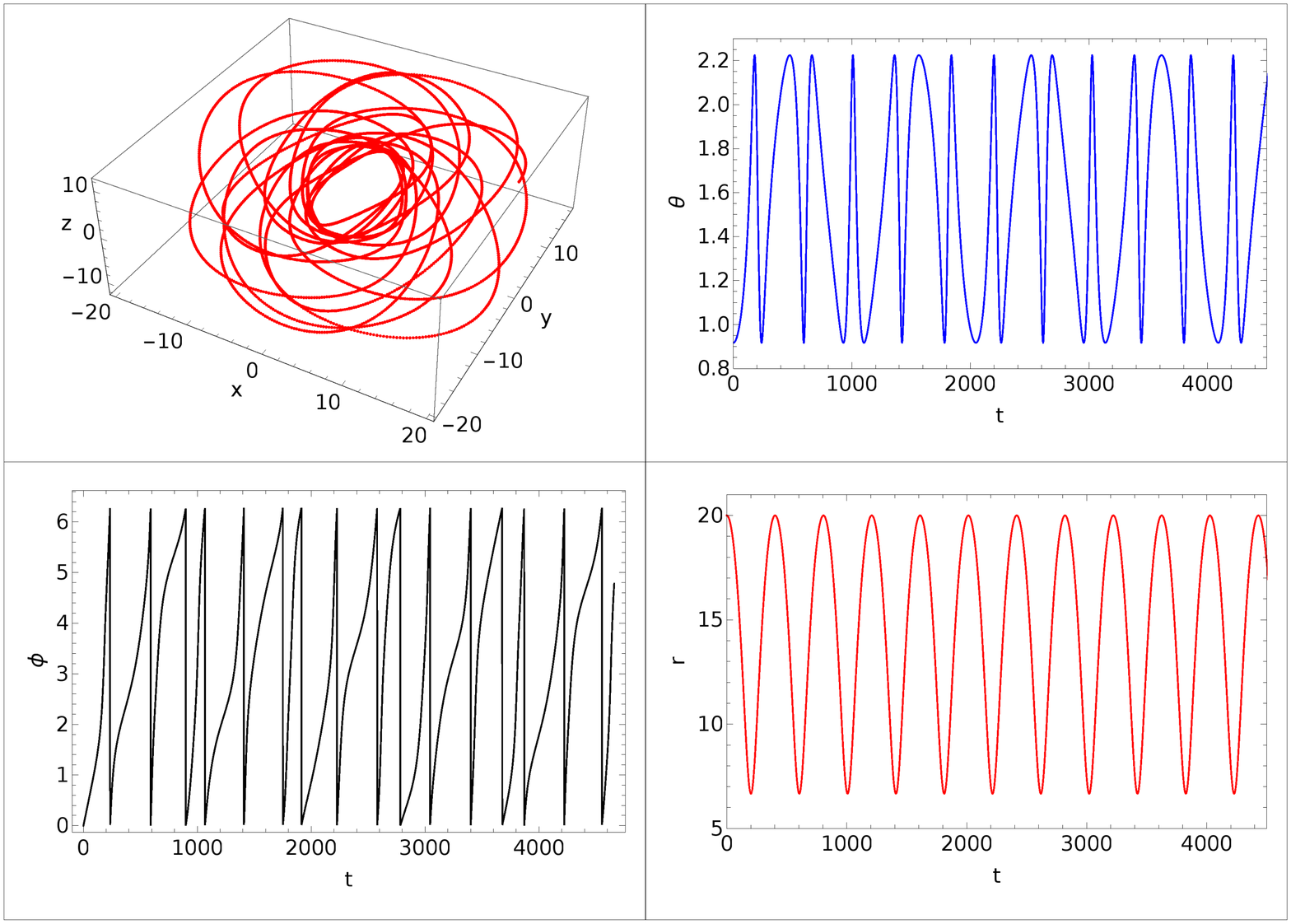}}} 
 \caption{\label{mu12}The figure shows the eccentric trajectories on the red curve of Fig. \ref{a5Q5} ($e=0.5$, $a=0.5$, $Q=5$) for (a) $\mu=0.05$, and (b) $\mu=0.1$.}
 \end{figure}
\begin{figure}
 \mbox{ \subfigure[]{
\includegraphics[scale=0.31]{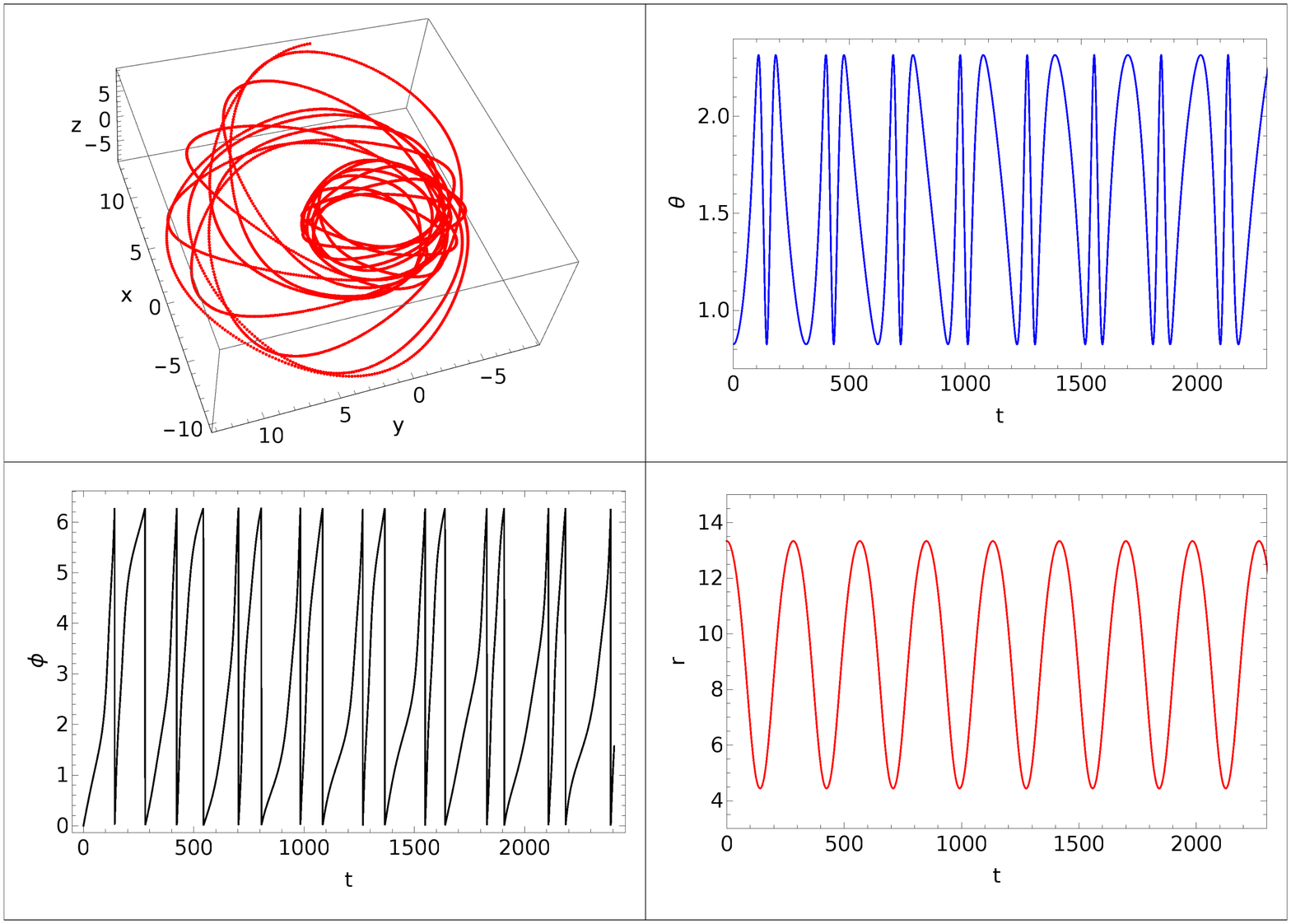} }
\hspace{-0.5cm}
\subfigure[]{
\includegraphics[scale=0.31]{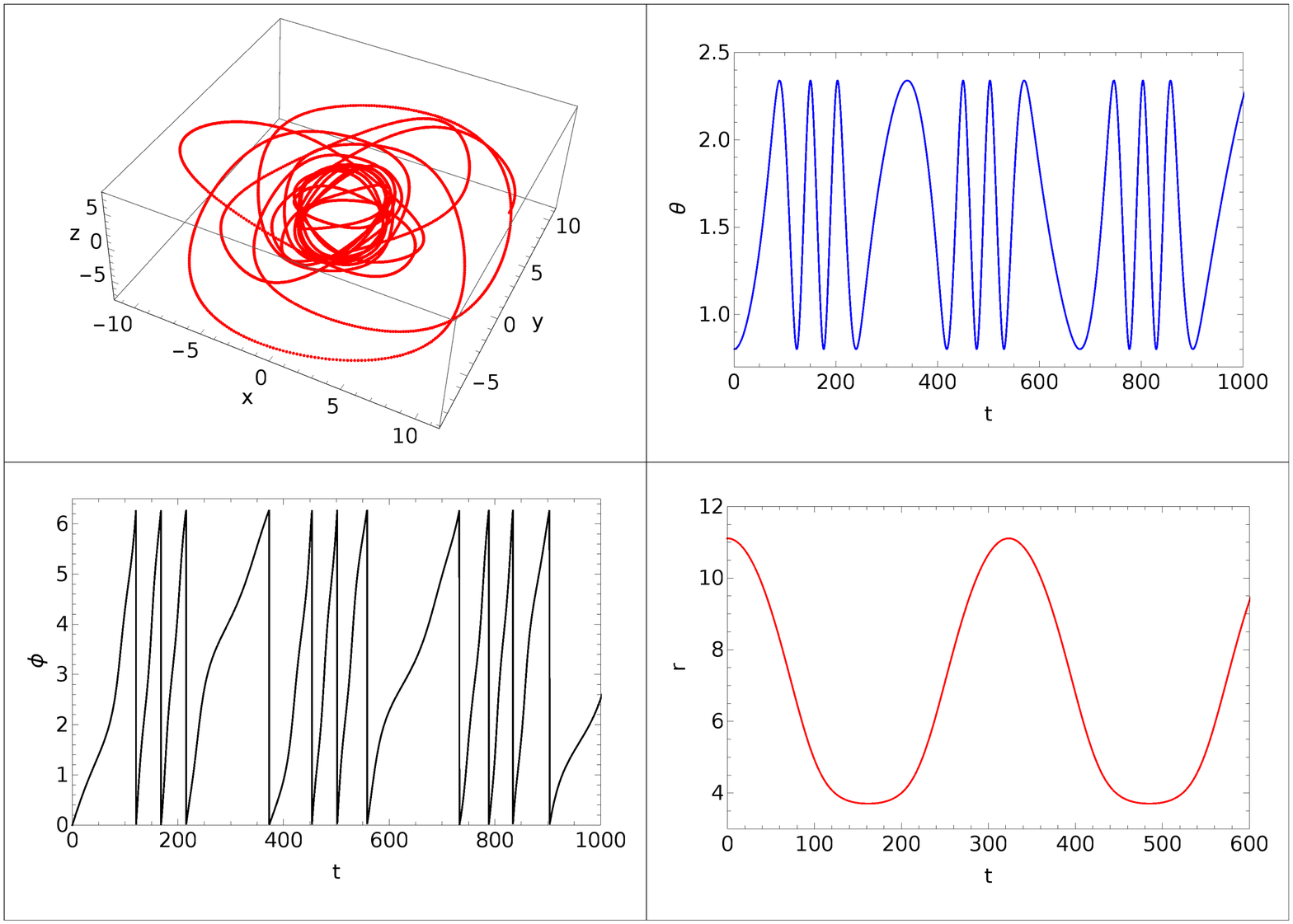}\label{mu34b}} } 
 \caption{\label{mu34}The figure shows the eccentric trajectories on the red curve of Fig. \ref{a5Q5} ($e=0.5$, $a=0.5$, $Q=5$) for (a) $\mu=0.15$, and (b) $\mu=0.18$. We see that the trajectory shown in (b) represent a zoom-whirl orbit.}
 \end{figure}

   We see from Figs. \ref{mu12} and \ref{mu34}, that for a fixed $e=0.5$, as $\mu$ is increased, the trajectory shows zoom-whirl behavior as it gets closer to the separatrix or homoclinic orbit for the corresponding $e$ value. It can be seen in the $t$-$r$ plot of Fig. \ref{mu34b} that the particle spends some time at the periastron which clearly depicts the zoom-whirl behavior. Hence, it can be said that zoom-whirl behavior is a near separatrix phenomenon and can occur at any eccentricity.

\section{Fundamental frequencies}
\label{frequencies}
In this section, we derive the expressions for fundamental frequencies ($\nu_{\phi}$, $\nu_{r}$, $\nu_{\theta}$) in terms of the integrals derived analytically in \S \ref{analyticsoln}. We take a long time average of Eq. \eqref{Rthetaint} on both the sides so that
\begin{subequations}
\begin{equation}
\lim_{T \to \infty }\frac{1}{T}\int_{r_{0}}^{r}  \frac{dr}{\sqrt{R}}=\lim_{T \to \infty } \frac{1}{T}\int_{\theta_{0}}^{\theta}  \frac{d \theta}{\sqrt{\Theta}}. \label{eqninf}
\end{equation}
As $T\rightarrow \infty$, there exists a large integer solutions, which can be found with arbitrary precision, so that $N_r t_r= N_{\theta}t_{\theta}=T$, where $N_r$ and $N_{\theta}$ are the number of radial and vertical oscillations; hence Eq. \eqref{eqninf} reduces to
\begin{equation}
\lim_{N_{r} \to \infty }\frac{2 N_{r} \int_{r_{p}}^{r_{a}} \frac{dr}{\sqrt{R}}}{N_{r} \cdot t_r}=\lim_{N_{\theta} \to \infty } \frac{2 N_{\theta} \int_{\theta_{-}}^{\pi-\theta_{-}} \frac{d \theta}{\sqrt{\Theta}}}{N_{\theta} \cdot t_{\theta}},
\end{equation}
where $r_{p}$ and $r_{a}$ are the periastron and apastron of the orbit and $\theta_{-}$ corresponds to the starting point of the vertical oscillation, and where $\theta_{-}= \arccos( z_{-})$ and $\pi-\theta_{-}=-\arccos( z_{-})$, which results in $\beta$ varying from $-\pi/2$ to $\pi/2$. Hence, using Eqs. (\ref{integral1}, \ref{integral2}) we find
\begin{equation}
\frac{\nu_{\theta}}{\nu_r}=\frac{\int_{r_{p}}^{r_{a}} \frac{dr}{\sqrt{R}}}{\int_{\theta_{-}}^{\pi-\theta_{-}} \frac{d \theta}{\sqrt{\Theta}}}=\frac{a \sqrt{1-E^2}z_{+} I_8 \left(\frac{\pi}{2} , e, \mu, a, Q \right) }{2 \cdot F\left( \frac{\pi}{2},\frac{z_{-}^{2}}{z_{+}^{2}}\right)  },  \label{nuthetanur}. 
\end{equation}
\end{subequations}
The similar expression can also be derived using formulae given in \cite{Fujita2009}. Again, we take a long time average of Eq. \eqref{tintmtn}, so that
\begin{subequations}
\begin{equation}
 \lim_{T \to \infty } \frac{t-t_0}{T}=\lim_{T \to \infty }\frac{1}{T}\left[ \frac{1}{2}\int_{r_{0}}^{r} \frac{1}{\Delta \sqrt{R}}\frac{\partial R}{\partial E} {\rm d} r^{'}  + \frac{1}{2}\int_{\theta_{0}}^{\theta} \frac{1}{\sqrt{\Theta}}\frac{\partial \Theta}{\partial E}{\rm d}\theta^{'}\right], \label{eqninf1}
\end{equation}
where using the same argument, again, of large possible integer solutions, so that $N_r t_r= N_{\theta}t_{\theta}=T$ to find
\begin{eqnarray}
1=&&\frac{2 N_r \int_{r_{p}}^{r_{a}} \frac{1}{\Delta \sqrt{R}}\frac{\partial R}{\partial E} {\rm d} r^{'} }{2N_r t_r} + \frac{2 N_{\theta} \int_{\theta_{-}}^{\pi-\theta_{-}} \frac{1}{\sqrt{\Theta}}\frac{\partial \Theta}{\partial E}{\rm d}\theta^{'} }{2N_{\theta} t_{\theta}}= \nu_{r} I_2 + \nu_{\theta} H_2, \label{equation1}
\end{eqnarray}
which gives
\begin{eqnarray}
\nu_r \left( e, \mu, a, Q \right) =&& \frac{1}{I_2 \left(\frac{\pi}{2} ,  e, \mu, a, Q \right) + \frac{\nu_{\theta}}{\nu_r} H_2 \left(- \frac{\pi}{2} , \frac{\pi}{2}, e, \mu, a, Q \right)}, \\
\nu_{\theta} \left(  e, \mu, a, Q \right)=&& \frac{1}{\frac{\nu_{r}}{\nu_{\theta}} I_2 \left(\frac{\pi}{2} , e, \mu, a, Q \right) + H_2 \left(- \frac{\pi}{2} , \frac{\pi}{2}, e, \mu,  a, Q \right)}.
\end{eqnarray}
\end{subequations}
 The limits of integral $I_2$ are $\alpha=\{0, \pi/2\}$, and that of $H_2$ are $\beta_{0}=\{\pi/2, -\pi/2\}$. The substitution  of $H_2\left(- \frac{\pi}{2} , \frac{\pi}{2}, e, \mu, a, Q \right)$ and $\displaystyle{\frac{\nu_{\theta}}{\nu_r}}$ from Eqs. \eqref{H2} and \eqref{nuthetanur} in the above equations give
 \begin{subequations}
\begin{equation}
\nu_r \left(   e, \mu, a, Q \right) = \frac{F\left( \frac{\pi}{2},\frac{z_{-}^{2}}{z_{+}^{2}}\right) }{\left\lbrace  \splitfrac{\left[ I_2 \left(\frac{\pi}{2} , e, \mu, a, Q \right)  + 2 a^2 z_{+}^2 E I_8 \left(\frac{\pi}{2} , e, \mu, a, Q \right)  \right] F\left( \frac{\pi}{2},\frac{z_{-}^{2}}{z_{+}^{2}}\right) }{
 - 2 a^2 z_{+}^2 E I_8 \left(\frac{\pi}{2} , e, \mu, a, Q \right) K\left( \frac{\pi}{2},\frac{z_{-}^{2}}{z_{+}^{2}}\right)  } \right\rbrace  }, \label{nur}
\end{equation}
 \begin{equation}
 \nu_{\theta} \left(  e, \mu, a, Q \right) =  \frac{a \sqrt{1- E^2} z_{+} I_8 \left(\frac{\pi}{2} ,  e, \mu, a, Q \right)}{2\left\lbrace \splitfrac{\left[ I_2\left(\frac{\pi}{2} ,  e, \mu, a, Q \right) + 2 a^2 z_{+}^2 E I_8\left(\frac{\pi}{2} , e, \mu, a, Q \right)  \right] F\left( \frac{\pi}{2},\frac{z_{-}^{2}}{z_{+}^{2}}\right)  }{
 - 2 a^2 z_{+}^2 E I_8 \left(\frac{\pi}{2} ,  e, \mu, a, Q \right) K\left( \frac{\pi}{2},\frac{z_{-}^{2}}{z_{+}^{2}}\right)  }\right\rbrace }. \label{nutheta}
 \end{equation}

Similarly, taking the long time average of Eq. \eqref{phiintmtn} and the substitution of $H_1$ and $H_2$ from Eqs. \eqref{H1} and \eqref{H2} yields
\begin{equation}
\nu_{\phi}\left(   e, \mu, a, Q \right)=\frac{\left\lbrace \splitfrac{ \left[ - I_1\left(\frac{\pi}{2} , e, \mu, a, Q \right) - 2 L I_8\left(\frac{\pi}{2} ,  e, \mu, a, Q \right)\right] F\left( \frac{\pi}{2},\frac{z_{-}^{2}}{z_{+}^{2}}\right)  }{+ 2 L I_8\left(\frac{\pi}{2} ,  e, \mu, a, Q \right) \Pi\left( z_{-}^2, \frac{\pi}{2},\frac{z_{-}^{2}}{z_{+}^{2}}\right)  }\right\rbrace }{2 \pi \left\lbrace  \splitfrac{ \left[I_2\left(\frac{\pi}{2} ,  e, \mu, a, Q \right) + 2 a^2 z_{+}^2 E I_8\left(\frac{\pi}{2} ,  e, \mu, a, Q \right) \right] F\left( \frac{\pi}{2},\frac{z_{-}^{2}}{z_{+}^{2}}\right) }{ -  2a^2 z_{+}^2 E I_8\left(\frac{\pi}{2} ,  e, \mu, a, Q \right) K\left( \frac{\pi}{2},\frac{z_{-}^{2}}{z_{+}^{2}}\right) } \right\rbrace }, \label{nuphi}
\end{equation}
\end{subequations}
where $I_1$, $I_2$, and $I_8$ are given by Eqs. \eqref{I1a}-\eqref{I7} and \eqref{integral1}. Hence, the fundamental frequencies are explicit functions of input parameters ($e$, $\mu$, $a$, $Q$), which can be chosen using the bound orbit conditions presented in \S \ref{bndcnd}. These frequency formulae also match with the quadrature formulae derived in \cite{Schmidt2002}; but here we have explicitly solved the integrals $I_1$, $I_2$ and $I_8$ in \S \ref{analyticsoln}. 

\begin{table}
\begin{center}
\caption{This table summarizes the fundamental frequency formulae derived using the long time average method in the Kerr geometry. The explicit expressions for integrals $I_1$, $I_2$ and $I_8$ are summarized in table \ref{integraltable}. }
\scalebox{0.85}{
\begin{tabular}{|c|c|}
\hline
\hline
& \\
$\nu_r \left(   e, \mu, a, Q \right)$ &$\frac{F\left( \frac{\pi}{2},\frac{z_{-}^{2}}{z_{+}^{2}}\right) }{M\left\lbrace \left[ I_2 \left(\frac{\pi}{2} ,  e, \mu, a, Q \right)  + 2  E I_8 \left(\frac{\pi}{2} ,  e, \mu, a, Q \right)  \right] F\left( \frac{\pi}{2},\frac{z_{-}^{2}}{z_{+}^{2}}\right) - 2 a^2 z_{+}^2  E I_8 \left(\frac{\pi}{2} ,  e, \mu, a, Q \right) K\left( \frac{\pi}{2},\frac{z_{-}^{2}}{z_{+}^{2}}\right)  \right\rbrace  }$\\
& \\
\hline
&\\
$\nu_{\theta}\left(  e, \mu, a, Q \right)$ &$\frac{a \sqrt{1- E^2} z_{+} I_8 \left(\frac{\pi}{2} , e, \mu, a, Q \right)}{2M\left\lbrace\left[ I_2 \left(\frac{\pi}{2} , e, \mu, a, Q \right)  + 2  E I_8 \left(\frac{\pi}{2} , e, \mu, a, Q \right)  \right] F\left( \frac{\pi}{2},\frac{z_{-}^{2}}{z_{+}^{2}}\right) - 2 a^2 z_{+}^2E I_8 \left(\frac{\pi}{2} , e, \mu, a, Q \right)  K\left( \frac{\pi}{2},\frac{z_{-}^{2}}{z_{+}^{2}}\right)   \right\rbrace  } $\\
& \\
\hline
& \\
$\nu_{\phi}\left( e, \mu, a, Q \right)$ &  $\frac{\left\lbrace \left[ - I_1\left(\frac{\pi}{2} , e, \mu, a, Q \right) - 2 L I_8\left(\frac{\pi}{2} , e, \mu, a, Q \right)\right] F\left( \frac{\pi}{2},\frac{z_{-}^{2}}{z_{+}^{2}}\right)  + 2 L I_8\left(\frac{\pi}{2} , e, \mu, a, Q \right) \Pi\left( z_{-}^2, \frac{\pi}{2},\frac{z_{-}^{2}}{z_{+}^{2}}\right)  \right\rbrace }{2 \pi M\left\lbrace \left[ I_2 \left(\frac{\pi}{2} , e, \mu, a, Q \right)  + 2  E I_8 \left(\frac{\pi}{2} , e, \mu, a, Q \right)  \right] F\left( \frac{\pi}{2},\frac{z_{-}^{2}}{z_{+}^{2}}\right) 
 - 2 a^2 z_{+}^2 E I_8 \left(\frac{\pi}{2} , e, \mu, a, Q \right) K\left( \frac{\pi}{2},\frac{z_{-}^{2}}{z_{+}^{2}}\right)  \right\rbrace  }$\\
 &\\
 \hline
 \hline
\end{tabular}
\label{freqformulae}
}

\end{center}

\end{table}

\section{Consistency check with previous results}
\label{reduction}
In order to verify our results, we have reduced our formulae for the non-equatorial separatrix trajectories, Eqs. (\ref{Qsepfinal}), to the case of equatorial separatrix orbits and found that they are consistent with earlier results derived in \cite{Levin2009}. We also found that the frequency ratio, $\nu_{\phi}/\nu_{\theta}$, from Eqs. (\ref{nutheta}, \ref{nuphi}), reduce to the case of maximally rotating black hole, $a=1$, for spherical orbits previously derived in \cite{Wilkins1972}. See \ref{dervconsistency} for these derivations.

 \section{Applications}
\label{app}
There are various important applications of our analytic solutions of the general non-equatorial trajectories and the fundamental frequencies for astrophysical studies as discussed below:
\begin{enumerate}
\item \underline{\textit{Gravitational waves}}: One of the crucial applications of our trajectory solution is the case of gravitational waves from the extreme-mass ratio inspirals (EMRIs). Our analytic formulae are directly applicable for the frequency domain calculation of the gravitational waves using the Teukolsky formalism, \cite{Teukolsky1973} or Kludge scheme \cite{Sopuerta2011}, and the orbits can be computed more accurately than the numerical calculations \cite{DrascoHughes2006}. Also, the homoclinic orbits, which are the separatrix between plunge and bound geodesics \cite{Levin2008,Perez-Giz2009}, have their importance to study the zoom-whirl behavior of inspirals near separatrix \cite{Glampedakis2002,HealyLevin2009}. In this paper, we provide the analytic formulae for eccentricity and inverse-latus rectum, ($e$, $\mu$), for non-equatorial separatrix orbits which are crucial for the selection of these orbits for the study of gravitational waveforms in the Kerr geometry. 

\item \underline{\textit{Relativistic precession}}: The exact analytic formula for azimuthal angle, $\phi-\phi_0$, is useful to find the precession of the orbits in the astrophysical systems like planets, black hole, and double pulsar systems. PSR J0737-3039 is one example of a double pulsar system having two pulsars,  PSR J0737-3039A and PSR J0737-3039B having 23 ms \cite{Burgay2003} and 2.8 s \cite{Lyne2004Sci} period respectively, which is useful to study the relativistic precession phenomenon valid in a strong gravitational field. The periastron advance was estimated in this source  using the first PK parameter, $\dot{\omega}$ \cite{Kramer2006Sci}. Our exact analytic results can be used to make a more accurate estimation of the relativistic advance of the periastron in pulsar systems where one component is having a major spin contribution. 

\item \underline{\textit{Quasi-periodic oscillations (QPOs)}}: QPOs are broad peaks seen in the Fourier power spectrum of the Neutron star X-ray binaries (NSXRB) and black hole X-ray binaries (BHXRB). The relativistic precession (RP) model was introduced \citep{Stella1999a} to explain the kHz QPOs in NSXRB and later applied to BHXRB \citep{Stella1999b}. The RP model can be used to calculate the black hole parameters assuming a circular or eccentric orbit is giving rise to a pair of observed high-frequency QPOs and a singular and nearly simultaneous corresponding Type-C QPO, \cite{Motta2014a,Motta2014b}, where our exact formulae for the fundamental frequencies are applicable.

\item \underline{\textit{Gyroscope precession}}: The calculation of precession of spin of a test gyroscope is another application for the test of general relativity. In previous studies, approximate expressions were used for the fundamental frequencies as a series expansion in terms of eccentricity up to order $e^{2}$ around a Kerr black hole for the stable bound orbits in the equatorial plane \cite{Bini2016a}. Our exact analytic results are useful to estimate more accurate results  which are useful to explain the reported results of geodetic drift rate and frame-dragging drift rate by the Gravity Probe B (GP-B) \cite{Everitt2015}.

\item \underline{\textit{Phase space study}}: Study of dynamics of Kerr orbits by Poincar\'e maps is also well discussed \cite{Levin2008}. Our closed-form solutions are directly applicable to the study of the extreme chaotic behavior of orbits like Zoom-whirl orbits, which are extreme forms of perihelion precession \cite{Glampedakis2002}, and their phase space structures.

\end{enumerate}
 
\section{Summary}
\label{summary}
The summary of this paper is given below:
\begin{enumerate}
\item
We first translate the parameters ($E$, $L$, $a$, $Q$) to ($e$, $\mu$, $a$, $Q$) using the translation formulae, Eqs. \eqref{transfinal} to completely describe the trajectory solution in the ($e$, $\mu$, $a$, $Q$) space. We then select the allowed bound orbit by choosing the parameters ($e$, $\mu$, $a$, $Q$) using the bound orbit conditions, Eqs. \eqref{deltaregion}.
  
\item
We have derived the closed-form analytic solutions of the general eccentric trajectory in the Kerr geometry as function of elliptic integrals, $\{ \phi \left( r, \theta \right) , t \left( r, \theta \right) , r \left( \theta \right) \}$, Eqs. (\ref{phifinal}\ref{rthetafinal}). These trajectories around a Kerr black hole were previously derived in terms of Mino time \cite{Fujita2009}, $\lambda$, subject to the initial conditions on $\mathrm{d}r \left( 0 \right) / \mathrm{d}\lambda $ and $\mathrm{d}\theta \left( 0 \right) / \mathrm{d}\lambda$. The application of our trajectory solution to the various possible studies is numerically faster and does not require any selection of initial conditions. We choose the starting point of the trajectory as the apastron of the orbit, $r_a$ or $\alpha=0$, and the initial polar angle, $\theta_0$ or $\beta_0$, is an extra parameter which can be arbitrarily chosen between maximum and minimum allowed $\theta$ range for a given $Q$.  The input variables for plotting the trajectories are $\alpha$ and $\beta$ which define the range of $r$ and $\theta$ for a fixed combination of ($e$, $\mu$, $a$, $Q$). These results are summarized in Table \ref{integraltable}.

\item We have derived the formulae for $E$ and $L$ for the spherical orbits as functions of radius $r_s$, $a$, and $Q$, given by Eqs. \eqref{ELsphfinal}. 

\item We have derived the equations for ISSO, MBSO, and spherical light radius, Eqs. (\ref{ISSOequation}-\ref{ISOrad}). The light radius derived is the same as that for the equatorial case.

\item We discussed the non-equatorial separatrix orbits, which asymptote to the unstable spherical radius sharing the same $E$ and $L$ values with the eccentric bound orbit. The radius of this unstable spherical radius for the separatrix orbit exists between $MBSO$ and $ISSO$. We write the exact forms for the eccentricity and inverse-latus rectum ($e_s$, $\mu_s$) for the non-equatorial separatrix orbits as functions of $r_s$, $a$, and $Q$, given by Eqs. \eqref{emusepfinal}.

\item
 We use our general trajectory solutions to derive the equations of motions for non-equatorial separatrix orbits, given by Eqs. (\ref{Qsepfinal}), and find that the radial part of the solutions can be completely reduced to the form containing only trigonometric and logarithmic functions. We also show the reduction of these trajectories to the equatorial case which is also a new and useful form and match the solutions with the previously known result derived in \cite{Levin2009}. Separatrix trajectories are essential in the study of gravitational waves from EMRIs, where our analytic solutions are directly applicable. These results are summarized in Table \ref{septable}.

\item We discuss families of allowed bound orbit trajectories like non-equatorial eccentric, non-equatorial separatrix, zoom whirl, and spherical orbits around a rotating black hole using our analytic solution for the trajectories. Homoclinic trajectories have their applications in the gravitational wave astronomy as these trajectories are the boundaries between bound eccentric and plunge orbits. Separatrix/homoclinic orbits were studied for the equatorial case in \cite{Levin2009,Perez-Giz2009}. In this paper, we describe non-equatorial homoclinic and zoom-whirl trajectories, which is the more generalized case for the application to gravitational astronomy.

\item We derived the closed-form expressions for the fundamental frequencies in terms of elliptic integrals, Eqs. (\ref{nur}, \ref{nutheta}, \ref{nuphi}), using the long time average method and without using Mino time, $\lambda$. We show that these expressions match with those derived in \cite{Schmidt2002} using Hamilton-Jacobi formulation, which were left in the quadrature form, and we have obtained a closed form using elliptic integrals. These expressions are summarized in Table \ref{freqformulae}. We present the consistency of our trajectory solution by reducing it to the equatorial separatrix case and also show that the frequency ratio, $\nu_{\phi} / \nu_{\theta}$, matches with the standard expression derived \cite{Wilkins1972} for the spherical orbits. 
\end{enumerate}
The results include novel aspects given in (i) and (iii)-(vii), listed above, and alternate new forms of the known formulae, given in the points (ii) and (viii) above. The equations and tables providing these results are indicated in the points above. 
\section{Discussion and Conclusions}
\label{disc}

There are several notable results in the vast literature discussing various aspects of dynamics in Kerr geometry such as the quadrature formulae for the trajectories (\cite{C1983book,Carter1968}), circular orbit formulae \cite{Bardeen1972}, conditions for spherical orbits \cite{Wilkins1972}, expressions in terms of quadratures for the oscillation frequencies \cite{Schmidt2002}, formulae for trajectories in terms of
quadratures for spherical polar motion \cite{Kraniotis2004}, trajectories for non-spherical polar motion \cite{Kraniotis2007}, and expressions for the
trajectories and oscillation frequencies \cite{Fujita2009} in terms of Mino time \cite{Mino2003}. Besides these key results there are other useful expressions reported for example on separatrix orbits \cite{Levin2009}, and on eccentric equatorial bound orbits (\cite{C1983book,Bini2016a}). 

We discuss below the utility of the results in our paper:

 The recipe for calculating frequencies and trajectories by \cite{Fujita2009} is as follows: The operative equations are $\{\phi(\lambda), t(\lambda), r(\lambda)\}$, Eqs. (6), (23)-(33), (35)-(45), which require linear combinations of many other equations. The analogy to $r(\chi)$ or $\chi(r)$ is $r(\lambda)$ or $\lambda(r)$ (Eqs. (26, 27)); the latter is non-trivial, whereas the former is simple. Given $\lambda$, ($\phi, t$) are calculated subsequently inverting linear combinations of many other elliptic integrals. We have numerically matched our frequency formulae with that given in \cite{Fujita2009} and we find that there is a minor typo in their expression of $\Gamma$ below Eq. (20) in section 3.3, where there is a factor of $E /2$ missing in the term $(r_1 - r_3 )(r_2 - r_4 )E(k_r)$. However, the correct factor has been applied to calculate the numbers in their Tables (1, 2, 3) given in \cite{Fujita2009}. By using the set of equations  in \cite{Fujita2009} and comparing with our expressions, it is found that our calculation is easier to implement and numerically faster by $\sim 20$, in the equatorial case, for example.
  
 The novel results listed in (i) and (iii)-(vii) of the summary: translation conditions of $\{E,L\} \rightarrow \{e, \mu\}$, bound orbit conditions, $\{E (r_s), L (r_s)\}$, ISSO, MBSO, and light radius formulae, besides new form for equatorial trajectories, are useful for various applications and simulations related to astrophysical scenarios involving relativistic precession like QPOs and accretion disks. We have also derived the locus of the $Q \neq 0$ separatrix curve in the $e-\mu$ plane besides providing the form of the trajectories. Using this, further studies can be carried for chaotic motion and study of gravitational waves from  zoom-whirl orbits which can be set-up by locating
them near the separatrix locus, in the same spirit, as was done for the equatorial case \cite{Levin2009,Glampedakis2002}.

The analytic results presented in this paper have direct applications in astrophysics for example, the study of non-equatorial separatrix orbits which has not been discussed before. They also help in understanding the highly eccentric behaviour of trajectories seen in numerical simulations \citep{Glampedakisetal2002} just before plunging onto the massive black hole in the case of EMRIs which is possibly related to the eccentric and inclined homoclinic orbits, besides relativistic precession in other astrophysical systems like binary pulsars and black holes, spin precession of gyroscopes around rotating black holes for the test of general relativity, and the study of chaotic orbits in the phase space. 
\appendix

\section{Deriving translation formulae between ($E$, $L$) and ($e$, $\mu$) parameters}
\label{trnsltnbndder}
 The turning points of radial motion around rotating black hole are derived from Eq. \eqref{quartic}, which can be factorized into two quadratics
\begin{equation}
u^4 + a^{'} u^3 +b^{'}u^2 +c^{'}u +d^{'}= \left( u^2 + {a_{1}}^{'}u + {b_{1}}^{'}\right) \left(u^2 + {a_{2}}^{'}u +{b_{2}}^{'} \right) , \label{quadratics}
\end{equation}
and the comparison of coefficients on both sides yield the following relations
\begin{subequations}
\begin{eqnarray}
{a_{1}}^{'}+{a_{2}}^{'}=a^{'}, \label{coeff1}\\
{a_{1}}^{'}{a_{2}}^{'} +{b_{1}}^{'}+{b_{2}}^{'}=b^{'}, \label{coeff2}
\end{eqnarray}
\begin{eqnarray}
{b_{1}}^{'}{a_{2}}^{'}+{b_{2}}^{'}{a_{1}}^{'}=c^{'}, \label{coeff3}\\
{b_{1}}^{'}{b_{2}}^{'}=d^{'}.\label{coeff4}
\end{eqnarray}
\end{subequations}
Assuming that the first quadratic in Eq. \eqref{quadratics} have the turning points of the orbit, $u_{1}=\mu \left( 1-e\right) $ and $u_{2}=\mu \left(1+e \right) $, implies that ${a_{1}}^{'}=-\left( u_{1}+u_{2}\right)=- 2\mu $ and ${b_{1}}^{'}=\mu^2 \left(1- e^2 \right) $. Using Eqs. \eqref{coeff4} and \eqref{coeff1} to replace ${a_{2}}^{'}$ and ${b_{2}}^{'}$ in Eqs. \eqref{coeff2} and \eqref{coeff3} and substituting for ${a_{1}}^{'}=-\left( u_{1}+u_{2}\right)=- 2\mu $ and ${b_{1}}^{'}=\mu^2 \left(1- e^2 \right) $ yields
\begin{subequations}
\begin{equation}
-2 \mu \left(a^{'} + 2 \mu \right) + \mu^{2}\left(1-e^2 \right) + \frac{d^{'} }{\mu^{2}\left(1-e^2 \right)}=b^{'},
\end{equation}
\begin{equation}
\mu^2 \left(1-e^2 \right) \left(a^{'} + 2 \mu \right) - \frac{2d^{'}}{\mu \left(1-e^2 \right)}= c^{'}.
\end{equation}
\end{subequations}
The substitution of $a^{'}$, $b^{'}$, $c^{'}$ and $d^{'}$ and rearrangement of the terms in above equations gives
\begin{subequations}
\begin{eqnarray}
 E^2\left( e, \mu, a, Q\right)&&=1-\mu^3 \left( 1-e^2\right)^2 \left( \mu a^2 Q -Q -x^2\right) - \mu  \left( 1-e^2\right), \label{EntransformQ2}\\
  E&&= \frac{1}{2a x}\left[ -x^2-a^2 +a^2 Q \mu^2 \left( 1-e^2\right)- Q + \frac{1}{\mu}- \left( 3+e^2\right) \mu \left( \mu a^2 Q -Q -x^2\right)  \right], \nonumber \label{Eexprsn1}\\
  &&= C_1 x+ \frac{C_2}{x}, \label{Eexprsn}
 \end{eqnarray}
 \end{subequations}
where 
\begin{subequations}
\begin{eqnarray}
C_1=&&\frac{1}{2a}\left[ \left(3+e^2 \right) \mu -1 \right], \\
C_2=&& \frac{1}{2a}\left[\frac{1}{\mu}-a^2 -Q +a^2 Q \mu^2 \left( 1-e^2 \right) - \mu \left( 3+e^2 \right)   \left( \mu a^2 Q - Q\right)   \right].
\end{eqnarray}
\end{subequations}
 Next, the substitution of Eq. \eqref{Eexprsn} in Eq. \eqref{EntransformQ2} gives 
 \begin{equation}
 x^4\left[C_1^2 -\mu^3 \left(1-e^2 \right)^2 \right] +x^2 \left[ \mu \left( 1-e^2\right)+ 2 C_1 C_2 +\mu^3 \left(1-e^2 \right)^2 \left( \mu a^2 Q -Q\right)-1 \right] +C_2^2=0,  
 \end{equation}
 which is further solved for $x^2$ to obtain Eqs. (\ref{xsqr}-\ref{Rvar}). These relations also reduce to the equatorial case, $Q=0$, which was first derived in \cite{Bini2016a}.

\section{Solution of integrals $I_1$-$I_8$}
\label{integrals}
In this appendix, we show the solutions of the integrals given in \S \ref{analyticsoln}. First, we derive the radial integrals $I_1-I_8$, given by the Eqs. (\ref{I1a}-\ref{integral1}). We make the substitution $1/r^{'}=\mu \left(1+e \cos \chi \right)$, which reduces the integrals to
\begin{subequations}
\begin{eqnarray}
I_1=&&-2 \mu \left( 1- e^2\right)  \int_{\pi}^{\chi} \frac{L- 2 \left(L-a E \right) \mu y}{\left[1- 2 \mu y +a^2 \mu^2 y^2 \right] \sqrt{A\cos^2 \chi +B \cos \chi +C}} {\rm d} \chi, \nonumber \\
\end{eqnarray}
\begin{eqnarray}
I_2=&&\frac{2\left( 1-e^2\right) }{\mu}\int_{\pi}^{\chi}\frac{E +a^2 E \mu^2 y^2 -2 a \left(L-a E \right) \mu^3  y^3 }{y^2 \left[1- 2 \mu y +a^2 \mu^2 y^2 \right] \sqrt{A\cos^2 \chi +B \cos \chi +C} } {\rm d} \chi, \nonumber \\
\end{eqnarray}
\end{subequations}
where $y=\left(1+e \cos \chi \right)$. Further, we implement the partial fraction method to reduce the integrals to
\begin{subequations}
\begin{eqnarray}
I_1=&& -\frac{2 \left( 1- e^2\right)}{\mu a^2 }\int_{\pi}^{\chi} \left[ \frac{A_1 }{y-y_{+}}+ \frac{B_1}{y-y_{-}}\right] \frac{1}{\sqrt{A\cos^2 \chi +B \cos \chi +C}}   {\rm d} \chi , \nonumber  \\
\end{eqnarray}
\begin{eqnarray}
I_2=&&\frac{2\left( 1-e^2\right) }{\mu^3 a^2}\int_{\pi}^{\chi} \left[ \frac{A_2}{y^2} +\frac{B_2}{y}+ \frac{C_2}{y-y_{+}}+\frac{D_2}{y-y_{-}}\right] \frac{1}{\sqrt{A\cos^2 \chi +B \cos \chi +C}}   {\rm d} \chi,  \nonumber \\
\end{eqnarray} 
\end{subequations}
\begin{subequations}
where
\begin{eqnarray}
y_{\pm}=&& \frac{r_{\pm}}{a^2 \mu}, \\
A_1=&&  \frac{L a^2 \mu -2 \left( L -a E \right) \mu r_{+} }{2\sqrt{1-a^2}}, \ \ \
B_1= \frac{-La^2 \mu +2 \left( L -a E \right) \mu r_{-} }{2\sqrt{1-a^2}}, \\
A_2=&& E a^2 \mu^2, \ \ \
B_2= 2 E a^2 \mu^3,  \ \ \
C_2= \frac{a^2 \mu^3 }{r_{-}\sqrt{1-a^2}} \left(-La + 2 E r_{-}\right) ,
\end{eqnarray}
\begin{eqnarray}
D_2=&& \frac{a \mu^3 }{r_{-}\sqrt{1-a^2}} \left( -2 L r_{-} \sqrt{1-a^2} -2 E a r_{-} +L a^2 \right).
\end{eqnarray}

\end{subequations}
Next, we make the substitution, $\displaystyle{\cos \chi =2 \cos^2 \frac{\chi}{2}-1}$ and $\displaystyle{\psi=\frac{\chi}{2}-\frac{\pi}{2}}$, which reduces the integrals to
\begin{subequations}
\begin{equation}
I_1 = -  \left[ C_3  \int_{0}^{\psi}  \frac{{\rm d} \psi}{\left( 1+ p_{2}^2 \sin^2 \psi \right) \sqrt{1-m^2 \sin^2 \psi} \sqrt{1-n^2 \sin^2 \psi}}+ \right. \nonumber 
\end{equation}
\begin{equation}
  \left. C_4 \int_{0}^{\psi}  \frac{{\rm d} \psi}{\left( 1+ p_{3}^2 \sin^2 \psi \right) \sqrt{1-m^2 \sin^2 \psi} \sqrt{1-n^2 \sin^2 \psi}} \right] , \nonumber 
\end{equation}
\begin{eqnarray} 
=&& - \left[ C_3 I_3 + C_4 I_4\right] , \label{I1} \\
I_2= &&\left[C_5 \int_{0}^{\psi} \frac{{\rm d} \psi}{\left( 1+ p_{1}^2 \sin^2 \psi \right) ^2 \sqrt{1-m^2 \sin^2 \psi} \sqrt{1-n^2 \sin^2 \psi}} + \right. \nonumber \\
&& C_6 \int_{0}^{\psi} \frac{{\rm d} \psi}{\left( 1+ p_{1}^2 \sin^2 \psi \right) \sqrt{1-m^2 \sin^2 \psi} \sqrt{1-n^2 \sin^2 \psi}} + \nonumber \\
&& C_7 \int_{0}^{\psi}  \frac{{\rm d} \psi}{\left( 1+ p_{2}^2 \sin^2 \psi \right) \sqrt{1-m^2 \sin^2 \psi} \sqrt{1-n^2 \sin^2 \psi}} + \nonumber \\
&& \left. C_8 \int_{0}^{\psi}\frac{{\rm d} \psi}{\left( 1+ p_{3}^2 \sin^2 \psi \right) \sqrt{1-m^2 \sin^2 \psi} \sqrt{1-n^2 \sin^2 \psi}}  \right], \nonumber \\
=&& C_5 I_5 + C_6 I_6 + C_7 I_3 + C_8 I_4.  \label{I2} 
\end{eqnarray}
\end{subequations}
where the constants, $C_3$-$C_8$, $n^2$, $m^2$, ${p_1}^2$, ${p_2}^2$, and ${p_3}^2$ are defined by Eq. \eqref{constants} in \S \ref{analyticsoln}. First, we solve the integrals $I_6$, $I_3$ or $I_4$ which are of the form given by 
\begin{equation}
I_a \equiv \int_{0}^{\psi} \frac{{\rm d} \psi}{\left( 1+ p^2 \sin^2 \psi \right) \sqrt{1-m^2 \sin^2 \psi} \sqrt{1-n^2 \sin^2 \psi}},  \nonumber
\end{equation}
\begin{equation}
= \int_{0}^{\psi} \frac{{\rm d} \psi }{ \sqrt{1-m^2 \sin^2 \psi} \sqrt{1-n^2 \sin^2 \psi}} -p^2  \int_{0}^{\psi}\frac{\sin^2 \psi }{\left( 1+ p^2 \sin^2 \psi \right)\sqrt{1-m^2 \sin^2 \psi} \sqrt{1-n^2 \sin^2 \psi}}{\rm d} \psi, \nonumber
\end{equation}
\begin{equation}
=\int_{0}^{\psi} \frac{{\rm d} \psi }{ \sqrt{1-m^2 \sin^2 \psi} \sqrt{1-n^2 \sin^2 \psi}} +\frac{p^2}{m^2}  \int_{0}^{\psi}\frac{1-m^2\sin^2 \psi -1}{\left( 1+ p^2 \sin^2 \psi \right)\sqrt{1-m^2 \sin^2 \psi} \sqrt{1-n^2 \sin^2 \psi}}{\rm d} \psi, \nonumber
\end{equation}
\begin{equation}
I_a =\frac{1}{\left( p^2 +m^2 \right) }\left[ m^2\int_{0}^{\psi} \frac{{\rm d} \psi }{ \sqrt{1-m^2 \sin^2 \psi} \sqrt{1-n^2 \sin^2 \psi}} + p^2 \int_{0}^{\psi}\frac{\sqrt{1-m^2 \sin^2 \psi}}{\left( 1+ p^2 \sin^2 \psi \right) \sqrt{1-n^2 \sin^2 \psi}}{\rm d} \psi\right] . 
\label{Ia}
\end{equation}

Now, the substitution given by
\begin{equation}
\sin \alpha = \frac{\sqrt{1-m^2} \sin \psi}{\sqrt{1-m^2 \sin^2 \psi}}, \label{substn}
\end{equation}
reduces the integrals in Eq. \eqref{Ia} to
\begin{equation}
I_a = \frac{1}{\sqrt{1-m^2}\left( p^2 +m^2 \right) }\left[ m^2 F\left( \alpha, \frac{n^2-m^2}{1-m^2}\right)  + p^2 \Pi\left( \frac{-p^2-m^2}{1-m^2}, \alpha, \frac{n^2-m^2}{1-m^2}\right)   \right]. \label{Iafinal}
\end{equation}
 Hence, integrals given by Eqs. (\ref{I3}, \ref{I4}, \ref{I6}) reduce to the forms given above. Next, we solve for $I_5$, which is of the form
 \begin{eqnarray}
I_b \equiv && \int_{0}^{\psi} \frac{{\rm d} \psi}{\left( 1+ p^2 \sin^2 \psi \right)^2 \sqrt{1-m^2 \sin^2 \psi} \sqrt{1-n^2 \sin^2 \psi}}, \nonumber  \\
=&&\int_{0}^{\psi} \frac{1+p^2 \sin^2 \psi - p^2 \sin^2 \psi}{\left( 1+ p^2 \sin^2 \psi \right)^2 \sqrt{1-m^2 \sin^2 \psi} \sqrt{1-n^2 \sin^2 \psi}}{\rm d} \psi, \nonumber \\
=&&I_a +  \frac{p^2}{m^2} \int_{0}^{\psi} \frac{\sqrt{1-m^2 \sin^2 \psi}}{\left( 1+ p^2 \sin^2 \psi \right)^2  \sqrt{1-n^2 \sin^2 \psi}} {\rm d} \psi - \frac{p^2}{m^2} I_b, \nonumber \\
I_b =&& \frac{1}{\left( m^2 +p^2\right)} \left\lbrace  m^2 I_a + p^2 \left[ \int_{0}^{\psi} \frac{\sqrt{1-m^2 \sin^2 \psi}}{\left( 1+ p^2 \sin^2 \psi \right) \sqrt{1-n^2 \sin^2 \psi}} {\rm d} \psi \right. \right. \nonumber \\
&&\left. \left. - p^2 \int_{0}^{\psi} \frac{\sin^2 \psi \sqrt{1-m^2 \sin^2 \psi} }{\left( 1+ p^2 \sin^2 \psi \right)^2 \sqrt{1-n^2 \sin^2 \psi}} {\rm d} \psi \right]  \right\rbrace,  
\end{eqnarray}
 where the substitution given by Eq. \eqref{substn} reduces the second integral in the above equation to
 \begin{equation}
 I_b= \frac{1}{\left( m^2 +p^2\right)} \left[  m^2 I_a + \frac{p^2}{\sqrt{1-m^2}} \Pi\left( \frac{-p^2-m^2}{1-m^2}, \alpha, \frac{n^2-m^2}{1-m^2}\right)   - p^4 I_c \right], 
 \end{equation}
 where
 \begin{equation}
I_c \equiv \int_{0}^{\psi} \frac{\sin^2 \psi \sqrt{1-m^2 \sin^2 \psi} }{\left( 1+ p^2 \sin^2 \psi \right)^2 \sqrt{1-n^2 \sin^2 \psi}} {\rm d} \psi = -\frac{1}{n^2}\int_{0}^{\psi} \frac{\left( 1- n^2 \sin^2 \psi -1 \right)  \sqrt{1-m^2 \sin^2 \psi} }{\left( 1+ p^2 \sin^2 \psi \right)^2 \sqrt{1-n^2 \sin^2 \psi}} {\rm d} \psi, \nonumber
\end{equation}
\begin{equation}
I_c \left(1 + \frac{p^2}{n^2}\right)= \frac{1}{n^2} \left[ -\int_{0}^{\psi} \frac{ \sqrt{1-n^2 \sin^2 \psi} \sqrt{1-m^2 \sin^2 \psi} }{\left( 1+ p^2 \sin^2 \psi \right)^2 } {\rm d} \psi  + \int_{0}^{\psi} \frac{ \sqrt{1-m^2 \sin^2 \psi} }{\left( 1+ p^2 \sin^2 \psi \right) \sqrt{1-n^2 \sin^2 \psi}} {\rm d} \psi \right]; \nonumber
 \end{equation} 
the substitution given by Eq. \eqref{substn} reduces the above expression to
 \begin{equation}
 I_c = \frac{1}{\left( n^2 + p^2 \right) \sqrt{1-m^2}}  \left[ - \int_{0}^{\alpha}  \frac{ \sqrt{1-k^2 \sin^2 \alpha}  }{\left[ 1+\frac{p^2+m^2}{1-m^2} \sin^2 \alpha \right]^2 } {\rm d} \alpha +  \Pi\left( \frac{-\left( p^2+m^2\right) }{1-m^2}, \alpha, k^2\right)  \right],
 \end{equation}
 and multiplying and dividing the first integral in above equation by $\sqrt{1-k^2 \sin^2 \alpha}$ gives
 \begin{equation}
 I_c = \frac{1}{\left( n^2 + p^2 \right) \sqrt{1-m^2}}\left[ - \int_{0}^{\alpha}  \frac{1}{\left[ 1+\frac{p^2 +m^2}{1-m^2} \sin^2 \alpha \right]^2 \sqrt{1-k^2 \sin^2 \alpha}} {\rm d} \alpha + \Pi\left( \frac{-\left( p^2+m^2\right) }{1-m^2}, \alpha, k^2\right)\right. \nonumber 
 \end{equation}
 \begin{equation}
 \left. +k^2 \int_{0}^{\alpha}  \frac{\sin^2 \alpha}{\left[ 1+\frac{p^2+m^2}{1-m^2} \sin^2 \alpha \right]^2 \sqrt{1-k^2 \sin^2 \alpha}} {\rm d} \alpha \right] , 
 \end{equation}

which can be further reduced to
\begin{eqnarray}
I_c = && \frac{1 }{\sqrt{1-m^2} \left(m^2 + p^2 \right)} \left\lbrace \Pi\left( \frac{-\left( p^2+m^2\right) }{1-m^2}, \alpha, \frac{n^2-m^2}{1-m^2}\right)  - I_d \right\rbrace,
\end{eqnarray}  
where
\begin{equation}
I_d=\int_{0}^{\alpha}  \frac{1}{\left[ 1+\frac{p^2+m^2}{1-m^2} \sin^2 \alpha \right]^2 \sqrt{1-k^2 \sin^2 \alpha}} {\rm d} \alpha.
\end{equation}
By defining integrals having a general form given by
\begin{equation}
T_{n}=\int \frac{{\rm d} y}{\left( h+g \sin^{2}y \right)^{n} \sqrt{1-w^{2} \sin^{2}y }},
\end{equation}
we use the following identity \cite{Grad}:
\begin{equation}
T_{n-3}= -\frac{g^{2}\sin y \cos y \sqrt{1-w^{2} \sin^{2}y }}{\left( h+g \sin^{2}y \right)^{n-1}\left(2 n -5 \right)w^{2} } -\frac{\left( 2 n -3\right) \left[ g^{2} + 2 h g \left( 1+ w^{2}\right) + 3 h^{2} w^{2} \right] T_{n-1}}{\left(2 n -5 \right)w^{2}}  \nonumber
\end{equation}
\begin{equation}
+  \frac{2 \left( n-2\right)\left[g\left( 1+ w^{2}\right) + 3 h w^{2}  \right] T_{n-2}}{\left(2 n -5 \right)w^{2}}+ \frac{2 \left(n-1 \right) h \left(g+h \right) \left( g+ h w^{2}\right)  T_{n}}{\left(2 n -5 \right)w^{2}}. \label{Tn3}
\end{equation}
The integral $I_d$ has a form similar to $T_{2}$ with $ \left(  h=1, g=\frac{p^2+m^2}{1-m^2}, w=k \right)$, which yields
\begin{equation}
I_d=T_{2}=\frac{1}{2\left(1+ g \right)\left( g + k^{2}\right) }\left\lbrace  \frac{g^2\sin \alpha \cos \alpha \sqrt{1-k^{2}\sin^{2}\alpha}}{ \left( 1+ g \sin^{2} \alpha \right)}+ \left[g^{2} + 2 g \left( 1+ k^{2}\right)+ 3 k^{2}  \right]T_{1} -k^2 T_{-1}\right\rbrace  ,  \label{Id}
\end{equation}
where
\begin{eqnarray}
T_{-1}=&&\int_{0}^{\alpha}\frac{\left(1+g \sin^{2}\alpha \right) {\rm  d} \alpha}{ \sqrt{1-k^{2}\sin^{2}\alpha}} = \int_{0}^{\alpha} \frac{{\rm  d} \alpha}{ \sqrt{1-k^{2}\sin^{2}\alpha}}- \frac{g}{k^{2}} \int_{0}^{\alpha} \frac{\left(  1-k^{2} \sin^{2}\alpha -1\right) \ {\rm  d} \alpha}{  \sqrt{1-k^{2}\sin^{2}\alpha}} \nonumber \\
&&=  \left(1+ \frac{g}{k^{2}} \right)F\left(\alpha, k^{2}\right) - \frac{g}{k^{2}} K\left( \alpha, k^{2}\right), \label{T-1} 
\end{eqnarray}
and 
\begin{equation}
T_{1}=\Pi\left( -g, \alpha, k^{2}\right), \label{T1}
\end{equation}
were substituted. See Table \ref{integraltable} for the summary of the final expressions of $I_1-I_7$ derived using the method given in this section.

\section{Reduction to the equatorial plane ($Q=0$)}
\label{Qeq0}
In this appendix, we reduce the integrals of motion to the case of equatorial plane. We start with the final expressions of $I_3$, $I_4$, $I_5$, and $I_6$ (in \S \ref{analyticsoln}) and take the limit $Q \rightarrow 0$. As shown in \S \ref{equatorialtrajec}, that for $Q \rightarrow 0$, we have $n^2 \rightarrow 0$ which gives $\displaystyle{k^2=\frac{-m^2}{1-m^2}}$. Now, we first reduce the expressions of $I_3$ and $I_4$, Eqs. (\ref{I3}, \ref{I4}) under the limit $Q \rightarrow 0$. We use the following identity (cf. \cite{Byrd}, Eq. 160.02) to write
\begin{equation}
\Pi\left( \alpha_{1}^2, \varphi, - k_{1}^2 \right) = \frac{k_{2}^{'}\left[ k_{2}^2 F \left( \beta_{1}, k_{2}^2\right) + {k_{2}^{'}}^2 \alpha_{1}^2 \Pi\left( \alpha_{2}^2, \beta_{1}, k_{2}^2\right)  \right] }{\left( \alpha_{1}^2 {k_{2}^{'}}^2 + k_{2}^2\right) }, \label{id}
\end{equation}
where
\begin{equation}
\sin \beta_{1} = \frac{\sqrt{1+k_{1}^2}\sin \varphi}{\sqrt{1+k_{1}^2 \sin ^2 \varphi}},
\end{equation}
 $\alpha_{2}^2 = \alpha_{1}^2 {k_{2}^{'}}^2 + k_{2}^2$, $k_{2}=\frac{k_{1}}{\sqrt{1+k_{1}^2}}$ and $k_{2}^{'}=\frac{k_{2}}{k_{1}}$, which reduces $I_3$ and $I_4$ to the forms
\begin{equation}
I_3 = \frac{1}{\sqrt{1-m^2} \left( m^2 + p_{2}^2\right) } \left[m^2 F \left(\alpha , k^2 \right)  + p_{2}^2 \Pi \left(\frac{- \left(p_{2}^2 + m^2\right) }{1-m^2} , \alpha , k^2 \right)  \right]  = \Pi \left( -p_{2}^2 , \psi , m^2 \right), \label{I3r}
\end{equation}
\begin{equation}
I_4 = \frac{1}{\sqrt{1-m^2} \left( m^2 + p_{3}^2\right) } \left[m^2 F \left(\alpha , k^2 \right)  + p_{3}^2 \Pi \left(\frac{- \left(p_{3}^2 + m^2\right) }{1-m^2} , \alpha , k^2 \right)  \right]  = \Pi \left( -p_{3}^2 , \psi , m^2 \right). \label{I4r}
\end{equation}
The above expressions can be directly obtained if $n^2=0$ is substituted in the definition of $I_3$ and $I_4$ at the intermediate step, Eq. \eqref{I1}. However, we aim to directly validate the final forms of ($I_3-I_7$). Hence, we apply the reduced expressions of $I_3$ and $I_4$ to  $\left(\phi-\phi_{0}\right)$ for the equatorial plane to obtain Eq. \eqref{eqphi}.

Next, we reduce the coordinate time integral. The expression for $I_6$ also reduces in similar way to $I_3$ and $I_4$ by applying the identity, Eq. \eqref{id}, to
\begin{equation}
I_6=\frac{1}{\sqrt{1-m^2} \left( m^2 + p_{1}^2\right) } \left[m^2 F \left(\alpha , k^2 \right)  + p_{1}^2 \Pi \left(\frac{- \left(p_{1}^2 + m^2\right) }{1-m^2} , \alpha , k^2 \right)  \right]  = \Pi \left( -p_{1}^2 , \psi , m^2 \right), \label{I6r}
\end{equation} 
which again can be directly obtained by substituting $n^2=0$ in the definition of $I_6$ in Eq. \eqref{I2}. Next, we reduce the expression of $I_5$, Eq. \eqref{I5}. 
We substitute for $\displaystyle{\sin \alpha = \frac{\sqrt{1-m^2}\sin \psi}{\sqrt{1-m^2 \sin^2 \psi}}}$, $\displaystyle{\cos \alpha = \frac{\cos \psi}{\sqrt{1-m^2 \sin^2 \psi}}}$, $\displaystyle{k^2 = \frac{-m^2}{1-m^2}}$ and $\displaystyle{s^2 = - \frac{p_{1}^2 +m^2}{1-m^2}}$ into the expression of $I_7$, Eq. \eqref{I7}, and use the following identities (cf. \cite{Byrd}, Eq. 160.02)
\begin{eqnarray}
F\left( \varphi , -k_{1}^2 \right) =&& k_{2}^{'} F \left( \beta_{1} , k_{2}^2\right) , \\ 
K\left( \varphi , -k_{1}^2 \right) =&& \frac{1}{k_{2}^{'}}\left[ K \left( \beta_{1} , k_{2}^2\right) - \frac{k_{2}^2 \sin \beta_{1} \cos \beta_{1}}{\sqrt{1- k_{2}^2 \sin^2 \beta_{1}}} \right] ,
\end{eqnarray}
and Eq. \eqref{id} to transform $ F\left(\alpha, k^2 \right)$, $K\left(\alpha, k^2\right)$, and $\Pi\left( s^2 , \alpha, k^2\right)$ to
\begin{eqnarray}
 F\left(\alpha, k^2 \right)=&&  \sqrt{1-m^2} F \left( \psi , m^2\right), \\
 K\left( \alpha , k^2\right) =&& \frac{1}{\sqrt{1-m^2}} \left[ K\left( \psi , m^2\right) - \frac{  m^2 \sin \psi \cos \psi}{\sqrt{1-m^2 \sin^2 \psi}}\right] , \\
  \Pi\left( s^2 , \alpha, k^2\right) =&& \frac{\sqrt{1-m^2}}{p_{1}^2}\left[ \left(p_{1}^2 +m^2 \right) \Pi \left( -p_{1}^2 , \psi , m^2\right) - m^2 F\left( \psi , m^2\right)  \right] .
\end{eqnarray} 
The substitution of the above equations reduces $I_7$ to
\begin{equation}
I_7 =\left\lbrace \frac{\left( p_{1}^2 + m^2\right)^2 \sqrt{1-m^2} \sin \psi \cos \psi }{2 p_{1}^2 \left(1+ p_{1}^2 \right) \sqrt{1-m^2 \sin^2 \psi} \left(1+ p_{1}^2 \sin^2 \psi \right)  }  +  \frac{\left( p_{1}^2 + m^2\right) \sqrt{1-m^2}}{2p_{1}^2 \left(1+ p_{1}^2 \right)} \left[ K\left( \psi , m^2\right) - \frac{ m^2 \sin \psi \cos \psi}{\sqrt{1-m^2 \sin^2 \psi}} \right] \right. \nonumber \\
\end{equation}
\begin{equation}
\left. +  \frac{\left(p_{1}^4 -2 p_{1}^2 m^2 + 2 p_{1}^2 -m^2 \right) \sqrt{1-m^2}}{2p_{1}^4 \left(1+ p_{1}^2 \right)  }\left[ \left( p_{1}^2 + m^2\right) \Pi \left( -p_{1}^2 , \psi , m^2\right) - m^2 F\left( \psi , m^2\right) \right] - \frac{\left( 1-m^2\right)^{3/2} F\left( \psi , m^2\right)}{2 \left(1+ p_{1}^2 \right)}\right\rbrace .
\end{equation}
Now, the substitution of $I_7$ from the above equation and $\sin \alpha$, $\cos \alpha$, $k^2$, $s^2$ into Eq. \eqref{I5} reduces $I_5$ to
\begin{eqnarray}
 I_5=&& \frac{p_{1}^4 \sin \psi \cos \psi \sqrt{1-m^2 \sin^2 \psi}}{2\left( 1+p_{1}^2\right) \left( m^2 +p_{1}^2\right) \left(1+p_{1}^2 \sin^2 \psi \right) } + \frac{\left[ p_{1}^4 +2p_{1}^2 \left( 1+m^2\right) + 3m^2\right] }{2\left( 1+p_{1}^2\right) \left( m^2 +p_{1}^2\right)} \Pi\left( -p_{1}^2 , \psi , m^2\right) - \nonumber \\
 && \frac{1}{2\left(1+p_{1}^2 \right) } F\left( \psi , m^2\right)  +\frac{p_{1}^2}{2 \left(1+p_{1}^2 \right) \left( m^2 + p_{1}^2 \right) } K\left( \psi , m^2\right) . \label{I5r}
\end{eqnarray} 
The above expression of $I_5$ can also be directly obtained by substituting $n^2=0$ in its definition given in Eq. \eqref{I2} and applying the identity given by Eq. \eqref{Tn3}. Hence, the substitution of $I_3$, $I_4$, $I_5$ and $I_6$ from Eqs. \eqref{I3r}, \eqref{I4r}, \eqref{I5r} and \eqref{I6r} in Eq. \eqref{tfinal} gives the expression for coordinate time, $t$, which simplifies to Eq. \eqref{eqtime}. 

See Table \ref{integraltable} for the final expressions of $(\phi, t)$ for the equatorial plane derived using the method given in this section.

\section{Innermost stable and marginally bound spherical radii}
\label{sphapp}
We discussed in \S \ref{emuseparatrix} that ISSO and MBSO radii represent the end points of the separatrix curve defined by Eq. \eqref{sepeq}. Hence, we use this to write equations for ISSO and MBSO.
\begin{enumerate}
\item For the case of ISSO, we substitute $e=0$ and $\mu=1/r_s$ into Eq. \eqref{sepeq}, which gives
\begin{equation}
r_{s}^3 - 3 r_s \left( Q + x^2 \right)  + 4 a^2 Q  =0,
\end{equation}
which further expands, by the substitution of $x^2$ for spherical orbits, to
\begin{eqnarray}
&&r_{s}^6-9r_{s}^5-3a^2r_{s}^4 + 18 r_{s}^4+ 4 a^2 Q r_{s}^3 - 7a^2 r_{s}^3-18 a^2 Q r_{s}^2 + 24 a^2 Q r_{s} -10 a^4 Q \nonumber \\
&&+ 6a \left( a^2 - 2r_{s} + r_{s}^2\right) \sqrt{a^2 Q^2 - Q \left(r_{s} -3 \right) r_{s}^3 + r_{s}^5}=0. \label{ISSOeqn} 
\end{eqnarray}
The above equation has a complicated explicit expression of order twelve in $r_s$,
which factorizes to
\begin{eqnarray} 
\left( \splitfrac{r_{s}^9 -12 r_{s}^8 - 6a^2 r_{s}^7 + 36 r_{s}^7 + 8 a^2 Q r_{s}^6 -28 a^2 r_{s}^6 -24 a^2 Q r_{s}^5 + 9 a^4 r_{s}^5 -24 a^4 Q r_{s}^4 }{ + 48 a^2 Q r_{s}^4 + 16 a^4 Q r_{s}^3 -8a^4 Q r_{s}^3 -48 a^4 Q^2 r_{s}^2 + 48 a^4 Q^2 r_{s} -16 a^6 Q^2}\right)&& \cdot \nonumber \\
\left(r_{s}^3 -6 r_{s}^2 + 9 r_{s} -4a^2 \right) =0,&& 
\end{eqnarray}
where the second factor corresponds to the light radius \cite{Bardeen1972} and the equation for ISSO is given by
\begin{eqnarray}
&&r_{s}^9 -12 r_{s}^8 - 6a^2 r_{s}^7 + 36 r_{s}^7 + 8 a^2 Q r_{s}^6 -28 a^2 r_{s}^6 -24 a^2 Q r_{s}^5 + 9 a^4 r_{s}^5 -24 a^4 Q r_{s}^4 + \nonumber \\
&&48 a^2 Q r_{s}^4 + 16 a^4 Q^2 r_{s}^3 -8a^4 Q r_{s}^3 -48 a^4 Q^2 r_{s}^2 + 48 a^4 Q^2 r_{s} -16 a^6 Q^2=0. \label{ISSOeqn2}
\end{eqnarray}
For the equatorial plane, $Q=0$ gives the following equation for the ISCO radius ($Z$)
\begin{equation}
Z^6-9Z^5-3a^2Z^4 + 18 Z^4- 7a^2 Z^3+ 6a \left( a^2 - 2Z + Z^2\right)Z^{5/2}=0;
\end{equation}
this equation can be factorized as
\begin{equation}
Z^{5/2}\left( Z^{3/2}- 3 Z^{1/2} - 2a \right) \left(Z^2 - 6Z + 8a Z^{1/2} -3a^2 \right) =0 , 
\end{equation} 
where the first bracket gives the solution for light radius and the second bracket,
\begin{equation}
Z^2 - 6Z + 8a Z^{1/2} -3a^2  =0,
\end{equation} 
gives the solution for ISCO \cite{Bardeen1972} given by 
\begin{subequations}
\begin{eqnarray}
Z=&&\left\lbrace 3+ Z_{2} - \left[\left(3- Z_{1}\right) \left( 3+ Z_{1}+ 2Z_{2}\right)  \right]^{1/2}  \right\rbrace , \label{ISCOrad} \\
Z_{1}=&& 1+\left( 1- a^{2}\right)^{1/3} \left[ \left(1+ a \right)^{1/3} + \left(1- a \right)^{1/3} \right], \\
Z_{2}=&& \left(3 a^{2} + Z_{1}^{2} \right)^{1/2}.
\end{eqnarray}
\end{subequations}

\item The condition for MBSO is derived by substituting $e=1$ and $\mu=1/(2Y)$ in Eq. \eqref{sepeq}, which yields
\begin{equation}
Y^3 - Y \left( Q + x^2 \right)  + a^2 Q  =0,
\end{equation}
where $r_{a}$ of the orbit reaches infinity and $r_s$ in the above equation corresponds to the unstable radius. By substituting $x^2$ for spherical orbits, the above equation reduces to 
\begin{eqnarray}
&&Y^6 - 7 Y^5 + 12Y^4 -a^2 Y^4 + a^2 Q Y^3 - 5 a^2 Y^3 - 4 a^2 Q Y^2 + 5 a^2 Q Y - 2a^4 Q, \nonumber \\
 &&+ 2a \left(a^2 -2 Y + Y^2 \right) \sqrt{a^2 Q^2 - Q \left( Y-3\right)Y^3 + Y^5 }=0.  \label{ISBOeqn}
\end{eqnarray} 
The above expression expands to an equation of eleventh order in $r_s$,
which factorizes to
\begin{eqnarray} 
\left( \splitfrac{Y^8 -8 Y^7 - 2a^2 Y^6 + 16 Y^6 + 2 a^2 Q Y^5 -8 a^2 Y^5 -6 a^2 Q Y^4 + a^4 Y^4 -2 a^4 Q Y^3 }{ + 8 a^2 Q Y^3 +  a^4 Q^2 Y^2 -2a^4 Q Y^2 -2 a^4 Q^2 Y + a^4 Q^2 }\right)&& \cdot \nonumber \\
\left(X^3 -6 X^2 + 9 X -4a^2 \right) =0,&& 
\end{eqnarray}
where the second factor corresponds to the light radius ($X$) \cite{Bardeen1972} and the equation for MBSO is given by
\begin{eqnarray}
&&Y^8 -8 Y^7 - 2a^2 Y^6 + 16 Y^6 + 2 a^2 Q Y^5 -8 a^2 Y^5 -6 a^2 Q Y^4 + a^4 Y^4 -2 a^4 Q Y^3 + \nonumber \\
&& 8 a^2 Q Y^3 +  a^4 Q^2 Y^2 -2a^4 Q Y^2 -2 a^4 Q^2 Y + a^4 Q^2 =0. \label{ISBOeqn2}
\end{eqnarray}
In the equatorial plane, $Q=0$, this reduces to
\begin{equation}
Y^6 - 7 Y^5 + 12Y^4 -a^2 Y^4- 5 a^2 Y^3+2a \left(a^2 -2 Y + Y^2 \right) Y^{5/2}=0,
\end{equation}  
which gets further factorized to
\begin{equation}
Y^{5/2}\left(Y + 2 Y^{1/2} -a \right) \left(Y - 2 Y^{1/2} +a \right) \left( X^{3/2}- 3 X^{1/2} - 2a\right) =0,
\end{equation}
where the last bracket in above equation gives the solution for light radius and the first two brackets give retrograde and prograde solutions for marginally bound orbit in the equatorial plane \cite{Bardeen1972}. According to the sign convention used in this paper ($-1<a<1$), both retrograde and prograde cases are covered by the formula given by
\begin{equation}
Y=2 -a + 2 \sqrt{1-a}. 
\label{ISBOrad}
\end{equation} 

\end{enumerate}

\section{Solution of ($e_s$, $\mu_s$) in the equatorial separatrix case}
\label{separatrixapp}
The expressions for eccentricity and inverse-latus rectum for the non-equatorial separatrix orbits, given by Eqs. \eqref{esepgen} and \eqref{musepgen} can be written in the form
\begin{eqnarray}
e_s=&&\frac{2u_s}{\left[ -\frac{a^{'}}{2}-\frac{1}{2}\sqrt{\left(a^{'}+ 2u_s \right)^2-\frac{4d^{'}}{u_{s}^2}}\right] }-1, \label{appc1}\\
\mu_s=&&\frac{1}{4}\left[-a^{'}-\sqrt{\left(a^{'}+ 2u_s \right)^2-\frac{4d^{'}}{u_{s}^2}} \right]. \label{appc2}
\end{eqnarray}
where the expressions for $a^{'}$ and $d^{'}$ are given by Eq. \eqref{adash}.
We solve for the factor in the denominator of $e_s$ by substituting $a^{'}$ and $d^{'}$ and taking the limit $Q\rightarrow 0$, we find
\begin{equation}
\left[-\frac{a^{'}}{2}-\frac{1}{2}\sqrt{\left(a^{'}+ 2u_s \right)^2-\frac{4d^{'}}{u_{s}^2}} \right]= \left[ \frac{x^2}{a^2 Q}+\frac{1}{a^2} -  \frac{x^2}{a^2 Q}\left( 1 + \frac{Q}{x^2} - \frac{u_s a^2 Q}{x^2}- \frac{a^2 Q}{2u_{s}^2 x^4} + \frac{E^2 a^2 Q}{2 u_{s}^2 x^4} +O[Q^2]\right) \right],
\end{equation}
\begin{eqnarray}
=&& u_s + \frac{1}{2x^2 u_s^2 }-\frac{E^2}{2x^2 u_{s}^2}. \label{expand}
\end{eqnarray}

Substitution of Eq. \eqref{expand} in (\ref{appc1}, \ref{appc2}) gives
\begin{eqnarray}
e_s =&& \frac{2u_{s}^3 x^2 -1 + E^2}{2 u_{s}^3x^2 +1 -E^2}, \ \
\mu_{s}= \frac{1}{4}\left[2 u_s + \frac{1- E^2}{x^2 u_{s}^2} \right].
\end{eqnarray}
Now, by substituting $x^2=\left(L-aE \right)^2$ and $E^2$ from Eq. \eqref{ELcircular} into the above equations, we find
\begin{eqnarray}
e_s=&& -\frac{r_{s}^{7/2} -9 r_{s}^{5/2} + 6 a r_{s}^{2}-3a^2 r_{s}^{3/2} + 18 r_{s}^{3/2}-12 a r_s - 7a^2 r_{s}^{1/2}+ 6a^3 }{r_{s}^{7/2} -5 r_{s}^{5/2} - 2 a r_{s}^{2}+ a^2 r_{s}^{3/2} + 6 r_{s}^{3/2}+ 4 a r_s - 3a^2 r_{s}^{1/2}-2a^3}, \nonumber \\
=&& -\frac{\left(r_{s}^{3/2} - 3 r_{s}^{1/2} -2 a \right) \left( r_{s}^2 - 6 r_s + 8 a r_{s}^{1/2} - 3 a^2 \right) }{ \left( r_{s}^{3/2} - 3 r_{s}^{1/2} -2 a \right)\left( r_{s}^2 +a^2 -2 r_s\right)  }= -\frac{r_{s}^2 - 6 r_s + 8 a r_{s}^{1/2} - 3 a^2}{r_{s}^2 +a^2 -2 r_s}. \nonumber \\
\end{eqnarray}
and 
\begin{eqnarray}
\mu_s =&&\frac{ r_{s}^4 - 5 r_{s}^3 -2 a r_{s}^{5/2} +a^2 r_{s}^2 + 6 r_{s}^2 + 4 a r_{s}^{3/2}  - 3 a^2 r_s - 2 a^3 r_{s}^{1/2}}{4 r_s\left(r_{s}^{3} - 3r_{s}^2 + a^2 r_{s}^2 + a^2 r_{s} - 2 a r_{s}^{5/2} + 4 a r_{s}^{3/2} - 2 a^3 r_{s}^{1/2}\right)  },\nonumber \\
=&& \frac{\left(r_{s }^2 + a^2 - 2 r_s\right) \left(  r_{s}^2 - 3 r_s - 2 a r_{s}^{1/2}\right) }{ 4 r_s\left(r_s +a^2 - 2 a {r_s}^{1/2}\right) \left(  r_{s}^2 - 3 r_s - 2 a r_{s}^{1/2}\right) }= \frac{r_{s }^2 + a^2 - 2 r_s}{4 r_s \left(r_{s}^{1/2} - a \right) ^2},
\end{eqnarray}
which are the expressions in Eq. (\ref{esepfor}) given in \S \ref{emuseparatrix}.
\section{Reducing the radial integrals for the case of non-equatorial separatrix trajectories}
\label{Qsepintegrals}
Here, we reduce the radial integrals in our general trajectory solutions presented in \S \ref{analyticsoln}, for the case of separatrix orbits with $Q\neq0$. As shown in \S \ref{separatrixtrajec} that $k^2=1$ for the separatrix orbits, we use the identities given by Eq. \eqref{sepred} \cite{Grad} to reduce the corresponding Elliptic integrals.

The substitution of these reduced form of the Elliptic integrals further reduces the integrals of the form $S_3$ and $S_4$, Eqs. (\ref{I3}, \ref{I4}), to the general form

\begin{equation}
S \equiv \frac{1}{\left(1+ p^{2} \right)} \left[ \frac{p^{2}}{\sqrt{-\left(p^{2} + m^2 \right) }  } \mathrm{ln} \sqrt{\frac{\sqrt{1-m^2} + \sqrt{- \left(p^{2} + m^2 \right)} \sin \alpha}{\sqrt{1-m^2} - \sqrt{- \left(p^{2} + m^2 \right)} \sin \alpha}}  +\frac{\mathrm{ln} \left( \tan \alpha + \sec \alpha \right)}{\sqrt{1-m^2}  }\right] , \label{I3sep} 
\end{equation}
where $p^2$ equals ${p_2}^2$ and ${p_3}^2$ for $S_3$ and $S_4$ respectively. Hence, from Eq. \eqref{I1a} the expression of $I_1$ reduces in terms of $S_3$ and $S_4$ for the separatrix trajectories to
\begin{equation}
I_1= -\frac{\sqrt{\mu \left( 1+e \right) \left( 3-e \right)}}{\sqrt{ e \left[ 1+ 2 a^2 \left( -1 + e^2 \right) Q \mu^3 \right]  \left(1-a^2 \right) }} \left[\frac{\left[L a^2  -2 x  r_{+}  \right]S_3}{\left( a^2 \mu - a^2 \mu e -r_{+}\right)} + \frac{\left[-L a^2  +2 x  r_{-}  \right]S_4}{\left( a^2 \mu - a^2 \mu e -r_{-}\right) }  \right]. \label{I1sep}
\end{equation}
Further, Eqs. \eqref{sepred} reduce the integrals $S_5-S_8$, Eqs. (\ref{I5}-\ref{integral1}), to

\begin{equation}
S_5= \frac{1}{\sqrt{1-m^2} \left( m^2 + p_{1}^2\right)^2} \left[ \frac{m^2 \left( m^2 - p_{1}^2 m^2 +2 p_1^2\right) }{\left( 1+p_1^2\right) }\mathrm{ln} \left( \tan \alpha + \sec \alpha \right) + p_{1}^4 S_7 + \right. \nonumber 
\end{equation}
\begin{equation}
  \left. \frac{2 p_{1}^2 m^2 \left(1-m^2 \right) }{\left( 1+p_1^2\right)} \mid s \mid \tan^{-1} \left[ \mid s \mid \sin \alpha \right] \right],  \label{I5sep}
\end{equation}
\begin{equation}
S_6=  \frac{\mathrm{ln} \left( \tan \alpha + \sec \alpha \right)}{\sqrt{1-m^2} \left(1+ p_1^{2} \right) } + \frac{p_{1}^2 \sqrt{1-m^2 } }{ \left( m^2 + p_{1}^2\right) \left( 1+ p_{1}^2\right) } \mid s \mid \tan^{-1} \left[ \mid s \mid \sin \alpha \right], \label{I6sep}
\end{equation}
\begin{equation}
S_7=\frac{1}{2 \left(1- s^2 \right)^2}\left[ \frac{s^4 \sin \alpha \cos^2 \alpha}{ \left( 1- s^2 \sin^2 \alpha \right)  } + 2 \ \mathrm{ln} \left( \tan \alpha + \sec \alpha \right) - s^2 \sin \alpha  + \left( 3-s^2\right) \mid s \mid \tan^{-1} \left( \mid s \mid \sin \alpha \right)  \right], \label{I7sep}
\end{equation}
\begin{equation}
S_8=\frac{2\mu \left( 1-e^2\right)}{\sqrt{C-A+\sqrt{{B}^2-4 AC}}}\mathrm{ln} \left( \tan \alpha + \sec \alpha \right).
\end{equation}
Finally, the expression of $I_2$, Eq. \eqref{I2a}, reduces for the separatrix orbits in terms of $S_3-S_6$ to
\begin{equation}
I_2= \frac{2\sqrt{\left( 1+e \right) \left( 3-e \right)}}{\sqrt{e \mu \left[ 1+ 2 a^2 \left( -1 + e^2 \right) Q \mu^3 \right] }}\left\lbrace \frac{ E  }{\mu \left( 1- e \right)^2   } S_5 + \frac{  a^2 \mu \left(-L a + 2 E r_{-} \right) }{r_{-}\sqrt{ \left(1-a^2 \right) }\left( a^2 \mu -a^2 \mu e -r_{+}\right) } S_3+ \frac{2 E }{\left( 1- e \right) }S_6 \right. \nonumber 
\end{equation}
\begin{equation}
 \left.    + \frac{  a \mu \left(-2 L r_{-}\sqrt{1-a^2} -2 E a r_{-} +L a^2 \right) }{r_{-}\sqrt{\left(1-a^2 \right) }\left( a^2 \mu -a^2 \mu e -r_{-}\right) } S_4 \right\rbrace. \label{I2sep}
\end{equation}
See Table \ref{septable} for the summary of radial integrals for the separatrix trajectories.

\section{Derivations for the consistency check with the previous results}
\label{dervconsistency}
\begin{enumerate}
\item \underline{\textit{Equatorial separatrix orbits}}:
Here, we show the reduction of the separatrix trajectory formulae to the equatorial case. We substitute $Q=0$ in the expression of the azimuthal angle given by Eq. \eqref{phisep}, which yields

\begin{equation}
\phi-\phi_{0}=  \frac{1}{2}  \sqrt{\frac{\mu\left( 1+e \right) \left( 3-e \right)}{e \left(1-a^2 \right)}}\left[\frac{\left[L a^2  -2 x  r_{+}  \right]}{\left( a^2 \mu - a^2 \mu e -r_{+}\right)} I_3+  \frac{\left[-L a^2  +2 x r_{-}  \right]}{\left( a^2 \mu - a^2 \mu e -r_{-}\right) } I_4  \right]. \label{phisepeq}
\end{equation}

We saw that $n^2=0$ for the equatorial orbits and $m^2$ reduces to 1 for separatrix orbits, which gives $\alpha=0$ by definition, given by Eq. \eqref{substn}. Hence, we solve the integrals in the form of $\psi$ variable. First, we solve integrals $I_3$ and $I_4$ for separatrix orbits using
\begin{eqnarray}
I_3=&& \int_{0}^{\psi} \frac{\mathrm{d}\psi}{\left( 1+ p_{2}^2 \sin^2 \psi\right) \cos \psi }, \ \ 
I_4= \int_{0}^{\psi} \frac{\mathrm{d}\psi}{\left( 1+ p_{3}^2 \sin^2 \psi\right) \cos \psi }.
\end{eqnarray}

Using the substitution $p_2 \sin \psi=z$ for $I_3$ and $p_3 \sin \psi =z$ for $I_4$, and applying the method of partial fractions, we find
\begin{subequations}
\begin{eqnarray}
I_3 =&& \frac{1}{\left( 1+p_{2}^2 \right) } \left[ p_2 \arctan \left( p_2 \sin \psi \right) + \mathrm{arctanh} \left( \sin \psi\right) \right], \\
I_4 =&& \frac{1}{\left( 1+p_{3}^2 \right) } \left[ p_3 \arctan \left( p_3 \sin \psi \right) + \mathrm{arctanh} \left( \sin \psi\right) \right].   
\end{eqnarray}
\end{subequations}
Hence, when the above equations for $I_3$ and $I_4$ are substituted into the expressions for azimuthal angle, Eq. \eqref{phisepeq}, yields
\begin{equation}
 \phi-\phi_{0}=  \sqrt{\frac{\mu\left( 1+e \right) \left( 3-e \right)}{e \left(1-a^2 \right)}} \left\lbrace \frac{\left( L a^2 -2 x r_{+}\right) p_2 \cdot \mathrm{arctan} \left( p_2 \sin \psi\right)}{2\left(a^2 \mu + a^2 \mu e - r_{+} \right)}  +  \right. \nonumber 
 \end{equation}
 \begin{equation}
 \left.\frac{\left( -L a^2 +2 x r_{-}\right) p_3 \cdot \mathrm{arctan} \left( p_3 \sin \psi\right)}{2\left(a^2 \mu + a^2 \mu e - r_{-} \right)} + \frac{ \sqrt{1-a^2}\left[L - 2 x \mu \left( 1+ e \right) \right]\mathrm{arctanh} \left( \sin \psi \right)}{\left[ \mu^2 a^2\left( 1+ e \right)^2 -2 \mu \left( 1+ e \right) +1 \right]}   \right\rbrace . \label{phisepeq1}
\end{equation}

Now, plugging in $Q=0$, reduces the expression for coordinate time of the separatrix orbits, Eq. \eqref{tsep}, to 
\begin{equation}
t-t_{0}= \sqrt{\frac{\left( 1+e \right) \left( 3-e \right)}{e \mu }}\left\lbrace \frac{ E  }{\mu \left( 1- e \right)^2   } I_5 + \frac{  a^2 \mu \left(-L a + 2 E r_{-} \right) }{r_{-}\sqrt{ \left(1-a^2 \right) }\left( a^2 \mu -a^2 \mu e -r_{+}\right) } I_3+ \frac{2 E }{\left( 1- e \right) }I_6 \right. \nonumber 
\end{equation}
\begin{equation}
 \left.   + \frac{  a \mu \left(-2 L r_{-}\sqrt{1-a^2} -2 E a r_{-} +L a^2 \right) }{r_{-}\sqrt{\left(1-a^2 \right) }\left( a^2 \mu -a^2 \mu e -r_{-}\right) } I_4 \right\rbrace. \label{tsepeq}
\end{equation}
And similarly, the integrals $I_5$ and $I_6$ reduce to
\begin{subequations}
\begin{eqnarray}
I_5&&=\int_{0}^{\psi} \frac{\mathrm{d}\psi}{\left( 1+ p_{1}^2 \sin^2 \psi\right)^2 \cos \psi } \nonumber \\ 
&&=\frac{1}{\left( 1+p_{1}^2 \right)^2 } \left[ \frac{p_1}{2} \left( 3+ p_1^2\right)  \arctan \left( p_1 \sin \psi \right) + \mathrm{arctanh} \left( \sin \psi\right)  + \frac{p_1^2 \sin \psi \left( 1+p_{1}^2 \right)}{2\left( 1+ p_1^2 \sin^2 \psi\right) }\right], \nonumber \\
\end{eqnarray}
\begin{eqnarray}
I_6 = \int_{0}^{\psi} \frac{\mathrm{d}\psi}{\left( 1+ p_{1}^2 \sin^2 \psi\right) \cos \psi }= \frac{1}{\left( 1+p_{1}^2 \right) } \left[ p_1 \arctan \left( p_1 \sin \psi \right) + \mathrm{arctanh} \left( \sin \psi\right) \right].\nonumber \\
\end{eqnarray}
\end{subequations}
Hence, the substitution of $I_3$, $I_4$, $I_5$, $I_6$ into Eq. \eqref{tsepeq} yields the expression for coordinate time for equatorial separatrix orbits to be
\begin{equation}
t-t_{0}= \sqrt{\frac{\left( 1+e \right) \left( 3-e \right)}{e \mu }}\left\lbrace \frac{E e \sin \psi}{\mu \left( 1- e^2 \right) \left ( 1-e + 2 e \sin^2 \psi \right)} +\frac{E \left[ 3 -e + 4 \mu \left( 1- e^2\right)\right] p_1 \cdot \mathrm{arctan}\left( p_1 \sin \psi\right)  }{2 \mu \left( 1- e^2 \right)\left( 1+e \right)} \right. \nonumber
\end{equation}
\begin{equation}
+\frac{a \mu}{r_{-}\sqrt{1-a^2}}\left[ \frac{\left(-L a^2 + 2 E a r_{-} \right) p_{2}\cdot \mathrm{arctan} \left(p_2\sin \psi \right)}{\left(a^2 \mu + a^2 \mu e - r_{+} \right) } + \frac{\splitfrac{\left(-2 L r_{-}\sqrt{1-a^2}- 2 E a r_{-} + L a^2 \right)\cdot}{ p_3 \cdot \mathrm{arctan} \left( p_3 \sin \psi \right)}}{\left(a^2 \mu + a^2 \mu e - r_{-} \right) }\right] \nonumber
\end{equation}
\begin{equation}
\left. +\left[ \frac{E \left[ 1+ 2 \mu \left(1+ e \right)\right] }{\mu \left(1+ e \right)^2 } + \frac{2 \mu \left[ - L a \mu \left(1+ e \right)+ 2 E \right] }{\left[ 1+ \mu^2 a^2 \left(1+ e \right)^2 -2 \mu \left(1+ e \right)\right] }\right] \mathrm{arctanh} \left( \sin \psi \right) \right\rbrace  \label{tsepeq1}
\end{equation}

We have numerically matched these expressions of the azimuthal angle and coordinate time for equatorial separatrix orbits, Eqs. \eqref{phisepeq1} and \eqref{tsepeq1}, with those derived in \cite{Levin2009}.

\item \underline{\textit{Spherical orbits}}:

Now, we verify our frequency formulae with the spherical orbit case. From Eqs. \eqref{nuphi} and \eqref{nutheta}, we have
\begin{equation}
\frac{\nu_{\phi}}{\nu_{\theta}}=\frac{2\left\lbrace \left[ - \displaystyle{\frac{I_1\left(\frac{\pi}{2} , e, \mu, 1, Q \right)}{2I_8\left(\frac{\pi}{2} , e, \mu, 1, Q \right)}} -  L \right] F\left( \frac{\pi}{2},\frac{z_{-}^{2}}{z_{+}^{2}}\right)  +  L \cdot \Pi\left( z_{-}^2, \frac{\pi}{2},\frac{z_{-}^{2}}{z_{+}^{2}}\right)  \right\rbrace}{\pi \sqrt{1-E^2}z_{+} }. \label{ratioour}
\end{equation}
 In \cite{Wilkins1972}, the azimuthal to polar motion frequency ratio, $\nu_{\phi}/\nu_{\theta}$, for maximally rotating black hole, $a=1$, was found to be
\begin{equation}
\frac{\nu_{\phi}}{\nu_{\theta}}=\frac{2\left\lbrace L \cdot \Pi\left( z_{-}^2, \frac{\pi}{2}, \frac{z_{-}^2}{z_{+}^2}\right) + \left(P \Delta^{-1}-E \right)  F\left( \frac{\pi}{2}, \frac{z_{-}^2}{z_{+}^2}\right) \right\rbrace }{\pi\sqrt{1-E^2}z_{+}}, \label{ratioWilkins}
\end{equation}
where $P=\left[ E \left( r^2 +1 \right) -L \right]\left( r-1\right)^2-E$ for $a=1$.

 For the case of spherical orbits, the limit $e \rightarrow 0$ reduces the ratio $\left(I_1/I_8\right)$ to the ratio of their integrands, which yields
 \begin{equation}
 \left[\frac{-I_1}{2I_8}-L\right]=-E + \frac{\left[E\left( r^2 +1 \right) - L\right]}{\left(r-1 \right)^2} = \left(P \Delta^{-1}-E \right);
 \end{equation}
 this reduces $\nu_{\phi}/ \nu_{\theta}$ from, Eq. \eqref{ratioour} to Eq. \eqref{ratioWilkins}, and establishes the consistency of our frequency formulae with the spherical orbits case.
 \end{enumerate}
 
\footnotesize

\end{document}